\documentclass[aps,epj,preprint,superscriptaddress,showpacs]{revtex4} 
\usepackage{graphicx}
\usepackage{fancyhdr}
\usepackage{amssymb} 

\newcommand{\be}{\begin{equation}}
\newcommand{\ee}{\end{equation}}
\newcommand{\bea}{\begin{eqnarray}}
\newcommand{\eea}{\end{eqnarray}}
\newcommand{\ba}{\begin{array}}
\newcommand{\ea}{\end{array}}

\newcommand{\kv}{{{k}}}
\newcommand{\qv}{q}
\newcommand{\Rv}{R}

\begin{document}
\draft
\title{Disordered $d$-wave superconductors with interactions}
\author{Luca Dell'Anna}
\address{International School for Advanced Studies (SISSA),
 I-34014 Trieste, Italy,\\
Dipartimento di Fisica, Universit\`a di Roma "La Sapienza", 
I-00185 Roma, Italy,\\
Max-Planck-Institut f\"ur Festk\"orperforschung, D-70569 
Stuttgart, Germany,\\
Institut f\"ur Theoretische Physik, 
Heinrich-Heine-Universit\"at, D-40225, D\"usseldorf, Germany}
%
%
\begin{abstract}
We study the localization properties of disordered $d$-wave
superconductors by means of the fermionic replica trick method. 
We derive the effective non-linear $\sigma$-model describing the 
diffusive modes related to spin transport which we 
analyze by the Wilson-Polyakov renormalization group. 
A lot of different symmetry classes are considered within the same framework. 
According to the presence or the absence of certain symmetries, 
we provide a detailed classification for the behavior of some physical 
quantities, like the density of states, 
the spin and the quasiparticle charge conductivities.
Following the original Finkel'stein approach, 
we finally extend the effective functional method to 
include residual quasiparticle interactions, at all orders in the scattering amplitudes. We consider both the 
superconducting and the normal phase, 
with and without chiral symmetry, which occurs in the so
called {\it two-sublattice models}.  

\end{abstract}

\maketitle

\section{Introduction}

After the discovery of superconductivity at high temperatures 
in doped cuprate materials \cite{Bednorz}, a lot of efforts have been 
devoted to find a 
theoretical justification of this amazing property and 
to explain the complex phenomenological aspects of these 
``high $T_c$ superconductors''. 

One of the novel features of these cuprates is the presence of 
gapless Landau-Bogoliubov quasiparticle excitations in the superconducting 
phase, due to the 
$d$-wave symmetry of the order parameter. 
This peculiarity makes these materials a suitable play ground for studying 
the role of disorder in gapless superconductors. 

Moreover, since cuprates are essentially two-dimensional, quantum interference 
and localization should strongly affect the quasiparticle transport and 
thermodynamic properties at low temperature 
\cite{e1,e2,e3,e4,e5,e6,e7,e8,e9,e10,e11,e12,e13,e14}. Indeed, starting 
from a $d$-wave BCS Hamiltonian in the presence of disorder a multitude of 
different regimes and crossovers has been predicted from quantum interference, 
depending on the specific symmetry of the quasiparticles and of the disorder 
\cite{Gorkov,hwe,Lee,mp,Balatsky94,Shura0,Shura,Balatsky95,Balatsky96,zhh,mcw,
mcw2,Altland,Joynt,Pepina,Balents,Fisher,FisherDOS,vsf,Fukui2,Lee2,ahm,ahmz,
yash,Pepina2,FDC,atkinson}.
%

The present work, which revisits and extends the analysis 
in Ref.\cite{FDC}, would 
like to contribute to the widely studied but 
somehow still controversial topic of localization effects
in these unconventional superconductors. 
There are several theoretical issues which are still open and motivate our 
investigation. 

The first issue concerns the charge transport. Within a BCS description of
the  superconducting phase, charge is not a conserved quantity due to the 
gauge symmetry breaking.
Therefore, although quasiparticles are gapless for a $d$-wave order parameter, 
nevertheless 
long-wavelength quasiparticle charge fluctuations are not 
diffusive in the presence of disorder, unlike in a normal metal. 
The diffusive modes just 
carry spin or energy, which remain conserved quantities. 
It is well known that, as disorder increases, 
quantum interference may lead to the Anderson localization in a normal metal \cite{Anderson,AALR,Lee&Rama}. 
While it is clear that such a phenomenon may suppress spin and 
thermal conductivities in a $d$-wave superconductor, 
the effects on the quasiparticle charge conductivity are not as settled yet.
For this reason we extend existing quantum field theory approaches built 
to deal with the truly diffusive modes 
in a $d$-wave superconductor to include the charge modes, which acquire 
a mass term by the onset of superconductivity. In this way we are able 
to calculate how the charge 
conductivity is modified by disorder in comparison with spin and thermal 
conductivities. We find the same corrections for all of them in all the 
cases under study but 
one, when time reversal symmetry is broken at half filling, 
although the conductivities are different at the Born level.

A second issue has to do with some controversial results 
about the quasiparticle density of states at the chemical potential in the 
presence of disorder. Within the self consistent T-matrix 
approximation scheme\cite{Lee,Lee2}, 
it was found that the density of states vanishes 
linearly as the Fermi 
energy is approached in the pure system, and acquires instead a finite value 
in the presence of disorder.  
This was later used as the starting point to build up a standard 
field-theoretical 
approach based on the non-linear $\sigma$-model to cope with the quantum 
interference corrections not included within the self-consistent 
T-matrix approximation\cite{Fisher}. This approach predicts localization and, eventually, a vanishing density of states. 

This standard perturbative technique was nevertheless somewhat unsatisfactory. 
It was argued \cite{Shura0,Shura} that systems with nodes in 
the spectrum require a more 
careful analysis. 
 
The quasiparticle spectrum in a $d$-wave superconductors can be 
described by 2-dimensional (2D) Dirac fermions, with conical spectrum. 
In the absence of 
interactions, the 2D quantum problem in the presence of disorder 
becomes effectively a 2D classical problem with the retarded-advanced 
frequency of the single particle Green's function playing 
the role of an external field. 
This is the non-linear $\sigma$-model description \cite{Wegner}.
On the other hand, a classical zero frequency model in 2-dimensions with 
conical spectrum 
is analogous to a quantum problem of Dirac fermions in 1+1 dimension. 
Within this scheme the DOS is found to vanish (with a different  
behavior from the approach above) and no localization is predicted 
\cite{Shura}. In the language of Dirac fermions in 1+1 dimension, 
the disorder average within the replica trick method 
generates an effective interaction among the 
one-dimensional (1D) fermions, with all the complications that are known 
to occur. For instance, translated in the 1D language, the self-consistent 
T-matrix approach which generates a finite density of states at the Fermi 
energy is analogous to the Hartree-Fock approximation, which 
always leads to density-wave order parameters. However, it is known 
that Hartree-Fock fails completely in 1D, which poses serious doubt 
about the validity of the T-matrix approach even as a starting point 
of a perturbative treatment. From this point of view the controversy 
concerns, more deeply, the question of which model correctly describes the 
quantum interference effects. 
Indeed in the peculiar case in which at most pairs of opposite nodes 
are coupled by disorder, the  
perturbation theory around the T-matrix saddle point solution does not contain 
any small parameter, like the inverse conductance in the conventional Anderson localization. 
However, in the most general case of disorder, when all four nodes are 
coupled, we find that a small parameter  
still exists being 
related to the anisotropy of the Dirac cones, 
suggesting a conventional field-theory treatment.
Looking more carefully at the problem we find that the above controversial 
results are related within the non-linear $\sigma$-model to the presence (when 
only opposite nodes are coupled by disorder) or the absence (when all four 
nodes are coupled) of a Wess-Zumino-Witten term  \cite{Wess,PolyWieg,Witten}.

A third interesting issue is the role of the nesting property in these 
kind of systems. 
Given a generic eigenfunction with energy $E$ and amplitude $\phi(i)_{E}$ at 
site $i=(n,m)$, the operator 
\begin{equation}
{\cal O}_\pi \phi(i)_{E}\equiv (-1)^{n+m}\phi(i)_{E},
\label{Opi}
\end{equation}
which shifts the momentum by $(\pi,\pi)$, generates the eigenfunction with 
energy $-E$ if nesting occurs. 
This implies an additional symmetry, which has the form of chiral symmetry, 
at $E=0$, when the two wavefunctions 
$(1\pm(-1)^{n+m})\phi_{E\to 0^+}$, defined on different sublattices, with 
$n+m$ even 
or odd, are both eigenvectors. The nesting property occurs when the operator 
${\cal O}_\pi$ anticommutes with the Hamiltonian, which is possible in models 
in which the Hamiltonian 
contains only terms which couple one sublattice with the other, so called 
{\sl two-sublattice} models. In addition, the chiral 
symmetry further requires half-filling. Both conditions are quite strict and 
do not represent a common physical situation.
Nevertheless chiral symmetry leads to quite different and somehow surprising 
scaling behaviors that are worth to be studied. It was seen, for instance, 
that this symmetry drastically changes the low energy density of states.
Several models presenting a chiral symmetry were found to have isolated 
delocalized states at the band center at low dimensions. 
It was argued \cite{Gade,Gade2} that these models corresponds to a particular 
class of non-linear 
$\sigma$-models and it was shown that quantum corrections to the $\beta$ 
function which controls 
the scaling behavior of conductivity vanish at the band center at all orders 
in the disorder strength, leading to a metallic behavior at that value of 
chemical potential. Moreover, the
$\beta$ function of the density of states was found to be finite, unlike 
in the standard Anderson localization.  
These scaling laws generate a divergent behavior at low energy of the density 
of states.
The anomalous terms in the action, when chiral symmetry holds, 
were found to be 
connected with fluctuations of the staggered density of states \cite{Fabrizio}.
The modes representing these fluctuations are massive in standard non-linear 
$\sigma$-models, while they become diffusive in two-sublattice cases. 
For this reason in the conductance channels with both positive and negative 
frequencies acquire diffusive poles and contribute to 
quantum interferences corrections.
This is what we find also in our two sublattices $d$-wave 
superconductive model that presents extended states at the band 
center which are associated with diffusive spin transport. The DOS diverges and shows a behavior similar to that of \cite{Fabrizio}.
Furthermore, we find an unexpected charge conductance behavior. 
As we have said before, although charge modes in $d$-wave superconductors 
are not diffusive, nevertheless quantum interference corrections affect 
the charge conductance. 
In particular, when chiral symmetry holds but time-reversal symmetry 
is broken, quasiparticle charge conductivity is suppressed, but 
spin and thermal conductivities stay finite,
leading to a spin-metal but charge-insulator quasiparticle behavior. 
Moreover we saw that, even though magnetic fields or magnetic impurities 
introduce on-site terms in the Hamiltonian that spoil the full 
sublattice symmetry, 
staggered fluctuations are not totally suppressed introducing other 
symmetries in the model under study. 
We saw, for instance, that the problem of $d$-wave superconductors with chiral 
symmetry and magnetic impurities can be mapped to a U(2$n$) non-linear 
$\sigma$-model and belongs accidentally to the same universality class 
as $d$-wave superconductors far from the nesting point 
embedded in a constant magnetic field.

From the point of view of the cuprate $d$-wave superconductors, it is not 
unlikely that chiral symmetry may play some role, especially in underdoped 
systems close to the half-filled Mott insulator. Indeed it is believed that 
the impurity potential is close 
to the unitary scattering limit \cite{Pepina} which, by taking out one site, 
reduces to a random 
nearest-neighbor hopping. 
Furthermore, although the band structure does not have a perfect nesting, 
the superexchange interaction which stabilizes a Ne\`el antiferromagnetic 
phase at half-filling may 
effectively reduce the energy scale at which deviations from perfect nesting 
get appreciable.  

The last issue that we will discuss is the role of the residual 
quasiparticle interaction and its effects on the conductivity and on the 
density of states.
Following the original Finkel'stein approach \cite{Finkelstein}, 
which extended the effective 
functional method to the disordered electron-electron interacting systems, 
we introduce the effective 
quasiparticle scattering amplitudes in different channels, firstly considering 
systems without sublattice symmetry. We find that, consistently with the 
charge not being a conserved quantity, 
the singlet particle-hole channel does not contribute while the triplet 
channel does. On the other hand, the scattering amplitude 
in the Cooper particle-particle channel acquires a factor 
1/2 with respect to the normal metal state, which corresponds to the fact that 
only the phase of the order parameter is massless. We see that the effective 
interaction, which we assumed being repulsive, has a delocalizing effect 
enhancing the density of states.
We extend the Finkel'stein model in order to include the nesting property, by 
introducing additional scattering amplitudes with $(\pi, \pi)$ 
momentum transferred, and we evaluate the new corrections to the density 
of states and to the conductivity. We also consider the metallic phase which 
can be of relevance for 2D metals or semimetals. 
Moreover, we notice an interesting fact which occurs at half-filling 
for a two-sublattice model in both the superconducting and normal phase. 
Depending on the sign of the interaction 
the staggered particle-hole fluctuations, 
being diffusive, can lead to a log-divergent staggered susceptibility, 
and a Stoner instability towards a spin or charge density wave.

\section{The model}

The characteristic feature of a $d$-wave superconductor is the existence of
four nodal points where the order parameter vanishes.
To study the low temperature transport properties 
of a $d$-wave superconductor, 
we consider the following model defined on a two-dimensional square lattice 
with lattice constant $a$:
\begin{eqnarray}
{\cal H} &=& -\sum_{\langle ij\rangle} 
\sum_\sigma \left( t_{ij}{\rm e}^{i\phi_{ij}}
c^\dagger_{i\sigma}c^{\phantom{\dagger}}_{j\sigma} + H.c.\right)
+ \sum_{\langle ij\rangle} \left[\Delta_{ij} 
\left(c^\dagger_{i\uparrow}c^\dagger_{j\downarrow} 
+ c^\dagger_{j\uparrow}c^\dagger_{i\downarrow}\right) + H.c.\right],
\label{Hamiltonian}
\end{eqnarray}
where $\langle ij \rangle$ means that the sum is restricted to 
nearest neighbor sites, $c^\dagger_{i\sigma}$ creates an electron with 
spin $\sigma=\uparrow,\downarrow$ at site $i$, while 
$c^{\phantom{\dagger}}_{i\sigma}$ annihilates it. We take a 
gap function $\Delta_{ij}$ of $d$-wave symmetry. 
The hopping 
matrix elements are independent random gaussian 
variables with average value $t$ and 
variance $ut$, and satisfy $t_{ij} = t_{ji}\in {\mathbb{R}}$, 
and $\phi_{ij}=-\phi_{ji}$, with $\phi_{ij}$ zero or finite depending 
whether or not time reversal invariance holds.
The spectrum of the Hamiltonian (\ref{Hamiltonian}) 
possesses a nesting property. This implies an additional symmetry (chiral 
symmetry) at half filling \cite{Fabrizio}. 
The localization properties are quite different 
whether chiral symmetry holds, which corresponds to the Fermi 
energy $E_F=0$ (half-filling), or broken, $E_F\not = 0$. 
In the latter case, the localization properties of   
(\ref{Hamiltonian}) are analogous to models in which   
on-site disorder is present or next-nearest neighbor hopping is 
included, which break chiral symmetry everywhere in the spectrum. 
For this reason while dealing with a bipartite lattice that induces 
an higher degree of symmetry we can reduce the problem to the 
standard $d$-wave case only by introducing an on site term in the 
Hamiltonian, thus spoiling chiral symmetry.

In the absence of randomness the quasiparticle spectrum has four nodes 
at $(\pm k_F,\pm k_F,)$.
In the vicinity of each gap 
node the Fourier transform of $-t_{ji}$, namely $\epsilon_{\kv} 
= -2t\cos(k_x a)-2t\cos(k_y a) $, 
varies linearly perpendicularly to the Fermi surface 
while the Fourier transform 
of $\Delta_{ij}$, that is $\Delta_\kv =2\Delta\left(\cos(k_x a) 
- \cos(k_y a)\right)$, varies linearly parallel to the Fermi surface. 
Let us rotate  
the axes from $k_x, k_y$ to $k_1, k_2$ by $\pi/4$ rotation, 
and define a Fermi velocity, ${\bf v}_{1}$ perpendicular to the Fermi 
surface, and a gap velocity ${\bf v}_{2}$ parallel to the same surface. 
Then, close to the nodes the quasiparticle spectrum is
\begin{equation}
E_{\kv}  \simeq \sqrt{v_{1}^{2} k_{1}^{2} + v_{2}^{2} k_{2}^{2}},
\label{eq:diracspectrum13}
\end{equation}
for nodes along $k_1$ axis and 
\begin{equation}
E_{\kv}  \simeq \sqrt{v_{2}^{2} k_{1}^{2} + v_{1}^{2} k_{2}^{2}},
\label{eq:diracspectrum24}
\end{equation}
for nodes along $k_2$ axis. 
The spectrum, in the vicinity of each gap node takes 
the form of a Dirac cone whose anisotropy is measured by the ratio of the 
two velocities. 
As we will see afterward a strong anisotropy brings in a weak coupling 
regime the non-linear $\sigma$-model representative of our disordered 
system, 
making the perturbation theory and the RG approach suitable tools 
of investigation.

We now analyze the disordered Hamiltonian (\ref{Hamiltonian}) by using the 
replica trick method within the path integral formalism \cite{EL&K}. 
We introduce the vector Grassmann variables  
$c_i$ and $\bar{c}_i$ with components 
$c_{i,\sigma,p,a}$ and 
$\bar{c}_{i,\sigma,p,a}$, where $i$ refers to a lattice site, $\sigma$ to the 
spin, $p=\pm$ is the index of positive ($+\omega$) 
and negative ($-\omega$) frequency components, 
and $a=1,\dots,n$ is the replica index, as well as the 
Nambu spinors 
\be
\label{nambu}
\Psi_i = \frac{1}{\sqrt{2}}
\left(
\begin{array}{c}
\bar{c}_i \\
i\sigma_y c_i\\
\end{array}
\right) \,\,\,\textrm{and }\,\,\,\bar{\Psi}_i=\left[C\Psi_i\right]^t,
\ee
with the charge conjugation matrix 
$C=i\sigma_y \tau_1$. Here and in the following, 
the Pauli 
matrices $\sigma_b$ ($b=x,y,z$) act on the spin components, 
$s_b$ ($b=1,2,3$) on the frequency (retarded/advanced) components, 
and $\tau_b$ ($b=1,2,3$) on the Nambu components $\bar{c}$ and 
$c$. The action corresponding 
to (\ref{Hamiltonian}) is 
\begin{eqnarray}
S &=& \sum_{ij} \bar{\Psi}_i
\left( -t_{ij}{\rm e}^{-i\phi_{ij}\tau_3} 
+ i\Delta_{ij}\tau_2 s_1 -i\delta_{ij} \omega s_3\right)
\Psi_j 
\label{S}
\end{eqnarray}
where the source term $\omega\bar{\Psi}_i s_3\Psi_i$ is introduced in order 
to reproduce positive and negative frequency propagators.
As in the standard Abrikosov-Gorkov-Dzyalozinskii approach \cite{AGD} to 
superconductivity, the gap function couples fermions with   
opposite frequency $\omega$. \\
If magnetic impurities are present, we must add to Eq. (\ref{S}) an 
additional spin-flip scattering term 
\be
-\sum_i \bar{\Psi}_i u_i\tau_3 \vec{\sigma}\cdot \vec{S}_i 
\Psi_i,
\ee
being $\vec{S}_i$ the impurity spin and $u_i$ the corresponding random 
potential. The same term, with 
$\vec{S}_i = \hat{B}$ and $u_i=B$, gives the Zeeman splitting in the 
presence of a constant magnetic 
field $\vec{B}$, which also breaks time-reversal invariance.\\ 
If an on site term is present or if we are far from half filling, we must add 
\be
-\sum_i \mu_i\bar{\Psi}_i \Psi_i,
\ee
which spoils chiral symmetry, $\mu_i$ being a random variable or $E_F$ 
respectively.

\section{Disorder average}

We derive the effective quantum field theory 
for the disordered $d$-wave superconducting model described in the 
previous section, following the work by Efetov, Larkin, 
and Khmel'nitsky \cite{EL&K}. As usually, this field theory is a non-linear 
$\sigma$-model where the broken gauge symmetry enters as a reduction of the 
symmetry of the $Q$-matrix fields with respect to a normal metal. 

The starting generating function is
\be
{\cal Z} = \int {\cal D}\overline{\Psi}
{\cal D}\Psi {\cal D}t P(t)\, {\rm e}^{-S},
\ee
where $P(t)$ is the gaussian probability distribution 
of the random bonds.
Within the replica trick method, we can average e$^{-S}$ over disorder, 
obtaining
\be
{\cal Z} = \int {\cal D}\overline{\Psi}
{\cal D}\Psi \, {\rm e}^{-S_{eff}}.
\ee
The effective action $S_{eff}=S_0+S_{imp}$ is the sum of a regular part
$S_0$ given by Eq.(\ref{S}) with $t_{ij}=t$ (and $\phi_{ij}\equiv 0$)
plus the impurity contribution    
\be
S_{imp} = -2u^2 t^2 \sum_{\langle ij \rangle} 
\left(\bar{\Psi}_i\Psi_j\right)^2 =  -2 u^2t^2\sum_{\langle ij\rangle}  
\left(\overline{\Psi}_{i}\Psi_{j}\right)
\left(\overline{\Psi}_{j}\Psi_{i}\right)
\ee
since $\overline{\Psi}_{i}\Psi_{j}=\overline{\Psi}_{j}\Psi_{i}$. 
By introducing the fields
\be
X^{\alpha\beta}_{i} = \Psi^\alpha_{i} \overline{\Psi}^\beta_{i},
\ee
where $\alpha$ and $\beta$ is a multilabel for Nambu, advanced-retarded and replica components, 
we can write
\be
S_{imp} = 2u^2 t^2 \sum_{\langle ij \rangle} 
X^{\alpha\beta}_{i} X^{\beta\alpha}_{j} 
= 2u^2 t^2 \sum_{\langle ij \rangle} Tr\left(X_i X_j\right).
\ee
In Fourier components it becomes
\begin{equation}
S_{imp} = \frac{1}{V}\sum_{\qv\in BZ} w_\qv Tr\left(X_\qv X_{-\qv}\right),
\label{Simp}
\end{equation}
where $BZ$ means the Brillouin zone, $V$ the volume and
\be 
w_\qv = 2u^2 t^2\left(\cos q_x a + \cos q_y a \right),
\ee
$a$ being the lattice spacing. We can decouple Eq. (\ref{Simp}) by an 
Hubbard-Stratonovich transformation, introducing an auxiliary field. 
However, since $w_\qv=-w_{\qv+(\pi,\pi)}$ and 
$w_\qv>0$ if $\qv$ is restricted to the magnetic Brillouin zone ($MBZ$), we 
need to introduce two auxiliary fields defined within the $MBZ$, 
$Q_{0\qv}=Q_{0-\qv}^\dagger$ and $Q_{3\qv}=Q_{3-\qv}^\dagger$ 
\cite{FDC,Fabrizio}, through which  
\begin{eqnarray}
S_{imp} &=& \frac{1}{V} \sum_{\qv\in MBZ} \frac{1}{4w_\qv} 
Tr\left[Q_{0\qv}Q_{0-\qv}+Q_{3\qv}Q_{3-\qv}\right] \nonumber\\
&-&\frac{i}{V}\sum_{\qv\in MBZ} Tr\left[ Q_{0\qv}X^t_{-\qv} 
+ i Q_{3\qv}X^t_{-\qv-(\pi,\pi)}\right].
\label{Simpurity}
\end{eqnarray}
The above expression shows that $Q_0$ corresponds to smooth 
fluctuations of the auxiliary field, while $Q_3$ to staggered fluctuations. 
Namely, in the long-wavelength limit, the auxiliary field in real space is
\be
Q_j = Q_{0j} + i(-1)^j Q_{3j}, 
\label{Qmatrix}
\ee
where $j$ is the site. 
However, to make sublattice symmetry more evident, it is convenient 
to use a unit cell which contains two sites, one 
for each sublattice. Indicating with $\Rv$ a new unit cell vector and 
with $A$ and $B$ the labels for the two sublattices, we introduce  
a two component field
\be
\Psi_\Rv =\left(
\begin{array}{c}
\Psi_{A\Rv}\\
\Psi_{B\Rv}\\
\end{array}
\right),
\label{PsiR}
\ee
through which we can rewrite Eq. (\ref{Qmatrix}) in the following way
\be
Q_\Rv = Q_{0\Rv}\,\gamma_0 + i Q_{3\Rv}\,\gamma_3,
\ee
where $\gamma_b$ ($b=1,2,3$) are Pauli matrices acting on the vector 
(\ref{PsiR}).
Notice that, differently from the non-chiral case, $Q$ here is not hermitian, 
in fact 
\be 
Q^{\dagger}_\Rv=Q_{0\Rv}\,\gamma_0 - i Q_{3\Rv}\,\gamma_3=\gamma_1 Q_\Rv \gamma_1=\gamma_2 Q_\Rv \gamma_2,
\ee
since $Q_0$ and $Q_3$ are both hermitian.

\section{Symmetries}
\label{sec:Symmetries}
From Eq (\ref{Simpurity}) we can deduce that a unitary transformation acting 
on the spinor $\Psi_R\rightarrow T\Psi_R$ results into a rotation of the 
matrix field
\be 
Q\rightarrow CT^tC^tQT.
\label{Qrotation}
\ee
Therefore, 
as originally done by Wegner \cite{Wegner}, we analyze the symmetries underling
the theory in order to distinguish 
the soft or transverse modes from the massive or 
longitudinal modes, once we will define the vacuum states for the $Q$ matrices 
by the saddle point approximation.
Since the Hamiltonian parameters couple sites of the 
two different sublattices, 
we can consider generally two different global unitary 
transformations, one for sublattice A and another for sublattice B 
\[
\Psi_A = T_A \Psi_A,\;\; \Psi_B = T_B \Psi_B.
\]
For frequency $\omega=0$, 
the action is invariant if
\be
 CT_A^t C^t H_{AB} T_B = H_{AB},
\ee
being $C$ is the charge conjugation matrix $i\sigma_y \tau_1$, that implies
\bea
\nonumber CT_A^tC^tT_B=1,\\
CT_B^tC^tT_A=1,
\label{sy1}
\eea
valid even for non-superconducting states, as well as 
\bea
\nonumber CT_A^tC^t\tau_2s_1T_B=\tau_2s_1,\\
CT_B^tC^t\tau_2s_1T_A=\tau_2s_1,
\label{sy2}
\eea
in the presence of a real superconducting order parameter. 
If time reversal symmetry is broken, namely if the hopping parameter acquires a phase e$^{i\phi_{ij}\tau_3}$, we must further impose that 
\bea
\nonumber CT_A^tC^t\tau_3T_B=\tau_3,\\
CT_B^tC^t\tau_3T_A=\tau_3.
\label{sy3}
\eea
In the presence of a constant magnetic field $B$, which introduces a 
Zeeman term $B_z\tau_3\sigma_z$ in the Hamiltonian, we add the following constrains, 
\bea
\nonumber CT_A^tC^t\tau_3T_B=\tau_3,\,\,\,\,CT_A^tC^t\tau_3\sigma_zT_A=\tau_3\sigma_z,\\
CT_B^tC^t\tau_3T_A=\tau_3,\,\,\,\,CT_B^tC^t\tau_3\sigma_zT_B=\tau_3\sigma_z,
\label{sy4}
\eea
and finally
\bea
\nonumber CT_A^tC^t\tau_3T_A=\tau_3,\,\,\,\,CT_A^tC^t\tau_3\vec{\sigma} T_A=\tau_3\vec{\sigma},\\
CT_B^tC^t\tau_3T_B=\tau_3,\,\,\,\,CT_B^tC^t\tau_3\vec{\sigma} T_B=\tau_3\vec{\sigma},
\label{sy5}
\eea
in the presence of magnetic impurities, represented by the term 
$\vec{S}\cdot\vec{\sigma}\tau_3$ in the Hamiltonian with $\vec{S}$ a 
random vector variable. \\
If we are not at half filling or there are on-site impurities, 
namely if sublattice symmetry does not hold,  
$T_A$ and $T_B$ have to satisfy 
\be
CT_A^tC^tT_A=1,\,\,\,\,CT_B^tC^tT_B=1,
\label{sy7}
\ee
in order to be symmetry transformations.
Finally, a finite frequency $\omega$, which acts as a symmetry breaking field, leads to the additional constrains
\be
CT_A^tC^ts_3T_A=s_3,\,\,\,\,CT_B^tC^ts_3T_B=s_3.\label{sy6}
\ee
All the above conditions imply that 
the unitary matrices $T$ belong to a group G if $\omega=0$ which is 
lowered to a group H at $\omega\not = 0$.\\
The unitary transformations, $T_A$ and $T_B$, can be written as
\be
T_A = \exp\frac{W_0+W_3}{2},\;\;\; 
T_B = \exp\frac{W_0-W_3}{2},
\label{TaTb}
\ee
with antihermitian $W$'s. Charge conjugacy invariance, 
through Eqs. (\ref{sy1}), implies
\bea
\label{seconda}
CW_0^t C^t = -W_0, \\
\label{seconda3}  
CW_3^t C^t = W_3, 
\eea
while the presence of superconducting term leads to the following constrains
\be
\label{terza}
\left[\tau_2 s_1, W_{0(3)}\right] =0.
\ee
If time reversal invariance is broken 
\be
\left[\tau_3, W_{0(3)}\right] =0.
\ee
If magnetic impurities are present, then 
\be
\left[\tau_3\vec{\sigma}, W_{0}\right] =
\left\{\tau_3\vec{\sigma}, W_{3}\right\} =0, 
\ee
while
\be
\left[\tau_3\vec{\sigma}\cdot\vec{B}, W_{0}\right] =
\left\{\tau_3\vec{\sigma}\cdot\vec{B}, W_{3}\right\} =0, 
\ee
in the presence of a constant magnetic field. 
Finally, if chiral symmetry is broken by terms which couple the 
same sublattice, 
we must set $W_3=0$. 
In the presence of finite frequency $\omega$ we have further to impose 
that 
\be
\left[ W_0,s_3\right] = 
\left\{W_3,s_3\right\} =0.
\ee

\section{Saddle Point}

The full action, after the Hubbard-Stratonovich decoupling,
\bea
S = -\sum_{\kv,\qv}
\overline{\Psi}_\kv \left(
 i\omega s_3\delta_{\qv0} 
- H^{0}_{\kv}\delta_{\qv0} +
\frac{i}{V}Q_{-\qv} 
\right)\Psi_{\kv+\qv}
+\frac{1}{V}\sum_\qv \frac{1}{2w_\qv}Tr\left[ Q_{\qv}^{\phantom{\dagger}}
Q_{\qv}^\dagger\right],
\label{FullAction}
\eea
by integrating over the Nambu spinors, transforms into  
\be
S[Q] = \frac{1}{V}\sum_\qv \frac{1}{2w_\qv} 
Tr\left[Q_{\qv}Q_{\qv}^\dagger\right] -
\frac{1}{2}Tr\ln\left[ 
 i{\omega}s_3 - H^{0} 
+i Q \right],
\ee
where $H^0$ is the regular part of the Hamiltonian.
In momentum space it reads
\be
H_\kv = \epsilon_\kv +i\Delta_\kv\tau_2 s_1 =
E_\kv {\rm e}^{2i\theta_\kv\tau_2 s_1},
\ee
where
\be
E_\kv = \sqrt{\epsilon_\kv^2 + \Delta_\kv^2}
\ee
and
\be
\cos 2\theta_\kv = \frac{\epsilon_\kv}{E_\kv}, \;\;\;
\sin 2\theta_\kv = \frac{\Delta_\kv}{E_\kv}.
\ee
Let us look for a saddle point $Q_{sp}\propto\sigma_0$ which has 
a $\tau_0 s_3$ component $\Sigma$ as well as 
a $\tau_2 s_1$ component $F$, both $k$ independent.
Therefore
\be
G^{-1}_\kv = i\omega s_3 -\epsilon_\kv -i \Delta_\kv\tau_2s_1 + i\Sigma s_3
+i F\tau_2 s_1,
\ee
where we introduce explicitly the symmetry breaking term, namely
$\omega s_3$.
We notice that the new pairing order parameter
is $\Delta_\kv - F$, so that, by defining
\be
\tilde{E}_\kv = \sqrt{\epsilon_\kv^2 + (\Delta_\kv-F)^2},
\ee
as well as a modified $\tilde{\theta}_\kv$, we find the self-consistency
equations
\begin{eqnarray*}
\Sigma &=& i\frac{u^2{t}^2}{8}\,\sum_\kv \, Tr\left(G_\kv s_3\right),\\
F &=& i\frac{u^2{t}^2}{8}\,\sum_\kv \, Tr\left(G_\kv \tau_2 s_1\right),
\end{eqnarray*}
where
\[
G_\kv = {\rm e}^{-i\tilde{\theta}_\kv\tau_2 s_1}
\frac{1}{-\tilde{E}_\kv + i(\omega + \Sigma)s_3}
{\rm e}^{-i\tilde{\theta}_\kv\tau_2 s_1}.
\]
Therefore,
\begin{eqnarray}
\Sigma &=& \frac{u^2{t}^2}{2}\left(\Sigma+\omega\right)
\,\sum_\kv \, \frac{1}{\tilde{E}_\kv^2 + \left(\Sigma+\omega\right)^2},\\
F &=& -\frac{u^2{t}^2}{2}\,\sum_\kv \,
\frac{\Delta_\kv - F}{\tilde{E}_\kv^2 + \Sigma^2}
=  \frac{u^2}{2} F \,\sum_\kv \,
\frac{1}{\tilde{E}_\kv^2 + \Sigma^2},
\end{eqnarray}
where the last identity holds for $d$-wave order parameter.
Notice that, for $s$-wave symmetry, these equations coincide with
those found by Abrikosov, Gorkov and Dzyalozinskii \cite{AGD}.
The first equation implies that
\be
\frac{\Sigma}{\Sigma+\omega} =  \frac{u^2{t}^2}{2}
\,\sum_\kv \, \frac{1}{\tilde{E}_\kv^2 + \left(\Sigma+\omega\right)^2},
\ee
which, inserted in the equation for $F$, leads to
\be
\frac{F\,\omega}{\Sigma+\omega}
= 0.
\ee
Being $\omega$ non zero, although infinitesimally small,
this equation has the solution $F=0$. Therefore, only
$\Sigma\not =0$ and such that
\be
1 =  \frac{u^2{t}^2}{2}\,\sum_\kv \, \frac{1}{E_\kv^2 + \Sigma^2}.
\ee
The above self-consistency equation leads to the Born approximation result
\be
\Sigma=\pi u^2 {t}^2 \nu=\frac{\pi}{4}w_0\,\nu \,\propto e^{-{\pi v_{_1}v_{_2}}/{ u^{2}t^{2}}},
\ee
with $\nu$ being the density of states at the chemical potential. Finally one gets the saddle point value
\be
Q_{sp}=\sigma_0\,\tau_0\, s_3\, \Sigma.
\ee

\section{Transverse modes}
\label{sec:transverse}
Now we will consider the transformations that leave the total Hamiltonian 
unchanged while rotating the saddle point, that is to say the transformations 
that allows to move from a vacuum state to another one. 
The degrees of freedom of these transformations are the Goldstone modes 
which are massless and whose number is equal to
$$\dim(\textrm{G/H})=\dim(\textrm{G})-\dim(\textrm{H}),$$
where G is the original symmetry group and H is the symmetry group that 
preserves the vacuum. The coset G/H tells us how many generators are broken. 
Since the saddle point has the same algebraic form of the frequency term in 
the action, the cosets related to Goldstone modes are obtained exactly 
by that transformations which are excluded 
in the lowering of symmetries due to finite frequency. 
In the following we will denote by $T$ only those kind of transformations, 
represented in sublattice space by 
\be
T=e^{\frac{W_0\gamma_0+W_3\gamma_3}{2}},
\ee
where the $\gamma$'s are $2\times 2$ matrices acting on vectors (\ref{PsiR}).
We adopt this exponential form, which is the adjoint representation 
of the group whose transverse modes belong to, because in this way small
fluctuations from the saddle point can be easily written by Taylor expansion.

For each $W_{0,3}$ we can separate the singlet term from the triplet one, 
writing
\be
W = W_S + i\vec{\sigma}\cdot\vec{W}_T,\label{Wspin}
\ee
where the Pauli matrices $\sigma_a$, $a=x,y,z$, act on spin space. 
In addition we rewrite $W$ in $\tau$-components (Nambu space)
\begin{eqnarray}
\label{W_s}
W_S &=& W_{S0}\tau_0 + i\sum_{j=1}^3 W_{Sj}\tau_j,\\ 
\vec{W}_T &=& \vec{W}_{T0}\tau_0 + i\sum_{j=1}^3 \vec{W}_{Tj}\tau_j.
\label{W_t}
\end{eqnarray}
Moreover, for each $\tau$-component, we write ($a=0,1,2,3$), 
\be
W_{S(T)a}=\sum_{\alpha=0}^3 W_{S(T)a\alpha}s_\alpha.
\label{W_a}
\ee
Each component of $W$ in Eq. (\ref{W_a}) is a $n\times n$ matrix 
in replica space. We denote by S and A  the symmetric and the antisymmetric 
matrices in replica space and by R and I the real and imaginary ones. These 
matrices are collected in Appendix A, together with the corresponding cosets.
The transverse components of 
the $W$-fields have to fulfill
\be
\left\{ W_0,s_3\right\} =
\left[W_3,s_3\right] =0.
\ee

\section{Non-linear $\sigma$-model}

Here we derive the effective field theory describing the
long wavelength transverse fluctuations of $Q(\Rv)$ around the saddle
point. In general terms we may parametrize the $Q$-matrix 
as follows (cfr Eq (\ref{Qrotation}))
\be
Q_P(\Rv) = \tilde{T}(\Rv)^{\dagger} \left[Q_{sp} + P(\Rv)\right] T(\Rv) \equiv
Q(\Rv) + \tilde{T}(\Rv)^{\dagger} P(\Rv) T(\Rv),
\label{QPR}
\ee
where $T$ involves transverse massless fluctuations and $P$ 
longitudinal massive ones, $Q_{sp}=\Sigma\,s_3$ being the saddle point. 
In Eq (\ref{QPR}), we used the short notation 
\be
\tilde{T}^{\dagger}\equiv CT^tC^t =\gamma_1 T^{\dagger}\gamma_1=\gamma_2 T^{\dagger}\gamma_2.
\ee
Since only the $T$'s are diffusive, at the moment we concentrate just on them, 
neglecting the $P$'s and writing the action in terms of 
$Q(\Rv) = \tilde{T}(\Rv)^{\dagger}Q_{sp}T(\Rv)$ alone, so that 
$QQ^\dagger=Q_{sp}$ is invariant, eventhough a term involving 
massless modes from integration over massive ones
might appear. Afterwards we will reconsider this point. 
By integrating (\ref{FullAction}) over the Grassmann
variables, we obtain the following action of $Q$:
\be
-S[Q] = -\frac{1}{V}\sum_\qv \frac{1}{2w_\qv}
Tr\left[Q_{\qv}Q_{\qv}^\dagger\right] +
\frac{1}{2}Tr\ln\left[
i{\omega}s_3 - H^{(0)}+i Q \right].
\ee
Since global transformations leave invariant the action with $\omega=0$, 
the effective theory in terms of $Q$-fields 
can be found by expanding to the second 
order in the gradients and to the first order in $\omega$. To this end we  
rewrite the second term of $S[Q]$ as
\bea
&&\frac{1}{2} Tr\ln\left(
i{\omega}\tilde{T}s_3T^\dagger
- \tilde{T}H^{(0)}T^\dagger + iQ_{sp} \right).
\label{intermedio}
\eea
$H^{(0)}_{\Rv \Rv'}$ involves either $\gamma_1$ and $\gamma_2$,
while $T$ involves $\gamma_0$ and $\gamma_3$. Then
\begin{eqnarray}
\label{eq:gradexpansion}
\tilde{T}(\Rv)H^{0}_{\Rv \Rv^{\prime}}T(\Rv^{\prime})^\dagger &=& H^{0}_{\Rv 
\Rv^{\prime}}
+\left(\tilde{T}(\Rv^{\prime})^\dagger-\tilde{T}(\Rv)^\dagger\right)H^{0}_{\Rv
  \Rv^{%
\prime}} \\
\nonumber&\simeq& H^{0}_{\Rv \Rv^{\prime}} -\tilde{T}(\Rv)\vec{\nabla}
\tilde{T}%
(\Rv)^\dagger\cdot \left({\vec{R}}-{\vec{R}^{\prime}}\right) H^{0}_{\Rv 
\Rv^{\prime}} \\
\nonumber&+&\frac{1}{2}\sum_{ij}\tilde{T}(\Rv)\partial_{ij}
\tilde{T}(\Rv)^\dagger
(R_i-R^{\prime}_i)(R_j-R^{\prime}_j)H^{0}_{\Rv \Rv^{\prime}}\equiv 
H^{0}_{\Rv \Rv^{\prime}}+U_{\Rv\Rv^{\prime}},
\end{eqnarray}
with $\partial_{ij}=\frac{\partial^2}{\partial R_{i}\partial R_{j}}$, $R_i$ being the components of ${\vec{R}}$. 
Differently from the tight-binding Hamiltonian case \cite{Fabrizio}, in which  
the term $(\vec{R}-\vec{R}^{\prime}) H^{0}_{\Rv\Rv^{\prime}}$ is related to 
the charge current vertex, in BCS Hamiltonian, since the charge is not  
a conserved quantity,   
the spin current vertex is involved, since the spin is conserved. 
This can be seen writing the continuity equation 
\be
i\nabla {\vec{J}}(\Rv)=[H^{0},\rho_{spin}(\Rv)],
\ee
with $\rho_{spin}(\Rv)=c^{\dagger}_{\Rv\uparrow}c_{\Rv\uparrow}-
c^{\dagger}_{\Rv\downarrow}c_{\Rv\downarrow}$, 
from which we obtain the following expression for 
spin current on the basis (\ref{nambu})
\be
{\vec{J}}(\Rv)=-i\sum_{\Rv_1}({\vec{R}}-{\vec{R}}_1)\bar{\Psi}_\Rv
H^{0}_{\Rv\Rv_1}\sigma_{z}\Psi_{\Rv_1}.
\ee
We have chosen $z$ as the spin quantization direction.
Using Eq. (\ref{eq:gradexpansion}) the term (\ref{intermedio}) in the action 
can be written as
\begin{eqnarray}
&&\frac{1}{2} Tr\ln\left(
i{\omega}\tilde{T}s_3T^\dagger
- U - H^{(0)} + iQ_{sp}\right)\nonumber \\
&=&  -\frac{1}{2}Tr\ln G +
\frac{1}{2} Tr\ln\left(1 
+ G\,i{\omega}\tilde{T}s_3T^\dagger
-G\, U \right),
\label{43bis}
\eea
where $G=(-H^{(0)}+iQ_{sp})^{-1}$ is the saddle point Green's function. 
By expanding 
 in $\omega$ the following term is found
\be
i\frac{\omega}{4}Tr\left(G\,\tilde{T}\hat{s}T^\dagger\right)
=\frac{\omega}{2w_0}Tr\left(s_3 Q\right).
\label{Exp:w}
\ee
The second order expansion in $U$ contains the terms:
\be
\label{Act:first}
-\frac{1}{2}Tr\left(G\,U\right)
\ee
and
\be
-\frac{1}{4}Tr\left(G\, U\, G\, U\right).
\label{Act:second}
\ee
Neglecting boundary terms coming from first derivatives 
and keeping in (\ref{Act:first}) the component of $U$ containing second
derivatives, we get
\be
-\frac{1}{4}Tr\left\{\tilde{T}(\Rv)\partial_{ij}\tilde{T}(\Rv)^{-1}
\left(R_i-R'_i\right)\left(R_j-R'_j\right)H^{(0)}_{\Rv\Rv'}G(\Rv',\Rv)\right\}.
\label{trUG}
\ee
Now let us consider the correlation function
\be
\chi_{\mu,i}(\Rv,\Rv';t,t_1,t_2)
=
\langle T\left[ c^\dagger_{\Rv_1}(t) J^\mu_{\Rv_1,\Rv_2}(\Rv)c^{\phantom{\dagger}}_{\Rv_2}(t)
c^\dagger_{\Rv_3}(t_1) J^i_{\Rv_3,\Rv_4}(\Rv')c^{\phantom{\dagger}}_{\Rv_4}(t_2)\right]\rangle,
\label{B4}
\ee
where $\mu=0,1,2$, and
$$J^0_{\Rv_1\Rv_2}(\Rv)=\delta_{\Rv\Rv_1}\delta_{\Rv\Rv_2}\sigma_{z}$$ 
the spin density matrix elements, while $J^i$, for $i=1,2$, are the matrix
element components of the spin current and read
\be
{\vec{J}}_{\Rv_1\Rv_2}(\Rv)=-i(%
\vec{R}_1-\vec{R}_2)H^{0}_{\Rv_1\Rv_2} \sigma_{z}\delta_{\Rv\Rv_2}.
\ee
By means of the continuity equation  
in the hydrodynamic limit we obtain the Ward identity
\be
\sum_{\Rv\Rv'} \chi_{j,i}(\Rv,\Rv';E)
= \sum_{\Rv\Rv'} \left(R_i-R'_i\right)\left(R_j-R'_j\right)
Tr\left( G(\Rv,\Rv';E)H^0_{\Rv',\Rv} \right).
\label{Ward3}
\ee
Through Eq (\ref{Ward3}), Eq. (\ref{trUG}) turns out to be
\be
-\frac{\chi^{++}_{ij}}{16}
Tr\left\{\tilde{T}(\Rv)\partial_{ij}\tilde{T}(\Rv)^{-1} \right\},
\ee
which, integrating by part, is also equal to
\bea
&-&\frac{\chi^{++}_{ij}}{16}
Tr\left\{\tilde{T}(\Rv)\partial_{i}\tilde{T}(\Rv)^{-1}
\tilde{T}(\Rv)\partial_{j}\tilde{T}(\Rv)^{-1}\right\}\nonumber\\
&=&-\frac{1}{16}\chi^{++}_{ij} Tr\left( D_i D_j \right).
\label{Act:diama1}
\eea
Here we have introduced the matrix ${\vec{D}}(\Rv)$ with the $i$-th component
\be
D_i(\Rv)
= D_{0,i}(\Rv)\gamma_0 + D_{3,i}(\Rv)\gamma_3 \equiv
\tilde{T}(\Rv)\partial_i \tilde{T}(\Rv)^{-1}.
\label{DR}
\ee
The second term of the expansion in $U$, given in Eq. (\ref{Act:second}), is
\be
-\frac{1}{4}Tr\left(G\,U\,G\,U\right) = \frac{1}{4}\sum_{\kv}\sum_\Rv \,
Tr\left\{{\vec D}(\Rv)\cdot {\vec J}_{\kv}\,\sigma_z G(\kv)\,{\vec D}(\Rv)\cdot
  {\vec J}_\kv\,\sigma_z G(\kv)\right\}.
\label{gugu}
\ee
Since the Fourier components of the Green's function and of the spin 
current operator in the long-wavelength limit, 
supposing $t_{ij}=t_{ji}\in \mathbb{R}$ and 
$\Delta_{ij}=\Delta_{ji}\in \mathbb{R}$, can be written in the following 
way  
\bea
\label{77}
&&G(\kv)=-\frac{\epsilon_\kv\gamma_1+ i\Delta_\kv \tau_2 s_1\gamma_1+i\Sigma
  s_3}{E_\kv^2+\Sigma^2},\\
&&{\vec J}_{\kv}=\vec{\nabla}_\kv\epsilon_\kv\gamma_1\sigma_z+ i\vec{\nabla}_\kv\Delta_\kv \tau_2 s_1\gamma_1\sigma_z,
\label{78}
\eea
and the positive and negative frequency Green's functions are related by
$G^+=s_1G^-s_1$, we have
\be
J^{i}_{\kv}\sigma_z G^+\,D_j=\frac{1}{2}(D_j+\gamma_1 s_3 D_j s_3 \gamma_1)J^{i}_{\kv}\sigma_z G^{+}+\frac{1%
}{2}(D_j-\gamma_1 s_3 D_j s_3 \gamma_1)J^{i}_{\kv}\sigma_z G^{-},
\ee
$J^{i}_{\kv}$ being a component of ${\vec J}_\kv$. 
Taking advantage of this result, Eq. (\ref{gugu}) becomes 
\begin{eqnarray}
\frac{1}{32}%
\chi^{++}_{ij}Tr\left[ D_i D_j +D_is_3\gamma_1 D_j s_3\gamma_1\right] 
+\frac{1}{32}\chi^{+-}_{ij}Tr\left[ D_i D_j - D_is_3\gamma_1 D_j
s_3\gamma_1\right],
\end{eqnarray}
which, summed to Eq. (\ref{Act:diama1}), gives
\be
\frac{1}{32}\,
{(\chi^{+-}_{ij}-\chi^{++}_{ij})}
{Tr[D_i D_j-D_is_3\gamma_1 D_j s_3\gamma_1]}
=-\frac{\pi\sigma_{ij}}{32\Sigma^2}%
Tr(\partial_i Q \partial_j Q^{\dagger}),
\label{sigmamodel}
\ee
where 
\be
\label{sigmaij}
\sigma_{ij}=-\frac{1}{4\pi}\sum_{\kv}Tr\left[J^{i}_{\kv}\left(G^{+}(\kv)-G^{-}
(\kv)\right)J^{j}_{\kv}\left(G^{+}(\kv)-G^{-}(\kv)\right)\right]
\ee
is the spin conductivity since ${\vec J}_\kv$ is the spin current vertex. 
For $i=j$, using Eqs. (\ref{77},\ref{78}), we obtain the following quantity
\be
\sigma = \frac{\Sigma^2}{\pi V}\sum_\kv \left[\frac{
\vec{\nabla}\epsilon_\kv\cdot \vec{\nabla}\epsilon_\kv
+ \vec{\nabla}\Delta_\kv\cdot \vec{\nabla}\Delta_\kv }
{\left(E_\kv^2+\Sigma^2\right)^2}\right]\simeq \frac{1}{\pi^2}
\frac{v_1^2+v_2^2}{v_1v_2},
\label{sigma}
\ee
where, as previously defined, $v_1$ and $v_2$ are the velocities 
perpendicular and parallel to the 
Fermi surface. This is the quasiparticle conductivity in the Drude 
approximation \cite{Lee,Lee2} 
which corresponds to the spin and the thermal conductivities since both 
energy and spin are conserved quantities, therefore fluctuations of energy and 
spin densities are diffusive. One can expect that in the limit in which 
pair order parameter goes to zero $\sigma$ corresponds also to the charge 
quasiparticle conductivity in Born approximation since in that case 
the charge turns to be a conserved quantity as well. This is what we will find
rigorously in Section IX.\\
Inserting Eq. (\ref{sigma}) in Eq. (\ref{sigmamodel}) and adding Eq. (\ref{Exp:w}), we arrive finally at the following non-linear $\sigma$-model action
\bea
\label{Salpha}
S[Q] = \frac{\pi}{32\Sigma^2}\sigma
\int d\Rv\, Tr\left(\partial_\mu Q(\Rv)\, \alpha_{\mu\nu}\,\partial_\nu Q^{\dagger}(\Rv)\right)
- \frac{\omega}{w_0}Tr\left(s_3 Q(\Rv)\right).
\eea
A particular metric $\alpha_{\mu\nu}$ appears, where $\mu,\nu=1,2$ denotes the 
directions $k_1$ and $k_2$, depending on which nodal contribution we include 
in Eq. (\ref{Salpha}) through Eq. (\ref{sigmaij}). Specifically
$\alpha_{\mu\nu}=\delta_{\mu\nu}$ for 4 nodes which is the model we have 
considered starting from Eq. (\ref{Hamiltonian}). If we assume to suitably 
modify the disorder such that it couples quasiparticles belonging to only one 
node or two opposite nodes, namely if we consider only forward scattering in 
the presence of a random potential or including also backscattering processes, 
the metric to be considered is $\alpha_{\mu\nu}=\delta_{\mu\nu}\frac{1}{2}\frac{v^2_\nu}{v_1^2+v_2^2}$ for one node or $\alpha_{\mu\nu}=\delta_{\mu\nu}\frac{v^2_\nu}{v_1^2+v_2^2}$ for two opposite nodes.

At this point it is important to discuss the differences which occur 
when disorder couples at most two opposite nodes, or 
the most generic case where all nodes are coupled together.  
We anticipate that the logarithmic terms, which appear upon 
integrating the gaussian propagator, derive from the expression  
\be
\frac{1}{\pi \sigma}\int \frac{d^2k}{4\pi^2} \frac{1}{k_\mu 
\alpha_{\mu\nu}k_\nu}\equiv g \ln(\dots), 
\ee
where the effective coupling constant controlling the 
perturbative expansion is gives by 
\bea
&&g=\frac{1}{2\pi^2\sigma}, \phantom{\frac{v_1^2+v_2^2}{2v_1v_2}}\,\,\,\,\,\,\,\textrm{for 4 nodes} ,\\
&&g=\frac{1}{\pi^2\sigma}\frac{v_1^2+v_2^2}{v_1v_2}, \,\,\,\,\,\,\,\,\,\,\textrm{for 1 node}.
\eea 
We readily see that, up to terms of order $u^4$, the disorder 
strength, 
for one node $g=1$ or two opposite nodes $g=0.5$, so that the system is never 
in a weak coupling regime and the saddle point physics with weak perturbative 
quantum interference effects is never realized not even as a crossover 
transient.
This is the situation analyzed in 
Ref. \cite{Shura} in which the authors found a vanishing density of states
approaching the chemical potential with a disorder strength dependent exponent
in the one node case while a universal exponent is found when two opposite 
nodes are coupled by the disorder. 
In Ref. \cite{Fukui1,Fendley,ASZ,Bocquet,Michele} it is pointed out 
that this results depend on the presence of a Wess-Zumino-Witten term. 
On the other hand, assuming for cuprates the generic 4-nodes case, since 
experimentally $v_2\simeq  v_1/15$, $g<<1$ and a perturbative 
expansion in $g$ is still meaningful. In the following we will consider 
the latter situation.   


The full expression of the $Q$-matrix is expressed by Eq. (\ref{QPR})
where the massive modes are
\be
P = \left(P_{00}s_0 + P_{03}s_3\right)\gamma_0
+ i\left(P_{31}s_1 + P_{32}s_2\right)\gamma_3,
\label{PR}
\ee
being all $P$'s a hermitian. Charge conjugation implies that
$
CP^tC^t = P
$.
Writing the free action of $Q_P(\Rv)$ and expanding $w_\qv$ 
we obtain two contributions,
a pure massive term and a term where 
massive and massless modes are mixed. Integrating over massive modes 
as in Ref.\cite{Fabrizio} we find another term which we have to add to the 
action for the transverse fluctuations
\be
-\frac{\pi}{8\cdot 32  \Sigma^4} \Pi
\int d\Rv\,
Tr\left[Q^\dagger(\Rv)\vec{\nabla}Q(\Rv)\gamma_3\right]
 Tr\left[Q^\dagger(\Rv)\vec{\nabla}Q(\Rv)\gamma_3\right].
\label{Anomalo}
\ee
$\Pi$ is a parameter related to staggered density of states fluctuations 
\cite{Fabrizio}. This term exists only if chiral symmetry is present.
Indeed a further term could appear, 
namely the Wess-Zumino-Witten term \cite{Fukui1,Fendley,ASZ,Bocquet,Michele}, 
which is calculated in detail in 
Ref.\cite{Michele} within the same approach.  
It is found that this term accidentally cancels out thanks to the 
four-fold symmetry of the Dirac nodes. Notice that 
if such a symmetry of the 
spectrum is broken the WZW would appear again and we 
would take it into account obtaining very different scaling behaviors. 
However, normally, the system flows to the strong 
coupling regime \cite{Fukui1,Fendley,ASZ,Bocquet,Michele}, 
where all the nodes are locked and again the WZW term is canceled.

\section{Renormalization group}
We now study the scaling behavior of our action 
by means of the Wilson-Polyakov renormalization 
group \cite{Wilson,Polyakov}.
We will also show how to evaluate the one loop 
correction to the conductivity, namely to the stiffness parameter of 
modes which acquire a mass, like the charge fluctuation inside 
the superconducting broken symmetry phase or the spin modes when spin 
isotropy is broken. 
According to the previous section 
the final expression of the action describing the
transverse massless modes in the long-wavelength limit is \cite{FDC}
\bea
&&S[Q] = \frac{\pi}{32}\sigma
\int d\Rv\, Tr\left(\vec{\nabla}Q(\Rv) \cdot\vec{\nabla} Q(\Rv)^\dagger\right)
- \frac{\omega}{w_0}\int d\Rv\,Tr\left(s_3 Q(\Rv)\right)\nonumber\\
&&-\frac{\pi}{8\cdot 32}\Pi
\int d\Rv\,
Tr\left[Q^\dagger(\Rv)\vec{\nabla}Q(\Rv)\gamma_3\right]
 Tr\left[Q^\dagger(\Rv)\vec{\nabla}Q(\Rv)\gamma_3\right],
\label{NLsM}
\eea
where we have rescaled the $Q$ matrices by a factor $\Sigma$ so that the new
$Q$ is normalized to unity, i.e. $Q^\dagger Q=I$ and $Q_{sp}=s_3$. 
Since $Q=\tilde{T}^\dagger s_3 T = s_3 T^2= s_3 e^{W}$,
at the gaussian level the first term in the action is simply 
\be
\frac{\pi\sigma}{32}\int d\Rv\;
Tr\left({\vec \nabla} Q^\dagger{\vec \nabla} Q\right)\simeq
-\frac{\pi \sigma}{32}\int d\Rv\;
Tr\left({\vec \nabla} W {\vec \nabla} W\right),
\ee
while the last term is
\bea
\nonumber&&-\frac{\pi \Pi}{32\cdot 8 }
\int d\Rv\,
Tr\left[Q^\dagger\vec{\nabla}Q\gamma_3\right] 
Tr\left[Q^\dagger\vec{\nabla}Q\gamma_3\right]\\
&&\simeq - \frac{\pi \Pi}{64}
\int d\Rv\,
Tr\left[\vec{\nabla}W_3\right]
 Tr\left[\vec{\nabla}W_3\right].
\label{anomal}
\eea 
Depending on whether the $W$ components defined by 
Eqs. (\ref{Wspin}-\ref{W_a}) are 
real or imaginary matrices in replica space,
which may be either symmetric or antisymmetric (see Appendix A),
we find the following 
gaussian propagators for the diffusive modes
\be
\langle W^{ab}_{q\,{\cal S} \,i\,\alpha}(\kv) W^{cd}_{q\,{\cal S}\, i\, \alpha}(-\kv) \rangle = \pm D(\kv)\left( \delta_{ac}\,\delta_{bd}
\pm \delta_{ad}\,\delta_{bc}\right)+D(\kv)\frac{\Pi}{\sigma+\Pi
n}\,\delta_{q 3}\,\delta_{{\cal S} S}\,\delta_{i 0}\,\delta_{\alpha 0}\,\delta_{ab}\,\delta_{cd},
\ee
where $q=0,3$ refers to the homogeneous or the staggered component, 
${\cal S}=S,T$ for singlet or triplet term, (\ref{Wspin}), 
$i=0,1,2,3$ refers to the $\tau$-components, (\ref{W_s}),(\ref{W_t}), 
$\alpha=0,1,2,3$ refers to the $s$-components, (\ref{W_a}), $a,b,c,d$ 
are replica indexes,
the $\pm$ sign in front refers to real (R) and imaginary (I) matrices,
while the $\pm$ sign inside the brackets refers to symmetric (S)
or antisymmetric (A) matrices, 
$n$ is the number of replicas and finally
\be
D(\kv)=\frac{1}{2\pi\sigma}\frac{1}{k^2}.
\label{D(k)}
\ee
This propagator in two-dimensions will induce logarithmic singularities 
within the perturbative expansion. A standard way to handle those 
divergences is provided by the Renormalization Group (RG). In 
particular we here  
apply the Wilson-Polyakov RG procedure 
\cite{Wilson,Polyakov,Amit}, which is particularly suitable 
to handle with the non-linear constraint $QQ^\dagger = I$. 
By this approach one assumes
\[
T(\Rv) = T_f(\Rv)T_s(\Rv),
\]
where $T_f$ involves fast modes with momentum $q\in [\Lambda/s,\Lambda]$,
while $T_s$ involves slow modes with $q\in [0,\Lambda/s]$,
being $\Lambda$ the high momentum cut-off, and $s>1$ the rescaling factor
(in terms of the elastic scattering time, $\tau= \frac{1}{2\Sigma}$, $\Lambda$ is given by $\frac{\tau\sigma}{2\nu}\Lambda^2\sim 1$, 
$\frac{\sigma}{2\nu}$ being the diffusion constant at the Born level).
The following equalities hold
\bea
&& Tr\left[\vec{\nabla}Q^\dagger \vec{\nabla} Q\right] =
Tr\left[\vec{\nabla}Q_f^\dagger \cdot \vec{\nabla} Q_f\right]
\nonumber\\
&& +2Tr\left[\vec{D}_s\gamma_1 Q_f \vec{D}_s Q_f^\dagger \gamma_1\right]
-2 Tr\left[\vec{D}_s \vec{D}_s \right]\nonumber\\
&&+4Tr\left[\vec{D}_s Q_f^\dagger \vec{\nabla} Q_f \right],
\label{RG:dQdQ}
\eea
where $Q_f = \tilde{T}_f^\dagger Q_{sp} T_f$ and
$\vec{D}_s = T_s\vec{\nabla}T_s^\dagger$, as well as 
\bea
&&Tr\left[Q^\dagger \vec{\nabla} Q\gamma_3\right]
\cdot Tr\left[Q^\dagger \vec{\nabla} Q\gamma_3\right] \nonumber\\
&&=
Tr\left[\left(\vec{\nabla}W_s + \vec{\nabla} W_f\right)\gamma_3\right]
\cdot Tr\left[\left(\vec{\nabla}W_s + \vec{\nabla} W_f\right)\gamma_3\right].
\label{RG:QdQ}
\eea
Since the fast and slow modes live in disconnected regions of momentum
space, only the stiffness 
$Tr\left[\vec{\nabla}Q^\dagger \vec{\nabla} Q\right]$ generates corrections.
By expanding in Eq. (\ref{RG:dQdQ}) the terms which 
couple slow and fast modes up to second order in $W_f$, after averaging over
the fast modes within a one loop expansion, the stiffness
generates the following contribution to the action for the slow modes 
\be
\frac{\pi\sigma}{32} \int d\Rv\,
Tr\left[ \vec{\nabla}Q_s^\dagger \vec{\nabla}Q_s \right]
+ \langle S_1\rangle_f - \frac{1}{2} \langle S_2^2 \rangle_f,
\ee
where
\bea
S_1 &=& \frac{\pi\sigma}{32} \int d\Rv\,
2 Tr\left[ {\vec D}\gamma_1 Q_{sp} {\vec D} Q_{sp}
{W}_f^2
\gamma_1\right]
\nonumber\\
&&-2 Tr\left[ {\vec D}\gamma_1 Q_{sp} W_f {\vec D} W_f Q_{sp}
\gamma_1\right]
\label{F1}
\eea
and 
\be
S_2 = 4 \frac{\pi\sigma}{32}
\int d\Rv\, Tr\left[{\vec D}W_f\vec{\nabla}W_f\right].
\label{F2}
\ee
After the effective action for the slow modes have been obtained, 
the slow modes momenta are then rescaled back according to :
\[
q\in \left[0,\frac{\Lambda}{s}\right] 
\rightarrow \frac{q'}{s},
\]
where $q'\in [0,\Lambda]$ runs over the original momentum space. In this 
way the model is mapped onto another model defined onto the same range of 
momenta with renormalized parameters $\sigma(s)$ and $\Pi(s)$. 
The logarithmic singularities are controlled by the 
following dimensionless coupling constants
\be
\label{def_g&c}
g=\frac{1}{2\pi^2\sigma},
\,\,\,\,\,\,c=\frac{1}{2\pi^2\Pi}.
\ee
When chiral symmetry holds, namely when $W^3$ is massless, the new coupling 
$c$ has to be included. However, the combination $\sigma+n\Pi$ can be shown 
\cite{Gade,Fabrizio} to represent the stiffness parameter of an  
abelian degree of freedom connected to the Tr$(W^3)$, which is 
finite and commutes with all other degrees of freedom. This implies that 
$\sigma+n\Pi$ is a constant of the RG flow, namely that 
\be
\label{beta_c}
\beta_c = -\frac{c^2}{g^2} \frac{\beta_g}{n}.
\ee
Since the theory is well behaved in the $n\to 0$ zero replica limit, this 
indirectly proves that 
\[
\lim_{n\to 0} \beta_g = 0,
\]
namely that when chiral symmetry holds and when Tr$(W^3)$ is massless 
the model stays metallic with a finite conductance.
   
The final results of the RG  are collected in the Table \ref{Table} 
in which the $\beta$ functions for $g$ ($\beta_g=dg/d\ln s$) 
and for the density of states (DOS) ($\beta_\rho=d\ln\rho/d\ln s$) 
are listed for the different symmetry classes.  
The density of states scaling behavior is obtained through  
its expression in the $Q$ matrix language 
\be
\rho=
\frac{\nu}{8}Tr(s_3Q),
\ee
which allows a very simple loop expansion. 

In Table \ref{Table} we also list the coset spaces G/H 
for the different classes
({\sl i}) 
time reversal invariance is preserved with chiral symmetry \cite{FDC,Fukui2} 
or without \cite{Fisher,SQHE,SQHE2};
({\sl ii}) time reversal symmetry is broken by introducing random phase with chiral symmetry \cite{FDC,Fukui2} 
or without \cite{Fisher};
({\sl iii}) a magnetic field is applied in the presence of chiral symmetry  
\cite{FDC} or without it \cite{Fisher,SQHE,SQHE2}; 
and finally ({\sl iv}) in the presence of magnetic impurities 
with chiral symmetry  \cite{FDC}
or in its absence \cite{Altland}. 

\begin{table}[!h]
\caption{Coset spaces and $\beta$ functions for the
coupling $g$ and for the DOS $\rho$ in the different universality classes. 
$\Gamma=\frac{g}{c+n\,g}$ and $\hat{T}$ is the time reversal invariance.}

\begin{tabular}{|l||c|c|c|}
 \hline\hline

~&  Coset space &  $\beta_g$       &   $\beta_\rho$  \\ 
 \hline\hline

Yes chiral, Yes $\hat{T}$ & U$(4n)\times$U$(4n)$/U$(4n)$
& $8ng^2$ & $(\Gamma/4-8n)g$  \\  \hline

Yes chiral, No $\hat{T}$ &  U$(4n)$/O$(4n)$
&         $4ng^2$       &   $(-1+\Gamma/4-4n)g$  \\  \hline

Yes chiral, magnetic field &  O$(4n)$/O$(2n)\times$O$(2n)$ &
$(2n-1)g^2$  & $-2ng$ \\ \hline

Yes chiral, spin flip & U$(2n)$/U$(n)\times$U$(n)$ &
$ng^2$ & $-ng$ \\ \hline

No chiral, Yes $\hat{T}$ & Sp$(2n)\times$Sp$(2n)$/Sp$(2n)$
& $2(2n+1)g^2$ & $(-1-4n)g$   \\ \hline

No chiral, No $\hat{T}$  & Sp$(2n)$/U$(2n)$
&    $(2n+1)g^2$         &   $(-1-2n)g$ \\ \hline

No chiral, magnetic field & U$(2n)$/U$(n)\times$U$(n)$
&      $ng^2$       & $-ng$  \\ \hline

No chiral, spin flip & O$(2n)$/U$(n)$
&   $(n-1)g^2/2$    & $(1-n)g/2$  \\\hline\hline
\end{tabular}
\label{Table}
\end{table}
According to Table \ref{Table}, in the zero replica limit we obtain that,
if chiral symmetry is absent and for non magnetic impurities, the
conductance vanishes,
and the DOS, which is finite within the simplest Born
approximation, is suppressed. As shown by Ref. \cite{FisherDOS},
in the localized phase the DOS vanishes as $|E|$ or
$E^2$ depending whether time reversal symmetry holds or not.
Quite surprisingly, magnetic impurities give a delocalization
correction to the conductance, as well as a DOS enhancement.
On the contrary, if chiral symmetry is present, the conductance stays
finite, or even increases in the presence of a
magnetic field. Without magnetic field and in the 
absence of spin flip scattering,
the DOS, according to the above $\beta$-function, diverges approximately like,
$\rho(E)\sim \frac{1}{E}\exp\left[-A\sqrt{-\ln E}\right]$, with $A>0$ 
a model dependent constant\cite{Gade,Fabrizio}. 
By a real space RG in the strong disorder regime 
\cite{LeDoussal, Damle, Horovitz} and 
through a supersymmetric \cite{Mudry} and replicated \cite{Yamada} 
random gauge field theory approach as well as  
within the standard non-linear $\sigma$-model \cite{Luca}, it has 
been recently argued  
that the correct asymptotic expression of the DOS is instead of the form 
$\rho(E)\sim \frac{1}{E}\exp\left[- A(-\ln E)^{2/3}\right] $.
In Ref. \cite{Luca} the origin of the disagreement is identified 
into the existence of an infinite chain of relevant operators which 
are related to moments of the $Q$-field  
and which are coupled together in the RG equations. 

%

\section{The action with vector potentials}

The quasiparticle charge modes, as well as the spin modes
when magnetic impurities or a magnetic field are present,
are not described by the non-linear $\sigma$-model (\ref{NLsM}),
which only
represents the truly massless diffusion modes. Nevertheless, charge and
spin conductivities,
$\sigma_c$ and $\sigma_s$, respectively,  can be still evaluated through
the stiffness of the
corresponding modes, although they acquire a mass term.
Alternatively, $\sigma_c$ and $\sigma_s$ can be determined by 
second derivatives of the action with respect to 
a source field which couples to the 
charge or to the spin current \cite{Castellani}.

As explained in Appendix B the source field which couples 
to the charge quasiparticle current is the vector potential 
\be
A_c=\lambda^s (A^0 \tau_3s_0+A^1 \tau_3s_1),\label{Ach_sim}
\ee
where $\lambda^s$ is a symmetric matrix in replica space, or, alternatively,
\be
A_c=\lambda^a (A^0 \tau_3\sigma_zs_0+A^1 \tau_3\sigma_zs_1),\label{Ach_ant}
\ee
with $\lambda^a$ an antisymmetric matrix. 
On the other hand the spin vector potential which couples to the 
spin current is given by 
\be
A_s=\lambda^s (A^0 \tau_0\sigma_zs_0+A^1 \tau_0\sigma_zs_1),\label{Aspin_sim}
\ee
or alternatively 
\be
A_s=\lambda^a (A^0 \tau_0 s_0+A^1 \tau_3 s_1).\label{Aspin_ant}
\ee
In the hydrodynamic limit the action in the presence of a vector 
potential acquires a new term which, up to second order is $A$, 
is (see Appendix C) 
\be
S_A=\frac{\pi}{32}\sigma_c Tr\left[ \left(
\nabla Q+i\frac{e}{c}[Q,A_c]\right) \left( \nabla Q^{\dagger }-i\frac{e}{c}[%
A_c,Q^{\dagger }]\right) -(\nabla Q\nabla Q^{\dagger })\right],
\label{SAc}
\ee
for a charge vector potential, where $\sigma_c$ is the bare charge conductivity
\be
\sigma_c =\frac{\Sigma^2}{\pi V}\sum_{k} \left[ \frac{(\nabla
_{k}\epsilon )^{2}}{(E_k^{2}+\Sigma ^{2})^{2}}\right],
\ee
while for the spin case
\be
S_A=\frac{\pi\sigma_s }{32}
Tr\left[ \left( \nabla Q+\frac{i}{2}[Q,A_s]\right) 
\left( \nabla Q^{\dagger }-\frac{i}{2}[A_s,Q^{\dagger }]\right)
-(\nabla Q\nabla Q^{\dagger })\right], 
\label{SAs}
\ee
where $\sigma_s$ is the bare spin conductivity, given by Eq.(\ref{sigma}). 
In the presence of these terms in the action, 
the generating function ${\mathcal{Z}}(A)$ depends now on $A$
\be
{\mathcal{Z}}(A)=\int DQe^{-S-S_A}.\label{ZA}
\ee
The Kubo formula both for charge and for spin conductivities is recovered by
\be
\left( \frac{\partial ^{2} ln{\mathcal{Z}}}{{\partial A^{0}}^{2}}-\frac{%
\partial ^{2} ln{\mathcal{Z}}}{{\partial A^{1}}^{2}}\right) \Bigg|_{A=0}.
\label{dZ}
\ee
We can now calculate, through the one-loop expansion, the corrections to 
the charge and the spin conductivities as responses to the source fields. 
To this end we expand to the second order in $A$  
the generating function 
\[
{\mathcal{Z}}(A)=\int DQ\,e^{-S-S_A}\simeq \int DQ\,e^{-S}\left(
1-S_{A1}-S_{A2}+\frac{1}{2}(S_{A1})^{2}\right),
\]
where $S_{A1}$ and $S_{A2}$ are 
\begin{eqnarray}
&&S_{A1}=i\frac{f}{t}\int dR\,\,Tr\left( \nabla
Q(R)\left[Q^{\dagger }(R),A\right]\right), \\
&&S_{A2}=\frac{f^{2}}{t}\int dR\,\,Tr\left(\left[A\phantom{^|},Q(R)\right]\left[Q^{\dagger
}(R),A\right]\right) 
\end{eqnarray}
and $t=\frac{32}{\pi\sigma }$, $f=\frac{1}{2}$ for the spin case and $t=\frac{32}{\pi\sigma_c}$,  $f=\frac{e}{c}$ for the charge case. 

Taking, for charge conductivity, the gauge (\ref{Ach_sim}), 
we can calculate 
the second derivatives of the generating function, 
$\frac{\partial^2 {\mathcal{Z}}(A)}{\partial {A^0}^2}$ and 
$\frac{\partial^2 {\mathcal{Z}}(A)}{\partial {A^1}^2}$, at one-loop level 
by expanding $Q$ in terms of $W$ and 
averaging with the quantum weight $e^{-S}$. 
The gauge (\ref{Ach_ant}) gives the same results. 
For spin conductivity we take the expression (\ref{Aspin_sim}) 
or alternatively (\ref{Aspin_ant}) as the vector potential. 
Through Eq. (\ref{dZ}) we find 
the one-loop quantum interference corrections 
for charge and for spin conductivity, which  
are summarized in the following Table \ref{Table2}.

\begin{table}[!h]
\caption{One loop corrections to to the spin and charge conductivity in $n=0$ 
replica limit.}

\begin{tabular}{|l||c|c|c|}
 \hline\hline

~&  Coset space &  $\delta \sigma_s/\sigma_s$  &  $\delta\sigma_c/\sigma_c$ \\
 \hline\hline

Yes chiral, Yes $\hat{T}$ & U$(4n)\times$U$(4n)$/U$(4n)$
& $0$ & $0$  \\  \hline

Yes chiral, No $\hat{T}$ &  U$(4n)$/O$(4n)$
&         $0$       &   $-2 g \ln{s}$  \\  \hline

Yes chiral, magnetic field &  O$(4n)$/O$(2n)\times$O$(2n)$ &
$0$  & $0$ \\ \hline

Yes chiral, spin flip & U$(2n)$/U$(n)\times$U$(n)$ &
$0$ & $0$ \\ \hline

No chiral, Yes $\hat{T}$ & Sp$(2n)\times$Sp$(2n)$/Sp$(2n)$
& $-2 g \ln{s}$ & $-2 g \ln{s}$   \\ \hline

No chiral, No $\hat{T}$  & Sp$(2n)$/U$(2n)$
&    $- g \ln{s}$         &   $- g \ln{s}$ \\ \hline

No chiral, magnetic field & U$(2n)$/U$(n)\times$U$(n)$
&      $0$       & $0$  \\ \hline

No chiral, spin flip & O$(2n)$/U$(n)$
&   $g \ln{s}/2$    & $ g \ln{s}/2$  \\\hline\hline
\end{tabular}
\label{Table2}
\end{table}

By this procedure we find that the one-loop 
corrections $\delta \sigma_c/\sigma_c$ and
$\delta \sigma_s/\sigma_s$ coincide with
$\delta \sigma/\sigma$ in the absence of sublattice symmetry.
When sublattice symmetry holds, quasiparticle charge
conductivity may behave differently from spin conductivity, as 
it happens when time reversal symmetry is broken. 
Notice that, according to Table \ref{Table2},
quantum interference corrections 
in the diffusive modes influence also the stiffness of modes which 
are, on the contrary, not diffusive. 

\section{The residual quasiparticle interaction}
So far we have dealt with disorder in d-wave superconductors 
modeled by a BCS Hamiltonian for free Landau-Bogoliubov quasiparticles. 
However strong correlation is a crucial ingredient of the cuprates. 
Therefore it is important to understand the effects of the residual 
quasiparticle interactions even within the superconducting phase. Besides 
producing dephasing scattering processes \cite{khve,bruno}, 
the interactions can renormalize 
both the density of states and the stiffness of the spin fluctuations.
In this Section we extend our analysis to 
include the quasiparticle interactions, following the original work  
by Finkel'stein \cite{Finkelstein,Fin84,Castel84} for disordered metals. 
Moreover, we will extend the Finkel'stein model by including the 
nesting property, which requires to add interaction amplitudes 
with $(\pi,\pi)$ momentum transfer.
Let us first consider the following interaction 
contributions to the free energy 
\be
{T}\sum_{|k|\ll k_F}\,\frac{\Gamma_1}{2} \,\overline{c}^{\alpha}_{n}(p_1)\,\overline{c}^{\beta}_{m}(p_2)\,c^{\beta}_{m-\omega}(p_2-k)\,c^{\alpha}_{n+\omega}(p_1+k),
\ee
\be
{T}\sum_{|k|\ll k_F}\,\frac{\Gamma_2}{2} \,\overline{c}^{\alpha}_{n}(p_1)\,\overline{c}^{\beta}_{m}(p_2)\,c^{\beta}_{n+\omega}(p_1+k)\,c^{\alpha}_{m-\omega}(p_2-k),
\ee
with $\alpha$ and $\beta$ spin indices and $n$, $m$ and $\omega$ 
Matsubara indices  
while $p_1$, $p_2$ $k$ are the momenta involved. $T$ is the temperature. 
The sum is performed 
both over momenta and frequencies. Indeed since interactions intermix the 
energies of the particles we should give up to a fixed energy description, 
used in the non-interacting case, introducing a discrete set of Matsubara 
frequencies. 

\begin{figure}[!htb]
\centering
\includegraphics[width=3cm]{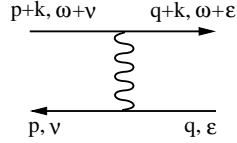}
\begin{center}
\caption{Diagram of interaction in particle-hole channel}
\label{fig:Gamma_ph}
\end{center}
\end{figure}
These interactions can be rewritten distinguishing the singlet from 
the triplet channel in this way
\be
{T}\sum_{|k|\ll k_F}\,\frac{\Gamma_s}{2} \,\overline{c}_{n}(p_1)\,\sigma_0\,c_{n+\omega}(p_1+k)\,\overline{c}_{m}(p_2)\,\sigma_0\,c_{m-\omega}(p_2-k),
\ee
\be
-{T}\sum_{|k|\ll k_F}\,\frac{\Gamma_t}{2} \,\overline{c}_{n}(p_1)\,\vec{\sigma}\,c_{n+\omega}(p_1+k)\,\overline{c}_{m}(p_2)\,\vec{\sigma}\,c_{m-\omega}(p_2-k),
\ee
with $\Gamma_t=\Gamma_2/2$ and $\Gamma_s=\Gamma_1-\Gamma_2/2$. 
By gaussian integration and using (\ref{nambu}) we have
\bea
\label{HSph}
&&e^{T\sum\,\frac{\Gamma}{2}\, \overline{c}_{n}(p_1)\,\sigma\,c_{n+\omega}(p_1+k)\,\overline{c}_{m}(p_2)\,\sigma\,c_{m-\omega}(p_2-k)}=\\
\nonumber&&\int dX e^{-\frac{1}{2}\sum_\omega \left(X_0(\omega)X_0(-\omega)-X_3(\omega)X_3(-\omega)\right)+2i\sum\sqrt{-T\Gamma}\left(X_0(\omega)(\overline{\Psi}_n\tau_0\sigma^t\Psi_{n+\omega})+X_3(\omega)(\overline{\Psi}_n\tau_3\sigma^t\Psi_{n+\omega})\right)},
\eea
being $X_0(-\omega)=X_0(\omega)$, $X_3(-\omega)=-X_3(\omega)$ 
auxiliary Hubbard-Stratonovich fields
and $\Gamma=-\Gamma_s$ for the singlet particle-hole channel with 
$\sigma=\sigma_0$ or
$\Gamma=\Gamma_t$ with $\sigma=\vec{\sigma}$ for the triplet particle-hole 
channel. 
In Eq. (\ref{HSph}) the positive-negative 
energy index of the Nambu spinors have 
been extended to label the positive and negative Matsubara frequencies.
By integrating the fermionic degrees of freedom, the full action 
including interaction is 
\be
\frac{1}{2} Tr\ln\left(
i\widetilde{U}\,\hat{\epsilon}\,U^\dagger
- \widetilde{U}H^{(0)}U^\dagger + iQ_{sp} + 2i\sqrt{-{T}\Gamma}X_0\widetilde{U}\tau_0\sigma^t U^\dagger \right)
\ee
where 
$\hat{\epsilon}_{nm}=\epsilon_{n}\delta_{nm}\equiv n\pi {T}\delta_{lk}$
with $n, m$ odd integers. 
$U$ is the unitary transformation, previously called $T$ 
in the non-interacting case. 
Then, by expanding in terms of the saddle point Green's functions, we find  
new terms in the action that represents the residual interactions in the 
p-h channels, namely 
\be
-\frac{1}{2}\sum\left(X_0(\omega)^2+X_3(\omega)^2\right)-\sum\frac{\sqrt{-{T}\Gamma}}{2}{\pi\nu}\left(X^{\phantom{\dagger}}_0(\omega)Tr(\tau_0\sigma Q_{n\,n+\omega})+X_3(\omega)Tr(\tau_3\sigma Q_{n,\,n+\omega})\right).
\ee
Finally, by integrating over the auxiliary fields $X_0$ and $X_3$ we get 
the following contributions to the free energy for the singlet channel
\be
-{T}\sum\frac{\pi^2\nu^2}{8}\Gamma_s\sum_{l=0,3}\left(Tr(Q^{\phantom{a}}_{n,\, n+\omega}\tau_l\,\sigma_0)\,Tr(Q^{\phantom{a}}_{m+\omega ,\, m}\tau_l\,\sigma_0)\right),
\label{phsinglet}
\ee
and for the triplet channel
\be
{T}\sum\frac{\pi^2\nu^2}{8}\Gamma_t\sum_{l=0,3}\left(Tr(Q^{\phantom{a}}_{n,\, n+\omega}\tau_l\,\vec{\sigma})\,Tr(Q^{\phantom{a}}_{m+\omega ,\, m}\tau_l\,\vec{\sigma})\right). 
\label{phtriplet}
\ee
In the replica space the $Q$ matrices contained in 
Eq. (\ref{phsinglet}) and in 
Eq. (\ref{phtriplet}) are diagonal and have the same indices since residual 
interactions is present at fixed disorder. 
For convenience we will put upper latin indices, like $Q^{ab}$, to denote 
replicas. 
In $d$-wave superconductors, from $[T,\tau_2s_1]=0$ we have
\be
\label{superQ}
\tau_2 s_1 Q \tau_2 s_1 = -Q
\ee
together with the condition 
\be
C^t Q^t C = Q,
\ee
where $s_i, i=0,1,2,3$, are Pauli matricies 
acting on the signs of Matsubara frequencies. 
For the singlet and $\tau_0$ and $\tau_3$ components, this means 
\bea
\label{QnmQnm}
&&Q_{S0,nm}^{ab} = Q_{S0,mn}^{ba} = - Q_{S0,-n-m}^{ab},\\
&&Q_{S3,nm}^{ab} = - Q_{S3,mn}^{ba} =   Q_{S3,-n-m}^{ab},
\label{QnmQnm3}
\eea
having defined
\be
Q=Q_{S}\sigma_0+i\vec{Q}_{T}\cdot\vec{\sigma}
\ee
in spin space and
\be
Q_{S}=Q_{S0}\tau_0+i\sum_{j=1,2,3} Q_{Sj}\tau_j, \;\;\;Q_{T}=Q_{T0}\tau_0+i\sum_{j=1,2,3} Q_{Tj}\tau_j
\ee
in particle-hole space. Because of Eqs. (\ref{QnmQnm}, \ref{QnmQnm3}) the interaction in the p-h singlet channel is therefore
\begin{eqnarray*}
&&\sum_{nm\omega}\sum_a 
Q_{S0,n,n+\omega}^{aa}
Q_{S0,m+\omega,m}^{aa}
+ Q_{S3,n,n+\omega}^{aa}Q_{S3,m+\omega,m}^{aa}\\
=&& -\sum_{nm\omega}\sum_a 
Q_{S0,-n-\omega,-n}^{aa}
Q_{S0,m+\omega,m}^{aa}
+Q_{S3,-n-\omega,-n}^{aa}Q_{S3,m+\omega,m}^{aa}.
\end{eqnarray*}
By setting $-n\to n+\omega$ in the last term, we recover the 
first with opposite sign, hence the sum is zero. 
More specifically, if we consider the transformation 
$Q_{n,m}\rightarrow Q_{-m,-n}$ we find that the interaction terms only involve 
the symmetric components of the $Q$'s because of energy conservation. On the 
other hand, only the antisymmetric $Q_S$ and the symmetric $Q_T$ stay massless 
in the superconducting phase due to Eq. (\ref{superQ}).
This means that the singlet term, with $\Gamma_s$, is suppress in $d$-wave 
superconductors by symmetry and only the triplet, with $\Gamma_t$, survives. 
This is physically expected being charge fluctuations not diffusive.

We now take into account also the interaction in the Cooper channel. 
The diffusive cooperon represents fluctuation in the 
particle-particle channel with $s$-wave symmetry. 
Since the real part of the order 
parameter is already finite, fluctuations in the $\tau_2 s_1$ channels are 
not diffusive, while only fluctuations in the $\tau_1 s_1$ channel, 
corresponding to fluctuations of an $is$ order parameter, stay massless. 
In the presence of residual interaction in the p-p channel, we must 
also consider the term
\be
{T}\sum_{|k|\ll k_F}\,\frac{\Gamma_c}{2} \,\overline{c}^{\alpha}_{n}
(p_1)\,\overline{c}^{\beta}_{\omega-n}(k-p_1)\,c^{\beta}_{m}(p_2)
\,c^{\alpha}_{\omega -m}(k-p_2).
\ee
\begin{figure}[!htb]
\centering
\includegraphics[width=3cm]{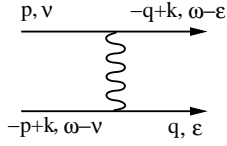}
\begin{center}
\caption{Diagram of interaction in particle-particle channel}
\label{fig:Gamma_pp}
\end{center}
\end{figure}
By introducing one more auxiliary field, $Y^{\alpha \beta}=(Y^{\beta\alpha})^*$
with $\alpha $ and $\beta$ spin indices, the quantum weight due to 
p-p interaction can be rewritten as
\be
\int dY e^{-\frac{1}{2}\sum Y_m^{\alpha \beta}(k)Y_m^{\beta \alpha}(-k)+i\sum\sqrt{T\Gamma_c}\left(\overline{c}^{\alpha}_{n}(p_1)Y_\omega^{\alpha \beta}(p_1+p_2)\overline{c}^{\alpha}_{\omega-n}(p_2)+c^{\beta}_{n}(p_1)Y_\omega^{\beta\alpha}(p_1+p_2)c^{\alpha}_{\omega-n}(p_2)\right)}
\label{cooperpath}
\ee
From Eq. (\ref{nambu}) the following equalities hold
\bea
\overline{\Psi}(\tau_1+i\tau_2)\Psi=-i\,c\,\sigma_y c,\\
\overline{\Psi}(\tau_1-i\tau_2)\Psi=-i\,\overline{c}\,\sigma_y\overline{c},
\eea
so that, calling $Y_R^{\alpha \beta}=\sum_\gamma Y^{\alpha \gamma} \sigma_y^{\gamma\beta} $ and  $Y_L^{\alpha \beta}=\sum_\gamma \sigma_y^{\alpha\gamma}Y^{\gamma\beta}$, which implies $Y_L^{\beta \alpha}=(Y_R^{\alpha \beta})^*$, 
Eq. (\ref{cooperpath}) becomes
\be
\int dY e^{-\frac{1}{2}\sum Y_{R\omega}^{\alpha \beta}Y_{L\omega}^{\beta \alpha}+\sum\sqrt{T\Gamma_c}(\overline{\Psi}^{\alpha}_n Y_{R\omega}^{\alpha \beta}\tau^{+}\Psi^{\beta}_{\omega-n}+\overline{\Psi}^{\beta}_n Y_{L\omega}^{\beta\alpha}\tau^{-}\Psi^{\alpha}_{\omega-n})},
\ee
where $\tau^{\pm}=\tau_1\pm i\tau_2$. Integrating over fermions we find
\be
\int dY e^{-\frac{1}{2}\sum Y_{R\omega}^{\alpha \beta} Y_{L\omega}^{\beta \alpha}+i\sum\frac{\pi\nu}{2}\sqrt{T\Gamma_c}(Y_{R\omega}^{\alpha \beta}Tr(Q^{\beta\alpha}_{\omega-n,\,n}\tau^{+})+Y_{L\omega}^{\beta\alpha}Tr(Q^{\alpha\beta}_{\omega-m,\,m}\tau^{-}))}
\ee
and finally, after integration over $Y_R$, we obtain the following additional 
term to the free energy, 
representing the interaction in the Cooper channel 
\be
{T}\sum \frac{\pi^2\nu^2}{4}\Gamma_c Tr_{spin}\{Tr(Q_{n+\omega,\,-n}\tau^+)Tr(Q_{m+\omega,\,-m}\tau^-)\}.
\label{coopertutto}
\ee
Also in this case the $Q$ matrices are diagonal in replica space and both of them have the same replica index. By the charge conjugacy relation $C^tQ^tC=Q$, the triplet terms don't contribute since 
\be
Q_{T1,nm}^{ab} = -Q_{T1,mn}^{ba},\;\;
Q_{T2,nm}^{ab} = -Q_{T2,mn}^{ba}
\ee
so, if in Eq. (\ref{coopertutto}) we transpose $Q_{n+\omega,\,-n}$ and put $-n\to
n+\omega$ we will have triplet terms with opposite sign. This means that at the end only the following term remains
\be
{T}\sum \frac{\pi^2\nu^2}{8}\Gamma_c \sum_{l=1,2}\left(Tr(Q^{aa}_{n+\omega,\,-n}\tau_l\,\sigma_0)\,Tr(Q^{aa}_{m+\omega,\,-m}\tau_l\,\sigma_0)\right).
\label{ppcooper}
\ee
This result is valid also in metal phase and reflects the local character of 
the $Q$ matrix.

\section{Renormalization group with interactions}
Now we calculate the corrections to the conductivity and 
to the density of states due to the interactions. 
Let us first consider the model without sublattice symmetry, for which the
relevant interactions are $\Gamma_t$ and $\Gamma_c$ defined above.

\subsection{Interactions with small momentum transfer}
\label{sec:RGint_q=0}
The properties of massless modes in the Matsubara frequency space are the following, having imposed antihermitianicity and conditions (\ref{seconda}), (\ref{terza}),
\begin{eqnarray*}
&&W_{S0,nm}^{ab} = \phantom{+}W_{S0,nm}^{ab*} = - W_{S0,mn}^{ba} = \phantom{+}W_{S0,-n-m}^{ab}
= -W_{S0,-m-n}^{ba},\\
&&W_{S1,nm}^{ab} = - W_{S1,nm}^{ab*}  = - W_{S1,mn}^{ba} = -W_{S1,-n-m}^{ab}
= \phantom{+}W_{S1,-m-n}^{ba},\\
&&W_{S2,nm}^{ab} = - W_{S2,nm}^{ab*}  = - W_{S2,mn}^{ba} = \phantom{+}W_{S2,-n-m}^{ab}
= -W_{S2,-m-n}^{ba},\\
&&W_{S3,nm}^{ab} = \phantom{+}W_{S3,nm}^{ab*}  =  \phantom{+}W_{S3,mn}^{ba} = -W_{S3,-n-m}^{ab} = -W_{S3,-m-n}^{ba},\\
&&\vec{W}_{T0,nm}^{ab} = \phantom{+}\vec{W}_{T0,nm}^{ab*}  = \phantom{+}\vec{W}_{T0,mn}^{ba} =
\phantom{+}\vec{W}_{T0,-n-m}^{ab} = \phantom{+}\vec{W}_{T0,-m-n}^{ba},\\
&&\vec{W}_{T1,nm}^{ab} = - \vec{W}_{T1,nm}^{ab*} = \phantom{+}\vec{W}_{T1,mn}^{ba} =
-\vec{W}_{T1,-n-m}^{ab} = -\vec{W}_{T1,-m-n}^{ba},\\
&&\vec{W}_{T2,nm}^{ab} = - \vec{W}_{T2,nm}^{ab*} = \phantom{+}\vec{W}_{T2,mn}^{ba} = \phantom{+}\vec{W}_{T2,-n-m}^{ab} = \phantom{+}\vec{W}_{T2,-m-n}^{ba},\\
&&\vec{W}_{T3,nm}^{ab} = \phantom{+}\vec{W}_{T3,nm}^{ab*} = - \vec{W}_{T3,mn}^{ba} = -\vec{W}_{T3,-n-m}^{ab} = \phantom{+}\vec{W}_{T3,-m-n}^{ba},
\end{eqnarray*}                                                               
where $a$ and $b$ are replica indices, while $n$ and $m$ are odd integer 
Matsubara indices with opposite sign. 
If we take $n=-m$ we recover the 
symmetry properties derived in the previous chapters and reported 
in Appendix A. By these relations we can write the gaussian propagator
\bea
\nonumber\left\langle W^{ab}_{{\cal S}i,nm}(k)W^{cd}_{{\cal S}i,rq}(k)\right\rangle&=&(\pm)\frac{1}{2}\Big( 1-\lambda_n\lambda_m \Big)D_{nm}(k)\\
&&\Big(\delta^{ac}_{nr}\,\delta^{bd}_{mq}\,[\pm]\,\delta^{ad}_{nq}\,\delta^{bc}_{mr}\,(-)^i\,\delta^{ac}_{n-r}\,\delta^{bd}_{m-q}\,(-)^i[\pm]\,\delta^{ad}_{n-q}\,\delta^{bc}_{m-r}\Big),
\eea
where  ${\cal{S}}$ means $S$ for singlet and $T$ for triplet components, $\delta^{ac}_{nr}=\delta_{ac}\delta_{nr}$, $(\pm)$ are related to real or imaginary matrix elements of $W$, 
$[\pm]$ for symmetric or antisymmetric matrix, 
$(-)^i$ the sign that $W$ acquires changing the signs of Matsubara 
frequencies and this occurs only for modes proportional 
to $\tau_1$ and $\tau_3$, and finally
\bea
D_{nm}(k)=\frac{1}{4\pi\nu}\frac{1}{Dk^2+z|\epsilon_n-\epsilon_m|},
\eea
with $\textrm{sign}(\epsilon_n)\equiv\lambda_n=-\lambda_m $ where
$\epsilon_n=n\pi {T}$ is a fermionic Matsubara frequency with odd
integer $n$. 
The factor $z$ is the frequency renormalization and $D=\sigma/(2\nu)$ the
diffusion coefficient. 
\begin{figure}[!htb]
\centering
\includegraphics[width=2.7cm]{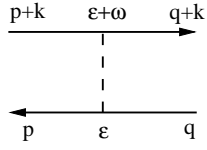}
\begin{center}
\caption{Diagram of the diffusion propagator}
\label{fig:Diffusion}
\end{center}
\end{figure}

Let us introduce slow and fast modes in the spirit of Wilson Polyakov 
procedure, as we have done before,
\be
Q=\widetilde{U}^{\dagger}_s Q_f U_s=\widetilde{U}^{\dagger}_s\widetilde{U}^{\dagger}_f Q_{sp} U_f U_s,
\ee
with $U=e^{\frac{W}{2}}$ (without chirality,
$\widetilde{U}={U}$) and finally 
${Q_{sp}}_{n\,m}= \lambda_n \delta_{nm}$. \\
$U_s$ contains only slow momentum fluctuations and 
\be
\label{slowU}
{U_s}_{nm}=\delta_{nm}, \,\,\,\, \textrm{if}\,\,\,\,\, (s_e\tau)^{-1}<|\epsilon_n|<\tau^{-1} \,\,\,\,\,\textrm{or}\,\,\,\,\,(s_e\tau)^{-1}<|\epsilon_m|<\tau^{-1},
\ee
where $\tau^{-1}$ acts as an energy cutoff and $s_e>1$ is the rescaling factor. 
The massless fast modes satisfy by definition
\be
{W_f}_{nm}(k)=0\,\,\,\, \textrm{if}\,\,\,\,\, \{Dk^2,|\epsilon_n|,|\epsilon_m|\}<(s_e\tau)^{-1}.
\ee 
Now let us expand the interaction contributions to the full action, namely Eqs.
(\ref{phsinglet},\ref{phtriplet},\ref{ppcooper}) multiplied by $T^{-1}$, 
in terms of $W_f$, leaving slow $U_s$ unexpanded. 
In this way, besides the terms (\ref{F1}) and (\ref{F2}), 
also the following contributions should be evaluated in the one-loop expansion
\bea
\nonumber&& S_{int}^{1}=\frac{\pi^2\nu}{8}\sum\nu\Gamma \,Tr\left(\widetilde{U}_{n_{1}m_{1}}^{\dagger de}\lambda_{m_{1}}{W_{m_{1}m_{2}}^{eg}}U_{m_{2}n_{2}}^{gd}\tau _{l}\sigma \right)
Tr\left(\widetilde{U}_{n_{3}m_{3}}^{\dagger df}\lambda_{m_{3}}{W_{m_{3}m_{4}}^{fh}}U_{m_{4}n_{4}}^{hd}\tau _{l}\sigma \right)\\
\label{inter1}
&&\phantom{S_{int}^{1}=}
\delta ({n_{1}}\mp {n_{2}}\pm {n_{3}}-{n_{4}}),\\
\nonumber&&\\
\nonumber&&  S_{int}^{2}=\frac{\pi^2\nu}{8}\sum\nu\Gamma \,
Tr\left(\widetilde{U}_{n_{1}m_{1}}^{\dagger de}\lambda_{m_{1}}
{W_{m_{1}m_{2}}^{eg}}{W_{m_{2}m_{3}}^{gh}}U_{m_{3}n_{2}}^{hd}\tau _{l}\sigma 
\right)Tr\left(Q^{dd}_{n_{3}n_{4}}\tau _{l}\sigma \right)\\
\label{inter2}
&&\phantom{S_{int}^{2}=}
\delta ({n_{1}}\mp {n_{2}}\pm {n_{3}}-{n_{4}}),
\eea
where the upper indices are in the replica space, 
$\Gamma=\Gamma_t$, $\sigma=\vec\sigma$, 
and we sum over $\tau_l=\tau_0, \tau_3$ and energies 
with constraint 
$\delta ({n_{1}}-{n_{2}}+{n_{3}}-{n_{4}})$
for p-h triplet channel and 
$\Gamma=\Gamma_c$, $\sigma=\sigma_0$, $\tau_l=\tau_1, \tau_2$, 
with energy conservation law 
$\delta ({n_{1}}+{n_{2}}-{n_{3}}-{n_{4}})$
for p-p Cooper channel 
(in metal phase we would have also the singlet channel with 
$\Gamma=-\Gamma_s$, $\sigma=\sigma_0$, $\tau_l=\tau_0, \tau_3$ and energy constraint $\delta ({n_{1}}-{n_{2}}+{n_{3}}-{n_{4}})$) and 
finally 
$Q=\widetilde{U}^\dagger Q_{sp} U$ is the slow mode matrix. The sums in Eqs. (\ref{inter1}, \ref{inter2}) are performed over all upper (replica) and lower (frequencies) indices and over $l$, according to the channel. 
Before starting the renormalization of all the parameters involved in the
theory we can easily take into account all the ladders diagrams 
sketched in Fig.\ref{fig:Ladder}, substituting the bare 
particle-hole triplet scattering amplitude with 
\be
\label{t-ladder}
\Gamma_t(q,\omega)=\Gamma_t\frac{D q^2+z|\omega|}{D q^2+(z+2\nu\Gamma_t)|\omega|},
\ee
which is the algebraic infinite ladder summation \cite{AltAr85}.\\

\begin{figure}[!htb]
\centering
\includegraphics[width=10.0cm]{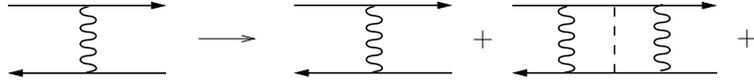}
\begin{center}
\caption{Ladder summation}
\label{fig:Ladder}
\end{center}
\end{figure}
\vspace{-0.5cm}
For Cooper amplitude we have instead, 
already after summing the first two diagrams, a logarithmic divergent 
term which does not depend on $g$, the small parameter of the theory.
To calculate one-loop corrections to conductivity due to the interactions, 
we have to consider the following averages over fast modes
\be
\langle S^1_{int}\rangle - \langle S_1 S^1_{int}\rangle - 
\langle S_2 S^1_{int}\rangle + \frac{1}{2}\langle (S_2)^2 S^1_{int}\rangle, 
\label{sumforg}
\ee
where $S^1$ and $S^2$ are defined by Eq. (\ref{F1}) and Eq. (\ref{F2}). 
For instance, contracting the fast modes, the contribution coming from 
\bea
&&\nonumber\left\langle S_{int}^{1}\right\rangle=-\frac{\pi^2\nu}{2}\sum
\nu\Gamma {({p},\epsilon_{n_1}\pm \epsilon_{n_2})}
D_{m_{1}m_{2}} ({ p+p_1})\Big(1-\lambda
_{m_{1}}\lambda _{m_{2}}\Big)\,\delta({n_{1}}\mp {n_{2}}\pm {n_{3}}-{n_{4}})\\
\nonumber&&\phantom{\left\langle S_{int}^{1}\right\rangle} \Big\{
Tr\Big(\lambda _{m_{2}} U_{m_{2}n_{2}}^{gd}({ p_1-p_2})\tau _{l}\sigma 
\widetilde{U}_{n_{1}m_{1}}^{\dagger de}({ p_2}) \lambda
_{m_{1}} U_{m_{1}n_{4}}^{ed}({ -p_3})\tau _l\sigma %
\widetilde{U}_{n_{3}m_{2}}^{\dagger dg}({ p_3-p_1})\Big)\\
&&\phantom{\left\langle S_{int}^{1}\right\rangle} 
-(-)^l Tr\Big(\lambda _{m_{2}} U_{m_{2}n_{2}}^{gd}({ p_1-p_2})
\tau _{l}\sigma 
\widetilde{U}_{n_{1}m_{1}}^{\dagger de} ({ p_2})\lambda
_{m_{1}} U_{m_{1}-n_{4}}^{ed}({ -p_3})\tau _{l}\sigma %
\widetilde{U}_{-n_{3}m_{2}}^{\dagger dg}({ p_3-p_1})\Big)\Big\},
\label{<S_int>}
\eea
which renormalizes the spin conductivity is obtained summing 
over fast momenta $m_1$ (the sum over slow momenta, instead, renormalizes the scattering amplitudes), giving simply
\bea
-{\pi^2\nu}\,n_{\sigma}\sum
\nu\Gamma {({ p},\epsilon_{m_1}\pm\epsilon_{n_2})}
D_{m_{1}m_2} ({ p+p_1})
Tr\Big(\widetilde{U}_{n_{2}m_{2}}^{\dagger dg}({ -p_1})
U_{m_{2}n_{2}}^{gd}({ p_1})\Big),
\label{<S_int>_simple}
\eea
with $n_{\sigma}=3$ in
the triplet p-h channel and $n_{\sigma}=1$ in the Cooper one. 
Expanding the propagator $D$ in terms of
slow ${ p_1}$, neglecting slow frequencies both in $\Gamma$ and in $D$, 
we have a factorized term
\be
-n_{\sigma} J\sum p_1^2 
Tr\Big(\widetilde{U}_{n_{2}m_{2}}^{\dagger dg}({ -p_1})
U_{m_{2}n_{2}}^{gd}({ p_1})\Big)=n_{\sigma} J \int dr Tr(AA),
\ee
where $A=\nabla \widetilde{U} \widetilde{U}^{\dagger}$ (notice that here, since chirality is not present, $\widetilde{U}=U$) and $J=\frac{1}{\pi}
\int dp\,
d\epsilon\, \Gamma({ p},\epsilon)\Big(\frac{D p}{(D p^2+z|\epsilon|)^2}-
\frac{2 D^2 p^3}{(D p^2+z|\epsilon|)^3}\Big)$. The other mean values in 
Eq. (\ref{sumforg}) give contributions of the same kind and also 
proportional to $\int
dr Tr(A\lambda A\lambda)$ (with chirality $\int
dr Tr(A\lambda \gamma_1 A\lambda \gamma_1)$). Reminding that 
\be
2 Tr(A\lambda A\lambda-AA)= Tr(\nabla Q\nabla Q^\dagger),
\ee
we obtain corrections to the stiffness coefficient.

To calculate corrections to the density of states, besides $\langle S_{\nu}\rangle$, we will consider
\be
\langle S^1_{int} S_{\nu}\rangle,
\label{fordos}
\ee
where
\be
S_{\nu}=\frac{1}{2} \sum Tr(\epsilon_n\widetilde{U}_{nm_1}^{\dagger ab}\lambda_{m_1} W_{m_1m_2}^{bc}W_{m_2m_3}^{cd}U_{m_3n}^{da}).
\ee

In order to calculate corrections to the interaction amplitudes 
at first order in $g$, we need to consider the sum
\be
\langle S_{int}\rangle 
-\frac{1}{2}\langle (S_{int})^2 \rangle 
+ \frac{1}{2}\langle (S^1_{int})^2 S^2_{int}\rangle 
+ \frac{1}{2}\langle S^1_{int} (S^2_{int})^2 \rangle 
- \frac{1}{4}\langle (S^1_{int} S^2_{int})^2\rangle,
\label{sumforamp}
\ee
where $S_{int}=S^1_{int}+S^2_{int}$.
Because of the different structures of the two terms 
$S^1_{int}$ and $S^2_{int}$, we notice that $\langle
S^2_{int}\rangle$ in the first term, $\langle S^1_{int} S^2_{int}\rangle$ 
in the second term and the third 
term of Eq. (\ref{sumforamp}) 
are vertex contributions (in Fig.\ref{fig:vertex},  
the corresponding diagrams are sketched  
using only the p-h triplet scattering amplitude) 
while the last two in Eq. (\ref{sumforamp}) and 
$\langle(S^2_{int})^2 \rangle$ in the second term are bubble corrections,
(see Fig. \ref{fig:bubble}). 
In the normal phase the the first two 
diagrams of Fig. \ref{fig:bubble} 
and the first diagram of Fig. \ref{fig:vertex} vanish. 
There are other diagrams which give corrections to
$\Gamma_t$ in which $\Gamma_c$ appears and viceversa, with 
the same topology but with different energy-momentum conservation laws.
\begin{figure}[!htb]
\begin{center}
\begin{minipage}{4.59cm}
  \hspace{1.0cm}
  \scalebox{1}[1]{\rotatebox{0}{\includegraphics[width=2.00cm]{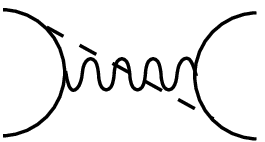}}}
\end{minipage}
\begin{minipage}{3.56cm}
  \scalebox{1}[1]{\rotatebox{0}{\includegraphics[width=2.00cm]{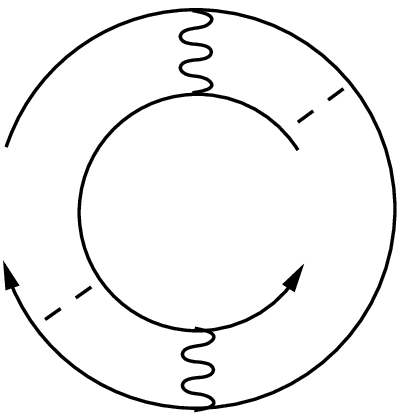}}}
\end{minipage}
\begin{minipage}{3.56cm}
  \scalebox{1}[1]{\rotatebox{-0}{\includegraphics[width=2.00cm]{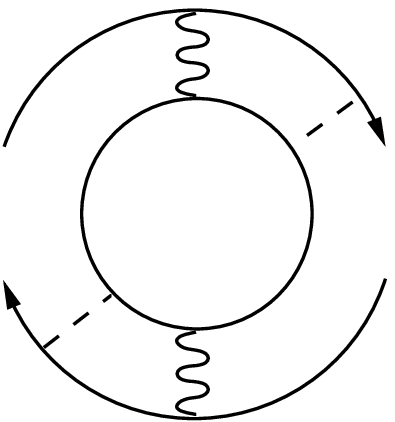}}}
\end{minipage}
\vspace{0.1cm}
\caption{Diagrams in $\langle S^1_{int}\rangle$ and 
$\langle (S^1_{int})^2\rangle$}
\label{fig:una}
\label{fig:vertex}
\vspace{0.5cm}
\begin{minipage}{3.3cm}
  \scalebox{1}[1]{\rotatebox{-0}{\includegraphics[width=2.00cm]{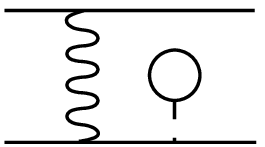}}}
\end{minipage}
\begin{minipage}{3.56cm}
  \scalebox{1}[1]{\rotatebox{-0}{\includegraphics[width=2.60cm]{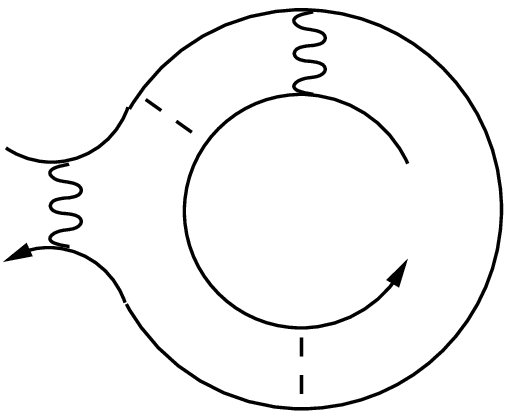}}}
\end{minipage}
\begin{minipage}{3.56cm}
  \scalebox{1}[1]{\rotatebox{-0}{\includegraphics[width=2.60cm]{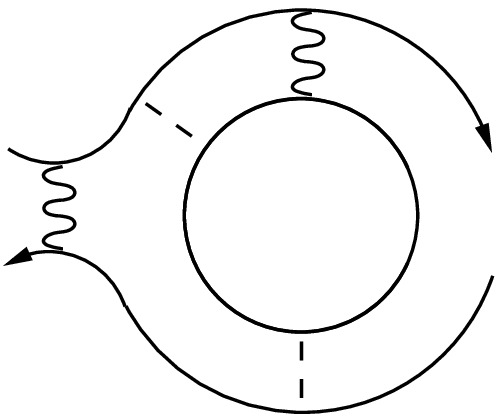}}}
\end{minipage}

\begin{minipage}{3.56cm}
  \scalebox{1}[1]{\rotatebox{-0}{\includegraphics[width=2.60cm]{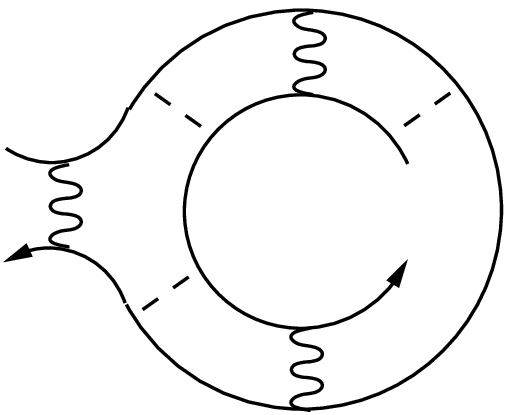}}}
\end{minipage}
\begin{minipage}{3.56cm}
  \scalebox{1}[1]{\rotatebox{-0}{\includegraphics[width=2.60cm]{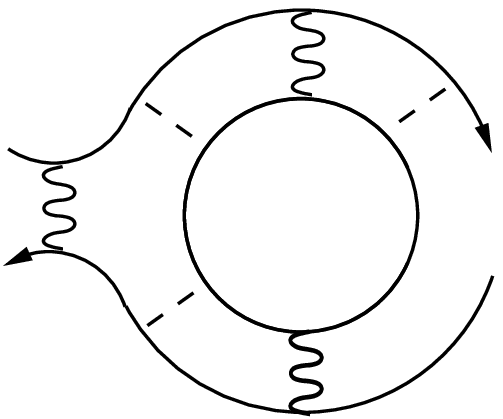}}}
\end{minipage}
\vspace{0.1cm}
\caption{Vertex contributions.}
\label{fig:vertex}
\vspace{0.5cm}
\begin{minipage}{3.56cm}
  \scalebox{1}[1]{\rotatebox{-0}{\includegraphics[width=3.20cm]{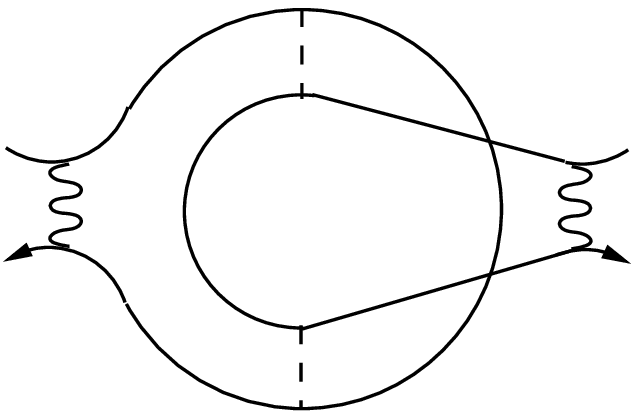}}}
\end{minipage}
\begin{minipage}{3.56cm}
  \scalebox{1}[1]{\rotatebox{-0}{\includegraphics[width=3.20cm]{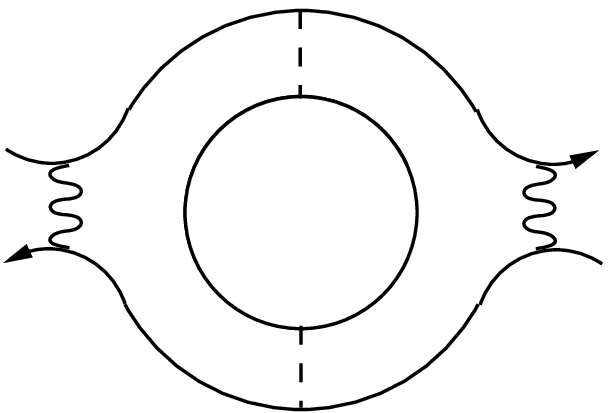}}}
\end{minipage}
\begin{minipage}{3.56cm}
  \scalebox{1}[1]{\rotatebox{-0}{\includegraphics[width=3.20cm]{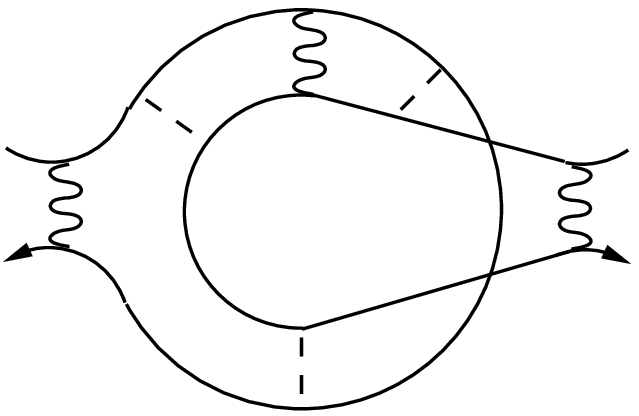}}}
\end{minipage}

\begin{minipage}{3.56cm}
  \scalebox{1}[1]{\rotatebox{-0}{\includegraphics[width=3.20cm]{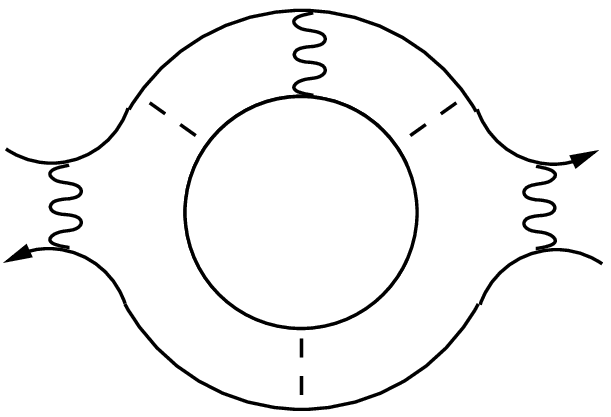}}}
\end{minipage}
\begin{minipage}{3.56cm}
  \scalebox{1}[1]{\rotatebox{-0}{\includegraphics[width=3.20cm]{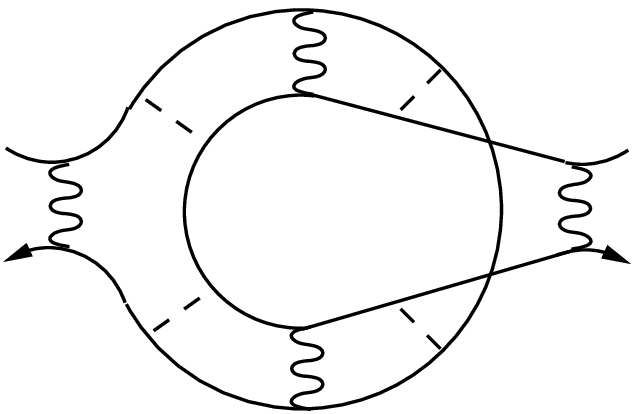}}}
\end{minipage}
\begin{minipage}{3.56cm}
  \scalebox{1}[1]{\rotatebox{-0}{\includegraphics[width=3.20cm]{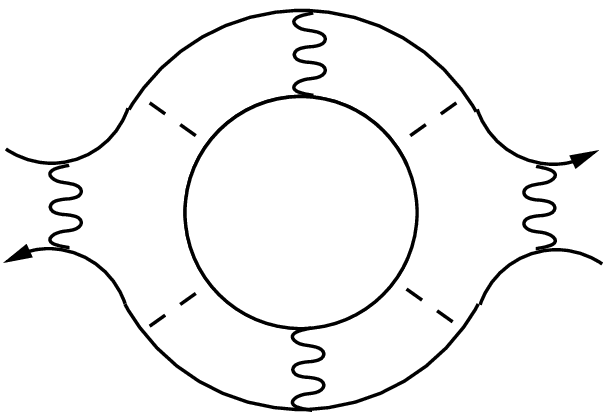}}}
\end{minipage}
\vspace{0.1cm}
\caption{Bubble contributions.}
\label{fig:bubble}
\end{center}
\end{figure}

Finally, the renormalization of $z$ is determined 
by three  contributions, the first two result from 
the corrections to the density 
of states $\langle S_\nu\rangle$ 
and from Eq. (\ref{fordos}) and the third from the expansion of 
$\langle S^1_{int}\rangle$,  
written explicitly in Eq.(\ref{<S_int>_simple}), in terms of the 
slow frequency $n_2$. This latter contribution cancels exactly all the 
terms of order higher than one in the interaction amplitudes 
coming from Eq. (\ref{fordos}).

\subsection{RG equations without chiral symmetry}

The final renormalization group equations for $d$-wave superconductors,
including residual quasiparticle interactions $\Gamma_t$ and $\Gamma_c$, 
in all the cases collected in Table \ref{Table} are the
following:

\paragraph{No chiral, Yes $\hat {T}$. }

If time-reversal symmetry is preserved we obtain
\begin{eqnarray}
\label{eq1}
d g/ d{l}{\phantom{\Gamma_p}}
&=&g^2\left\{-3\nu\Gamma_tf_2(z,z,z_2)-\nu\Gamma_c/z  +1 \right\},\\
\label{eq2}
d z/ d{l} {\phantom{\Gamma_p}} &=& g\left\{3\nu\Gamma_t+\nu\Gamma_c -
  z/2\right\},\\
\label{eq3}
d\nu\Gamma_t/d
{l}&=&g\left\{\nu\left(\Gamma_t+\Gamma_c\right)/2+4\nu^2\,\left(\Gamma_t\Gamma_t+\Gamma_t\Gamma_c\right)/z+4\nu^3\left(\Gamma_t\Gamma_c\Gamma_t\right)/z^2-\nu\Gamma_t-\nu^2\Gamma_t\Gamma_t/z\right\},\\
\label{eq4}
d\nu\Gamma_c/d {l}&=&g\left\{\nu\left(3\Gamma_t-\Gamma_c\right)/2+2\nu^2\left(3\Gamma_t\Gamma_cz f_1(z,z_2)+\Gamma_c\Gamma_c\right)/z-\nu\Gamma_c\right\}-2(\nu\Gamma_c)^2/z,
\end{eqnarray}
where, as usual, we have introduced the functions $f_1$ and $f_2$ which result from integrations of products of diffusion propagators, defined by
\bea
&&f_1(a,b)=\ln(a/b)/(a-b),\\
&&f_2(a,b,c)=2\Big(b\,f_1(a,b)-c\,f_1(a,c)\Big)/(b-c),
\eea
and $z_2=z+2\nu\Gamma_t$. Notice that in Eq. (\ref{eq1}) the last term is 
taken from Table \ref{Table} with $n=0$ and divided by $2$ since here we have
adopted an energy scaling factor instead of a momentum scaling factor, being $l=\ln{s_e}$ with $s_e$ defined in Eq. (\ref{slowU}). 
The contributions due to the non-vanishing density of states in the 
non-interacting case are singled out, they are the last two terms of 
Eq.(\ref{eq3}) (corresponding to the first diagram of Fig. \ref{fig:vertex} 
and to the first two of Fig. \ref{fig:bubble}) 
and the last term in the curly brackets of Eq. (\ref{eq4}). 
In the normal phase, they disappear meanwhile
the singlet particle-hole channel is turned on \cite{Finkelstein,Fin84}. 
The last term of Eq. (\ref{eq4}) is due to the ladder summation.
The scaling behavior of the density of states becomes
\bea
\label{eq5}
d \nu/ d{l}&=&g\nu\left\{3\nu\Gamma_tf_1(z,z_2)+\nu\Gamma_c/z-1/2\right\}.
\eea
Notice that the theory has only four parameters, $g, z, \nu\Gamma_t, \nu
\Gamma_c$ since the density of states appears always multiplied with the
scattering amplitudes. 
Summing and writing in terms of $ u_t\equiv 2\nu\Gamma_t/z$ and 
$ u_c\equiv 2\nu\Gamma_c/z$ 
we can separate from the set of equations the $z$-equation, obtaining
\begin{eqnarray}
\label{eq:159}
d g/ d{l} &=&g^2\left\{3(1-(1+ u_{t})
\ln(1+ u_t)/ u_t)- u_c/2  +1 \right\},\\
d z/ d{l}  &=& g\,z\left\{3 u_t + u_c  - 1\right\}/2,\\
\label{eq:161}
d u_t/d {l}&=&g\left\{ u_c/2+3\, u_t u_c/2+ u_t u_c u_t\right\},\\
\label{eq:162}
d u_c/d {l}&=&g\left\{\left(3 u_t-2 u_c\right)/2+3\,\left( u_c \ln(1+
    u_t)- u_t u_c\right)/2\right\}-( u_c)^2,\\
\label{eq:163}
d \nu/ d{l}&=&g\nu\left\{3\ln(1+ u_t)+ u_c-1
\right\}/2.
\eea
Supposing $T^{-1}$, the inverse of the temperature, 
being the coherence time for the quasiparticles, the integration of 
the equations above will run from $T$ to $\tau^{-1}$, 
the elastic scattering rate. Strictly speaking, half of the localization correction in the last term of Eq. (\ref{eq:159}), is cut off not by the temperature but by the inelastic scattering rate $\tau_{in}^{-1}$. However, $\tau_{in}^{-1}$, for some source of inelastic processes, it is found to be linear in $T$ \cite{bruno}. At the first order in the scattering amplitudes, supposing them 
so small, 
at the moment, that they do not flow fast, we have the following variation in 
the conductivity induced both by the disorder and by the interactions  
\be 
\delta\sigma=\frac{1}{2\pi^2}\left(1-\frac{3}{2}u_t- \frac{1}{2}u_c\right)\ln (T\tau) .
\ee
Solving now the whole set of equations (\ref{eq:159}-\ref{eq:162}) 
we find that, starting
with a small and positive $ u_{t0}$ (in Fig.\ref{fig:gcoop} 
$ u_{t0}= u_t(T\simeq \tau^{-1})=0.15$, 
$ u_{c0}= u_c(T\simeq \tau^{-1})=0$ and 
$g_0=g(T\simeq \tau^{-1})=0.03$), $g$ increases for the presence of the
quantum interferences but when the interactions become strong enough, after
reaching a maximum, it decreases at low temperature, 
as shown in Fig. \ref{fig:gcoop}. The temperature of crossover depends 
on the interaction amplitudes. 
The smaller is the value of the positive $ u_{t0}$, 
the higher is the peak of $g$, while   
$ u_{t}$
diverges as in the normal phase {\cite{Fin84,Castel84}}.
A negative starting value of $ u_{t0}$, instead, reinforces 
the localization.
\begin{figure}[!htb]
\centering
  \scalebox{1}[1]{\rotatebox{-0}{\includegraphics[width=8cm]{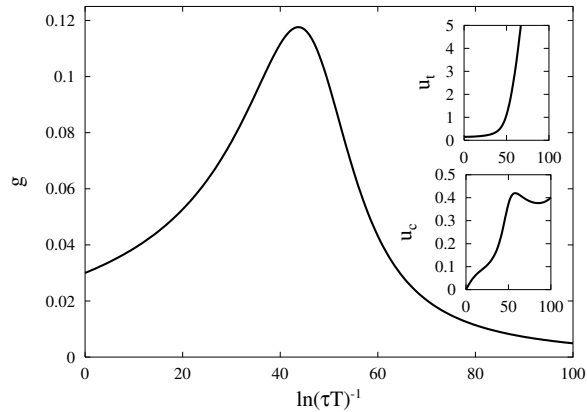}}}
\begin{center}
\caption{Scaling behavior of $g$ of paragraph {\it a}, in the inserts $ u_t$ and $ u_c$.}
\label{fig:gcoop}
\end{center}
\end{figure}

\paragraph{No chiral, No $\hat {T}$. }

If time reversal symmetry is broken, $\Gamma_c$ disappears from 
Eqs.(\ref{eq1}-\ref{eq3}, \ref{eq5}) and the last term of Eq. (\ref{eq1}), 
which is the contribution due only to the disorder, is 
halved (see Table \ref{Table}).  $ u_t$ remains unrenormalized 
at first order in $g$ (Eq.(\ref{eq:161}) with $ u_c=0$), namely  
$d u_t/d {l} = O(g^2)$, which is a result obtained also 
by Keldysh technique \cite{Jeng}, 
so  $ u_t= u_{t0}$ at the one-loop level. 
The equation for $g$, exact in the interaction, becomes simply
\begin{equation}
d g/ d{l}
=g^2\left\{3\left(1-(1+ u_{t0})\ln(1+ u_{t0})/ u_{t0}\right)+1/2
\right\},
\end{equation}
implying that there is a positive critical value of $ u_{t0}$  
($ u_{t0}^* \simeq 0.37$) 
for which $d g/ d{l} =0$ at lowest order in $g$. Above $ u_{t0}^*$
the resistivity $g$ decreases, below $ u_{t0}^*$ it increases. 

\paragraph{No chiral, magnetic field. }

If also the spin rotational invariance is broken, we have
\begin{eqnarray}
\label{165}
d g/ d{l}{\phantom{\Gamma_p}} &=&-g^2\nu\Gamma_tf_2(z,z,z_2),\\
d z/ d{l} {\phantom{\Gamma_p}} &=&\phantom{+} g\nu\Gamma_t,\\
d\nu\Gamma_t/d {l}&=&-g\nu\Gamma_t/2,\\
\label{168}
d \nu/ d{l}{\phantom{\Gamma_p}}&=&g\nu\Gamma_tf_1(z,z_2).
\end{eqnarray}
Also in this case, as in the previous ones, we can eliminate $z$ introducing  
$ u_t$, obtaining
\bea
&&d g/ d{l} = g^2\left\{1-(1+ u_{t})\ln(1+ u_t)/ u_t\right\},\\ 
&&d u_t/d {l} = -g u_t(1+ u_t)/2.
\eea
The solution is a 
transient since the scattering amplitude flows to zero and the conductivity 
changes value only in the meanwhile. For a small value of
$ u_{t0}= u_{t}(T\simeq \tau^{-1})$, we can retain only lowest order in 
$ u_{t}$ so that, given $g_0=g(T\simeq \tau^{-1})$,  
the renormalized resistivity is given by
\begin{eqnarray}
\label{easysolution}
g=g_0 \,e^{- u_{t0}}.
\end{eqnarray}
\paragraph{No chiral, magnetic impurities. }

In the presence of magnetic impurities there are no corrections due to
the interactions. Indeed both cooperons and triplet diffusons become massive.

\subsection{Interactions with $(\pi,\pi)$ momentum transfer}
At the nesting point where the staggered fluctuations become diffusive 
(i.e when the chiral symmetry holds), also quasiparticle interactions with 
$(\pi,\pi)$ momentum transfer must be included in the effective action.
First notice that in the sublattice representation, the the contribution 
due to the interactions with small momentum 
transfer can be rewritten as
\be
\sum \,\frac{\pi^2\nu}{32}\nu\Gamma^{^0}\,
Tr(Q^{aa}_{n_1 n_2}\tau_l\sigma\gamma_0)
Tr(Q^{aa}_{n_3 n_4}\tau_l\sigma\gamma_0) \,
\delta({n_1} \mp {n_2} \pm {n_3} -{n_4}),
\label{inter_0}
\ee
where $\gamma_0$ is the identity in the sublattice space and $\Gamma^{^0}$ 
can be $-\Gamma^{^0}_s$, $\Gamma^{^0}_t$ and $\Gamma^{^0}_c$.\\
Let us now consider interactions whose transferred momentum 
is $q_\pi=(\pm\pi,\pm\pi)$, which also involve quasiparticles at the 
Fermi energy in the case of half filling, 
\be
{T}\sum_{|k|\ll k_F}\,\frac{\Gamma_1^{^3}}{2} \,\overline{c}^{\alpha}_{n}(p_1)\,\overline{c}^{\beta}_{m}(p_2)\,c^{\beta}_{m-\omega}(p_2-k-q_\pi)\,c^{\alpha}_{n+\omega}(p_1+k+q_\pi),
\ee
\be
{T}\sum_{|k|\ll k_F}\,\frac{\Gamma_2^{^3}}{2} \,\overline{c}^{\alpha}_{n}(p_1)\,\overline{c}^{\beta}_{m}(p_2)\,c^{\beta}_{n+\omega}(p_1+k+q_\pi)\,c^{\alpha}_{m-\omega}(p_2-k-q_\pi),
\ee
\be
{T}\sum_{|k|\ll k_F}\,\frac{\Gamma_c^{^3}}{2} \,\overline{c}^{\alpha}_{n}
(p_1)\,\overline{c}^{\beta}_{\omega-n}(k-p_1+q_\pi)\,c^{\beta}_{m}(p_2)
\,c^{\alpha}_{\omega -m}(k-p_2+q_\pi).
\ee
In this case $\Gamma^{^3}$ indicates the amplitude related to 
staggered modes. Repeating the steps which led to Eq. (\ref{phsinglet}), 
Eq. (\ref{phtriplet}) and Eq. (\ref{ppcooper}), we obtain the following additional contribution to the action
\be
\sum\,\frac{\pi^2\nu}{32}\nu\Gamma^{^3}\,
Tr(Q^{aa}_{n_1 n_2}\tau_l\sigma\gamma_3)
Tr(Q^{aa}_{n_3 n_4}\tau_l\sigma\gamma_3)\, \delta({n_1} \mp 
{n_2} \pm {n_3} -{n_4}),
\label{inter_0_3}
\ee
with $\Gamma^{^3}$ taking the values $-\Gamma^{^3}_s\equiv\Gamma^{^3}_2 /2-\Gamma^{^3}_1$ for the singlet p-h staggered channel, $\Gamma^{^3}_t\equiv\Gamma^{^3}_2/2$ for the triplet and $\Gamma^{^3}_c$ for the p-p Cooper staggered channel.
In the superconducting phase only $\Gamma^{^3}_t$ and $\Gamma^{^3}_c$ are 
relevant. The properties of $W^3$ massless modes in energy space derive from 
the conditions (\ref{seconda3}), (\ref{terza}) and antihermitianicity, 
leading to
\begin{eqnarray*}
&&W_{S0,nm}^{ab} = - W_{S0,nm}^{ab*} = \phantom{+} W_{S0,mn}^{ba} = \phantom{+}W_{S0,-n-m}^{ab}
= \phantom{+}W_{S0,-m-n}^{ba},\\
&&W_{S1,nm}^{ab} = \phantom{+}W_{S1,nm}^{ab*} = \phantom{+} W_{S1,mn}^{ba} = - W_{S1,-n-m}^{ab}
=  - W_{S1,-m-n}^{ba},\\
&&W_{S2,nm}^{ab} = \phantom{+}W_{S2,nm}^{ab*} = \phantom{+} W_{S2,mn}^{ba} = \phantom{+}W_{S2,-n-m}^{ab}
= \phantom{+}W_{S2,-m-n}^{ba},\\
&&W_{S3,nm}^{ab} = - W_{S3,nm}^{ab*} = - W_{S3,mn}^{ba} = -W_{S3,-n-m}^{ab}
= \phantom{+}W_{S3,-m-n}^{ba},\\
&&\vec{W}_{T0,nm}^{ab} = - \vec{W}_{T0,nm}^{ab*} =  - \vec{W}_{T0,mn}^{ba} =
\phantom{+}\vec{W}_{T0,-n-m}^{ab} = - \vec{W}_{T0,-m-n}^{ba},\\
&&\vec{W}_{T1,nm}^{ab} = \phantom{+}\vec{W}_{T1,nm}^{ab*} = - \vec{W}_{T1,mn}^{ba} =
-\vec{W}_{T1,-n-m}^{ab} = \phantom{+}\vec{W}_{T1,-m-n}^{ba},\\
&&\vec{W}_{T2,nm}^{ab} = \phantom{+}\vec{W}_{T2,nm}^{ab*} = - \vec{W}_{T2,mn}^{ba} =
\phantom{+}\vec{W}_{T2,-n-m}^{ab} = - \vec{W}_{T2,-m-n}^{ba},\\
&&\vec{W}_{T3,nm}^{ab} = - \vec{W}_{T3,nm}^{ab*} = \phantom{+}\vec{W}_{T3,mn}^{ba} =
-\vec{W}_{T3,-n-m}^{ab} = - \vec{W}_{T3,-m-n}^{ba},
\end{eqnarray*}
where now $n$ and $m$ have same signs.
By these properties we can write down the propagators, reported in Appendix D, 
and evaluate Eq. (\ref{sumforg}), Eq. (\ref{fordos}) and Eq. (\ref{sumforamp}).

\subsection{RG equations with chiral symmetry}

The renormalization group equations in different symmetry classes in the
presence of chirality in the sublattice space are the following:

\paragraph{Yes chiral, Yes $\hat {T}$. } 
If both time reversal symmetry and spin rotation invariance are preserved, we
have, together with $\Gamma=g/c$ and $d c/ d{l}=-4c^2$ (see Eqs. (\ref{def_g&c}, \ref{beta_c})), the following equations 
\begin{eqnarray}
\label{g_sc_stag}
d g/ d{l}{\phantom{\Gamma_p}} &=&g^2\left\{-3\nu\Gamma_t^{^0}f_2(z,z,z_2)+3\nu\Gamma_t^{^3}/z-\nu\Gamma_c^{^0}/z+\nu\Gamma_c^{^3}f_2(z,z,z_c)\right\},\\
d z/ d{l} {\phantom{\Gamma_p}} &=& g\left\{3\nu\Gamma_t^{^0}-3\nu\Gamma_t^{^3}+\nu\Gamma_c^{^0}-\nu\Gamma_c^{^3}
%
+z\Gamma/8\right\},\\
\nonumber d\nu\Gamma^{^0}_t/d {l}&=&g\left\{\nu\left(\Gamma_t^{^0}+\Gamma_t^{^3}+\Gamma_c^{^0}+\Gamma_c^{^3}\right)/2+4\nu^2\,\left(\Gamma_t^{^0}\Gamma_t^{^0}+\Gamma_t^{^0}\Gamma_c^{^0}-\Gamma_t^{^3}\Gamma_t^{^3}-\Gamma_t^{^0}\Gamma_t^{^3}\right)/z  \right.\\
\nonumber &&\left.+4\nu^3\left(\Gamma_t^{^0}\Gamma_c^{^0}\Gamma_t^{^0}
-\Gamma_t^{^0}\Gamma_t^{^3}\Gamma_t^{^0}\right)/z^2\right.\\
%
&&\left.+\Gamma\nu\Gamma^{^0}_t/4
+\Gamma\nu^2\Gamma^{^0}_t\Gamma^{^0}_t/4z 
+\Gamma\nu\Gamma^{^3}_t\left(1+2\nu\Gamma^{^0}_t/z\right)^2 /4
\right\},\\
\label{eq:176}
\nonumber d\nu\Gamma_t^{^3}/d {l}&=&g\left\{\nu\left(\Gamma_t^{^0}+\Gamma_t^{^3}+\Gamma_c^{^0}+\Gamma_c^{^3}\right)/2-2\nu^2\left(\Gamma_t^{^0}\Gamma_t^{^3} z f_1(z,z_2)+3\,\Gamma_t^{^3}\Gamma_t^{^3}-\Gamma_c^{^0}\Gamma_t^{^3}\right.\right.\\
&&\left.\left.+\Gamma_c^{^3}\Gamma_t^{^3} z f_1(z,z_c)\right)/z
%
+\Gamma\nu(\Gamma^{^3}_t+\Gamma^{^0}_t)/4
\right\}+2(\nu\Gamma_t^{^3})^2/z,\\
\label{eq:177}
\nonumber d\nu\Gamma_c^{^0}/d {l}&=&g\left\{\nu\left(3\Gamma_t^{^0}+3\Gamma_t^{^3}-\Gamma_c^{^0}-\Gamma_c^{^3}\right)/2+2\nu^2\left(3\Gamma_t^{^0}\Gamma_c^{^0}z f_1(z,z_2)-3\Gamma_t^{^3}\Gamma_c^{^0}\right.\right.\\
&&\left.\left.+\Gamma_c^{^0}\Gamma_c^{^0}-\Gamma_c^{^3}\Gamma_c^{^0}z f_1(z,z_c)\right)/z
%
+\Gamma\nu(\Gamma^{^0}_c+\Gamma^{^3}_c)/4
\right\}-2(\nu\Gamma_c^{^0})^2/z,\\
\nonumber d\nu\Gamma_c^{^3}/d {l}&=&g\left\{\nu\left(3\Gamma_t^{^0}+3\Gamma_t^{^3}-\Gamma_c^{^0}-\Gamma_c^{^3}\right)/2+4\nu^2\left(\Gamma_c^{^0}\Gamma_c^{^3}-3\Gamma_t^{^3}\Gamma_c^{^3}\right)/z +\right.\\
\nonumber &&\left.4\nu^3\left(3\Gamma_c^{^3}\Gamma_t^{^3}\Gamma_c^{^3}-\Gamma_c^{^3}\Gamma_c^{^0}\Gamma_c^{^3}\right)/z^2\right.\\
%
&&\left.+\Gamma\nu\Gamma^{^3}_c/4      
-\Gamma\nu^2\Gamma^{^3}_c\Gamma^{^3}_c/4z 
+\Gamma\nu\Gamma^{^0}_c\left(1-2\nu\Gamma^{^3}_c/z\right)^2 /4
\right\},\\
d \nu/ d{l} {\phantom{\Gamma_p}} &=&
g\nu\left\{3\nu\Gamma_t^{^0}f_1(z,z_2)-3\nu\Gamma_t^{^3}/z+\nu\Gamma_c^{^0}/z
-\nu\Gamma_c^{^3}f_1(z,z_c)  
%
+z\Gamma/8\right\},
\label{gammac3_sc}
\end{eqnarray}
where $z_2=z+2\nu\Gamma_t^{^0}$ and $z_c=z-2\nu\Gamma_c^{^3}$, which come from ladder summations in the triplet p-h channel, Eq. (\ref{t-ladder}), and in the 
staggered p-p Cooper channel,
\be
\Gamma_c^{^3}(q,\omega)=\Gamma_c^{^3}\frac{D q^2+z|\omega|}{D q^2+(z-2\nu\Gamma^{^3}_c)|\omega|}.
\ee
The last term of Eq.(\ref{eq:176}) comes also from the ladder summation in the 
staggered p-h triplet channel that present the same log-divergence 
of the standard Cooper 
channel, Eq.(\ref{eq:177}), namely it does not depend on $g$. As a result, for repulsive interaction, the Stoner instability towards a spin-density wave 
is not destroyed by disorder. 
At the first order in the scattering amplitudes, introducing
$u^{_0}_t=2\Gamma^{^0}_t/z$, $u^{_0}_c=2\Gamma^{^0}_c/z$,  
$u^{_3}_t=2\Gamma^{^3}_t/z$ and $u^{_3}_c=2\Gamma^{^3}_c/z$ 
the conductivity and the density of states are corrected in this way
\bea
\label{deltasigma_int}
&&\delta \sigma/\sigma = g\left[{3}( u^{_0}_t- u^{_3}_t)/2+
( u^{_0}_c- u^{_3}_c)/2\right]\ln (T\tau)^{-1},\\
\label{deltanu_int}
&&\delta \nu/\nu = g\left[{3}( u^{_0}_t- u^{_3}_t)/2+
( u^{_0}_c- u^{_3}_c)/2+{\Gamma}/{8}\right]\ln (T\tau)^{-1}.
\eea
%
%
%
\paragraph{Yes chiral, No $\hat {T}$. }

If time reversal invariance symmetry is broken both $\Gamma_c^{^0}$ and
$\Gamma_c^{^3}$ disappear in the equations above, $d c/ d{l}=-2 c^2$, and
\begin{eqnarray}
 d g/ d{l}\phantom{\Gamma_p}
 &=&g^2\nu\left\{-3\Gamma_t^{^0}f_2(z,z,z_2)+3\Gamma_t^{^3}/z\right\}/2,\\ 
 d z/ d{l} {\phantom{\Gamma_p}}
 &=&g\nu\left\{3\Gamma_t^{^0}-3\Gamma_t^{^3}
+z(\Gamma/4-1)/2\right\},\\
 \nonumber d\nu\Gamma^{^0}_t/d {l}&=&g\left\{\nu\left(\Gamma_t^{^0}+\Gamma_t^{^3}\right)/2+4\nu^2\,\left(\Gamma_t^{^0}\Gamma_t^{^0}-\Gamma_t^{^3}\Gamma_t^{^3}-\Gamma_t^{^0}\Gamma_t^{^3}\right)/z\right.
\left.-4\nu^3\Gamma_t^{^0}\Gamma_t^{^3}\Gamma_t^{^0}/z^2\right.\\
%
 &&\left.+(\Gamma/4-1)\nu\Gamma^{^0}_t
+(\Gamma/4-1)\nu^2\Gamma^{^0}_t\Gamma^{^0}_t/z
+\Gamma\nu\Gamma^{^3}_t\left(1+2\nu\Gamma^{^0}_t/z\right)^2 /4
\right\},\\
\nonumber d\nu\Gamma_t^{^3}/d {l}&=&g\left\{\nu\left(\Gamma_t^{^0}+\Gamma_t^{^3}\right)/2-2\nu^2\left(\Gamma_t^{^0}\Gamma_t^{^3} z f_1(z,z_2)+3\,\Gamma_t^{^3}\Gamma_t^{^3}\right)/z\right.\\
%
&&\left.+(\Gamma/4-1)\nu \Gamma^{^3}_t+\Gamma\nu \Gamma^{^0}_t/4
\right\}+2(\nu\Gamma_t^{^3})^2/z,\\
d \nu/ d{l} {\phantom{\Gamma_p}} &=&
g\nu\left\{3\nu\Gamma_t^{^0}f_1(z,z_2)-3\nu\Gamma_t^{^3}/z
%
+(\Gamma/4-1)/2\right\}.
\end{eqnarray}
%
\paragraph{Yes chiral, magnetic field. }

If the $d$-wave superconductor is 
embedded in a constant magnetic field, namely the spin rotation invariance is
broken, we have
\begin{eqnarray}
 d g/ d{l} {\phantom{\Gamma_p}} &=&g^2\left\{-\nu\Gamma_t^{^0}f_2(z,z,z_2)+2\nu\Gamma_t^{^3}/z-1/2\right\},\\
 d z/ d{l} {\phantom{\Gamma_p}} &=&g\left\{\nu\Gamma_t^{^0}-2\nu\Gamma_t^{^3}\right\},\\
d\nu\Gamma_t^{^0}/d {l}&=&g\left\{\nu\left(-\Gamma_t^{^0}/2+\Gamma_t^{^3}\right)-4\nu^2\Gamma_t^{^3}\Gamma_t^{^3}/z\right\},\\
d\nu\Gamma_t^{^3}/d {l}&=&g\left\{\nu\Gamma_t^{^0}/2-2\nu^2\left(3\Gamma_t^{^0}\Gamma_t^{^3}zf_1(z,z_2)+2\Gamma_t^{^3}\Gamma_t^{^3}\right)/z\right\}+2(\nu\Gamma_t^{^3})^2/z,\\
 d \nu/ d{l} {\phantom{\Gamma_p}} &=&g\nu\left\{\nu\Gamma_t^{^0}
f_1(z,z_2)-2\nu\Gamma_t^{^3}\right\}.
\end{eqnarray}
An effect of the SU($2$) symmetry breaking is to reduce the triplet scattering amplitudes as $\Gamma_t^{^0}\rightarrow \Gamma_t^{^0}/3$ and 
$\Gamma_t^{^3}\rightarrow 2\Gamma_t^{^3}/3$ in the conductance and in the DOS. 
Indeed for the standard modes the massless ones are those which commute with the Zeeman term, i.e. only one component in the triplet sector survives, while the staggered massless modes anticommute with the Zeeman term, i.e. there are two components in the triplet sector. Solving the equations above we
obtain that for moderate values of the interactions $g$ decreases, $u_t^{_0}$
is finite and $u_t^{_3}$ does not diverge if the starting 
value $u_{t0}^{_3}\sim -g^{1/2}$ or if $u_t^{_0}$ is positive and 
$u_{t0}^{_3}\sim 0$.
%
%
%
%
%
%
\paragraph{Yes chiral, magnetic impurities. }

In the presence of magnetic impurities we have 
\begin{eqnarray}
d g/ d{l} {\phantom{\Gamma_p}} &=&g^2\nu\Gamma_c^{^3}f_2(z,z,z_c),\\
d z/ d{l} {\phantom{\Gamma_p}} &=&g\nu\Gamma_c^{^3},\\
d\nu\Gamma_c^{^3}/d {l}&=&-g\nu\Gamma_c^{^3}/2,\\
d \nu/ d{l} {\phantom{\Gamma_p}} &=&g\nu\Gamma_c^{^3} f_1(z,z_c).
\end{eqnarray}
These equations are formally the same of Eq.(\ref{165}-\ref{168}), 
consistently to the fact that the soft modes live in the same manifold,
U($2n$)/U($n$)$\times$U($n$), where now $n=$ \#replicas $\times$
\#positive frequencies. At low interaction regime we have the 
solution $g=g_0\,e^{u_{c0}^{_3}}$, similar to Eq. (\ref{easysolution}) 
except for the sign in front of the interaction amplitude.

\subsection{In normal phase}
The role of staggered fluctuations can also be considered in the metallic
phase, extending the Finkel'stein model to include nesting effects. 
If superconductive order parameter is turned off and 
if staggered fluctuations are
supposed to be massless, the singlet contributions stay relevant and the
renormalization group equations become the following:

\paragraph{Yes chiral, Yes $\hat {T}$. }
In the presence of time reversal symmetry we obtain
\begin{eqnarray}
\nonumber d g/ d{l} {\phantom{\Gamma_p}} &=&g^2\left\{\nu\Gamma_s^{^0}f_2(z,z,z_1)-\nu\Gamma_s^{^3}/z-3\nu\Gamma_t^{^0}f_2(z,z,z_2)+3\nu\Gamma_t^{^3}/z-2\nu\Gamma_c^{^0}/z+2\nu\Gamma_c^{^3}f_2(z,z,z_c)\right\},\\
\nonumber d z/ d{l} {\phantom{\Gamma_p}} &=&g\left\{\nu\Gamma_s^{^3}-\nu\Gamma_s^{^0}+3\nu\Gamma_t^{^0}-3\nu\Gamma_t^{^3}+2\nu\Gamma_c^{^0}-2\nu\Gamma_c^{^3}
%
+z\left(1+\Gamma/8\right)      
\right\},\\
\nonumber d\nu\Gamma^{^0}_t/d {l}&=&g\left\{\nu\left(\Gamma_s^{^0}+\Gamma_s^{^3}+\Gamma_t^{^0}+\Gamma_t^{^3}+2\Gamma_c^{^0}+2\Gamma_c^{^3}\right)/2+4\nu^2\,\left(\Gamma_s^{^3}\Gamma_t^{^0}+\Gamma_t^{^0}\Gamma_t^{^0}+2\Gamma_t^{^0}\Gamma_c^{^0}-\Gamma_t^{^3}\Gamma_t^{^3}\right.\right.\\
\nonumber&&\left.\left.-\Gamma_t^{^0}\Gamma_t^{^3}\right)/z+4\nu^3\left(2\Gamma_t^{^0}\Gamma_c^{^0}\Gamma_t^{^0}-\Gamma_t^{^0}\Gamma_t^{^3}\Gamma_t^{^0}+\Gamma_t^{^0}\Gamma_s^{^3}\Gamma_t^{^0}\right)/z^2\right.\\
%
\nonumber&&\left.+\left(2+\Gamma/4\right) 
\left(\nu\Gamma^{^0}_t+\nu^2\Gamma^{^0}_t\Gamma^{^0}_t/z\right)+
\Gamma\nu\Gamma^{^3}_t \left(1+2\nu\Gamma^{^0}_t/z\right)^2/4
\right\},\\
\nonumber d\nu\Gamma_t^{^3}/d {l}&=&g\left\{\nu\left(\Gamma_s^{^0}+\Gamma_s^{^3}+\Gamma_t^{^0}+\Gamma_t^{^3}+2\Gamma_c^{^0}+2\Gamma_c^{^3}\right)/2-2\nu^2\left(\Gamma_t^{^0}\Gamma_t^{^3} z f_1(z,z_2)+3\,\Gamma_t^{^3}\Gamma_t^{^3}\right.\right.\\
\nonumber&&\left.\left.-2\Gamma_c^{^0}\Gamma_t^{^3}+2\Gamma_c^{^3}\Gamma_t^{^3} z f_1(z,z_c)+\Gamma_s^{^0}\Gamma_t^{^3} z f_1(z,z_1)-\Gamma_s^{^3}\Gamma_t^{^3}\right)/z \right.\\
%
\nonumber&&\left.+\left(2+\Gamma/4\right) \nu\Gamma^{^3}_t+\Gamma\nu \Gamma^{^0}_t /4
\right\}+2(\nu\Gamma_t^{^3})^2/z,\\
\nonumber d\nu\Gamma_s^{^0}/d {l}&=&g\left\{\nu\left(-\Gamma_s^{^0}-\Gamma_s^{^3}+3\Gamma_t^{^0}+3\Gamma_t^{^3}+2\Gamma_c^{^0}+2\Gamma_c^{^3}\right)/2
+4\nu^2\left(\Gamma_s^{^3}\Gamma_s^{^0}-3\Gamma_t^{^3}\Gamma_s^{^0}\right.\right.\\
\nonumber&&\left.+\Gamma_c^{^0}\Gamma_c^{^0}+\Gamma_c^{^3}\Gamma_c^{^3} z /z_c\right)/z 
+ 4\nu^3\left(-\Gamma_s^{^0}\Gamma_s^{^3}\Gamma_s^{^0}+3\Gamma_s^{^0}\Gamma_t^{^3}\Gamma_s^{^0}\right)/z^2 \\
\nonumber&&\left.-4\nu^2\left(2\Gamma_s^{^0}\Gamma_c^{^3}/z_c+\Gamma_s^{^0}\Gamma_s^{^0}\left(\frac{1}{z}-\frac{1}{z_c}\right)\right)\right.\\
%
\nonumber&&\left.+\left(2+\Gamma/4\right)
  \left(\nu\Gamma^{^0}_s-\nu^2\Gamma^{^0}_s\Gamma^{^0}_s/z\right)+\Gamma\nu
  \Gamma^{^3}_s \left(1-2\nu\Gamma^{^0}_s/z\right)^2 /4 
\right\},\\
\nonumber d\nu\Gamma_s^{^3}/d {l}&=&g\left\{\nu\left(-\Gamma_s^{^0}-\Gamma_s^{^3}+3\Gamma_t^{^0}+3\Gamma_t^{^3}+2\Gamma_c^{^0}+2\Gamma_c^{^3}\right)/2
+2\nu^2\left(-\Gamma_s^{^0}\Gamma_s^{^3}z f_1(z,z_1)\right.\right.\\
\nonumber&&+\Gamma_s^{^3}\Gamma_s^{^3}+3\Gamma_t^{^0}\Gamma_s^{^3}z f_1(z,z_2)-3\Gamma_t^{^3}\Gamma_s^{^3}+2\Gamma_c^{^0}\Gamma_s^{^3}-2\Gamma_c^{^3}\Gamma_s^{^3}z f_1(z,z_c)\\
\nonumber&& \left.\left.+4\Gamma_c^{^0}\Gamma_c^{^3}zf_1(z,z_c)\right)/z
%
+\left(2+\Gamma/4\right) \nu\Gamma^{^3}_s+\Gamma\nu \Gamma^{^0}_s /4
\right\}-2(\nu\Gamma_s^{^3})^2/z,\\
\nonumber d\nu\Gamma_c^{^0}/d {l}&=&g\left\{\nu\left(\Gamma_s^{^0}+\Gamma_s^{^3}+3\Gamma_t^{^0}+3\Gamma_t^{^3}\right)/2+2\nu^2\left(3\Gamma_t^{^0}\Gamma_c^{^0}z f_1(z,z_2)-3\Gamma_t^{^3}\Gamma_c^{^0}+\Gamma_s^{^3}\Gamma_c^{^0}\right.\right.\\
\nonumber&&-\Gamma_s^{^0}\Gamma_c^{^0}zf_1(z,z_1)+2\Gamma_c^{^0}\Gamma_c^{^0}-2\Gamma_c^{^3}\Gamma_c^{^0}z f_1(z,z_c)
+2\Gamma_c^{^0}\Gamma_s^{^0}zf_1(z,z_1)\\
\nonumber&&\left.\left.+2\Gamma_c^{^3}\Gamma_s^{^3}zf_1(z,z_c)\right)/z
%
+\left(2+\Gamma/4\right) \nu\Gamma^{^0}_c+\Gamma \nu \Gamma^{^3}_c /4
\right\}-2(\nu\Gamma_c^{^0})^2/z,\\
\nonumber d\nu\Gamma_c^{^3}/d {l}&=&g\left\{\nu\left(\Gamma_s^{^0}+\Gamma_s^{^3}+3\Gamma_t^{^0}+3\Gamma_t^{^3}\right)/2+4\nu^2\left(\Gamma_c^{^0}\Gamma_c^{^3}+\Gamma_s^{^0}\Gamma_c^{^3}z(f_1(z_c,z_1)-f_1(z,z_1))\right.\right.\\
\nonumber &&\left.-3\Gamma_t^{^3}\Gamma_c^{^3}+\Gamma_s^{^3}\Gamma_c^{^0}\right)/z+4\nu^3\left(3\Gamma_c^{^3}\Gamma_t^{^3}\Gamma_c^{^3}-2\Gamma_c^{^3}\Gamma_c^{^0}\Gamma_c^{^3}+\Gamma_c^{^3}\Gamma_s^{^3}\Gamma_c^{^3}\right)/z^2\\
\nonumber &&\left.
-4\nu^2\Gamma_c^{^3}\Gamma_c^{^3}\left(f_1(z_1,z_c)+f_1(z,z_1)-2f_1(z,z_c)+\frac{1}{z}\right)\right.\\
%
\nonumber&&\left.+\left(2+\Gamma/4\right)
 \left(\nu\Gamma^{^3}_c-\nu^2\Gamma^{^3}_c\Gamma^{^3}_c/z\right)+\Gamma \nu
 \Gamma^{^0}_c \left(1-2\nu\Gamma^{^3}_c/z\right)^2 /4 
\right\},\\
\nonumber d \nu/ d{l} {\phantom{\Gamma_p}} &=&g\nu^2\left\{\Gamma_s^{^3}-\Gamma_s^{^0}zf_1(z,z_1)+3\Gamma_t^{^0}zf_1(z,z_2)-3\Gamma_t^{^3}+2\Gamma_c^{^0}-2\Gamma_c^{^3}zf_1(z,z_c)
%
+z\left(1+\Gamma/8\right) 
\right\}/z,
\end{eqnarray}
where $z_1=z-2\nu\Gamma_s^{^0}$ and $z_2$, $z_c$ as before.
$z_1$ is related to the compressibility \cite{Castel84}, here it appears 
in the diffusion propagators, modified by singlet p-h channel ladder summation 
\[
\Gamma_s^{^0}(q,\omega)=\Gamma_s^{^{0}}\frac{D q^2+z|\omega|}{D q^2+(z-2\nu\Gamma_s^{^0})|\omega|}.
\] 
If chiral symmetry is broken, the last contributions in the curly
brackets proportional to $(2+\Gamma/4)$ and to $\Gamma$ together with all the
contributions proportions to the $\Gamma^{^3}$s vanish and 
we obtain again the standard Finkel'stein equations \cite{Finkelstein,Fin84}.
The equations above, although rather complex, hide a very amazing property, they in fact are symmetric with respect to the following transformation
\begin{eqnarray*}
\Gamma_s^{^0}=\Gamma_c^{^3}\longleftrightarrow -\Gamma_t^{^0}, \\
\Gamma_s^{^3}=\Gamma_c^{^0}\longleftrightarrow -\Gamma_t^{^3}.
\end{eqnarray*}
This symmetric property represents the invariance with respect to the particle-hole symmetry transformation, 
$c_{i\uparrow}=
d_{i\uparrow}$, 
$c_{i\downarrow}=(-)^id^{\dagger}_{i\downarrow}$, 
that maps the charge to the spin and viceversa. 

\paragraph{Yes chiral, No $\hat {T}$. }
If time reversal symmetry is broken the renormalization group equations are
found to be 
formally the same obtained in the superconductive case, when time reversal
symmetry holds, Eqs.(\ref{g_sc_stag}-\ref{gammac3_sc}), 
%
%
with the substitutions
\begin{eqnarray*}
\Gamma_s^{^0}\rightarrow \Gamma_c^{^3}\\
\Gamma_s^{^3}\rightarrow \Gamma_c^{^0}
\end{eqnarray*}
This accidentally simple equivalence 
of the equations is 
consistent with the fact that in both the cases the soft modes take values in
the same coset U($4n$)$\times$U($4n$)/U($4n$).

\paragraph{Yes chiral, magnetic field. }

In the presence of a constant magnetic field we have
\begin{eqnarray*}
\nonumber d g/ d{l} {\phantom{\Gamma_p}} &=&g^2\left\{\nu\Gamma_s^{^0}f_2(z,z,z_1)-\nu\Gamma_t^{^0}f_2(z,z,z_2)+2\nu\Gamma_t^{^3}/z\right\},\\
\nonumber d z/ d{l} {\phantom{\Gamma_p}} &=&g\left\{-\nu\Gamma_s^{^0}+\nu\Gamma_t^{^0}-2\nu\Gamma_t^{^3}\right\},\\
d\nu\Gamma_t^{^0}/d {l}&=&g\left\{\nu\left(\Gamma_s^{^0}/2-\Gamma_t^{^0}/2+\Gamma_t^{^3}\right)-4\nu^2\Gamma_t^{^3}\Gamma_t^{^3}/z\right\},\\
d\nu\Gamma_t^{^3}/d {l}&=&g\left\{\nu\Gamma_t^{^0}/2-2\nu^2\left(3\Gamma_t^{^0}\Gamma_t^{^3}zf_1(z,z_2)+2\Gamma_t^{^3}\Gamma_t^{^3}+\Gamma_s^{^0}\Gamma_t^{^3}zf_1(z,z_1)\right)/z\right\}+2(\nu\Gamma_t^{^3})^2,\\
d\nu\Gamma_s^{^0}/d {l}&=&g\left\{\nu(-\Gamma_s^{^0}+\Gamma_t^{^0}+2\Gamma_t^{^3})/2-8\nu^2\Gamma_t^{^3}\Gamma_s^{^0}/z+4\nu^3\Gamma_s^{^0}\Gamma_t^{^3}\Gamma_s^{^0}/z^2\right\},\\
\nonumber d z/ d{l} {\phantom{\Gamma_p}}
&=&g\nu\left\{-\nu\Gamma_s^{^0}f_1(z,z_1)+\nu\Gamma_t^{^0}f_1(z,z_2)
-2\nu\Gamma_t^{^3}\right\}.
\end{eqnarray*}
\paragraph{Yes chiral, magnetic impurities. }
While with magnetic impurities we have
\begin{eqnarray*}
\nonumber d g/ d{l} {\phantom{\Gamma_p}} &=&{\phantom{+}}g^2\left\{\nu\Gamma_s^{^0}f_2(z,z,z_1)+2\nu\Gamma_c^{^3}f_2(z,z,z_c)+1/2\right\},\\
\nonumber d z/ d{l} {\phantom{\Gamma_p}} &=&-g\left\{\nu\Gamma_s^{^0} + 2\nu\Gamma_c^{^3} + z/2\right\},\\
d\nu\Gamma_s^{^0}/d {l}&=&g\left\{\nu\Gamma_c^{^3}-\frac{3}{2}\nu\Gamma_s^{^0} 
+ \nu^2\Gamma_s^{^0}\Gamma_s^{^0}/z  +4\nu^2\Gamma_c^{^3}\Gamma_c^{^3}/z_c -4\nu^2\left(2\Gamma_s^{^0}\Gamma_c^{^3}/z_c+\Gamma_s^{^0}\Gamma_s^{^0}\left(\frac{1}{z}-\frac{1}{z_c}\right)\right)\right\},\\
\nonumber d\nu\Gamma_c^{^3}/d {l}&=&g\left\{\nu\Gamma_s^{^0}/2+4\nu^2\Gamma_s^{^0}\Gamma_c^{^3}\left(f_1(z_c,z_1)-f_1(z,z_1)\right)-\nu\Gamma^{^3}_c+\nu^2\Gamma^{^3}_c\Gamma^{^3}_c/z\right.\\
\nonumber &&\left.
-4\nu^2\Gamma_c^{^3}\Gamma_c^{^3}\left(f_1(z_1,z_c)+f_1(z,z_1)-2f_1(z,z_c)+\frac{1}{z}\right)\right\},\\
\nonumber d \nu/ d{l} {\phantom{\Gamma_p}} &=&-g\nu\left\{\nu\Gamma_s^{^0}f_1(z,z_1) + 2\nu\Gamma_c^{^3}f_1(z,z_c) + 1/2\right\}.
\end{eqnarray*}
Notice that in this case 
by the transformation 
$\Gamma_s^{^0}=\Gamma_c^{^3}\rightarrow -\Gamma_t^{^0}$ one  
obtains the same equations of the superconductive case with broken time reversal
and chiral symmetries, 
analyzed in paragraph $b$, since 
also in the latter case the soft modes take values
in the same manifold Sp($2n$)/U($2n$).  \\

Now we can also study the scaling behavior of the quantities
$z_1=z-2\nu\Gamma_s^{^0}$ and $z_2=z+2\nu\Gamma_t^{^0}$ 
that are related by Ward identities 
to the compressibility and to the static spin susceptibility
\cite{Castel84}. As in \cite{Finkelstein,Castel84}, because the ladder terms
do not involve the small parameter $g$,
we can consider the situation in which only the lowest order with respect to
$\Gamma_c^{^0}$ and $\Gamma_t^{^3}$ (by choosing $\Gamma_{c0}^{^0}> g^{1/2}$
and $\Gamma_{t0}^{^3}< -g^{1/2}$) 
are retained in all the equations, consequently, we have 
for $z_1$ and $z_2$ the following equations:\\  
($i$) if time reversal symmetry holds
\begin{eqnarray*}
d\left(z-2\nu\Gamma_s^{^0}\right)/d{l}&=&2g\nu\left(\Gamma_s^{^3}
-3\Gamma_t^{^3}-2\Gamma_c^{^3}\frac{z}{z_c}+\frac{z}{2}
\left(1+\frac{\Gamma}{8}\right)
-\frac{\Gamma}{4}\Gamma_s^{^3} 
\right)\left(z-2\nu\Gamma_s^{^0}\right)^2/z^2,\\
d\left(z+2\nu\Gamma_t^{^0}\right)/d
{l}&=&2g\nu\left(\Gamma_s^{^3}-\Gamma_t^{^3}+2\Gamma_c^{^0}+2\Gamma_t^{^0}
\frac{z}{z_2}+\frac{z}{2}
\left(1+\frac{\Gamma}{8}\right)+ 
\frac{\Gamma}{4}\Gamma_t^{^3}
\right)\left(z+2\nu\Gamma_t^{^0}\right)^2/z^2,
\end{eqnarray*}
($j$) if time reversal symmetry is broken
\begin{eqnarray*}
d\left(z-2\nu\Gamma_s^{^0}\right)/d {l}&=&2g\nu\left(\Gamma_s^{^3}-3\Gamma_t^{^3}
+{z\Gamma}/{16}-{\Gamma}\Gamma_s^{^3}/4 \right)
\left(z-2\nu\Gamma_s^{^0}\right)^2/z^2,\\
d\left(z+2\nu\Gamma_t^{^0}\right)/d
{l}&=&2g\nu\left(\Gamma_s^{^3}-\Gamma_t^{^3}+2\Gamma_t^{^0}\frac{z}{z_2}
+{z\Gamma}/{16}+{\Gamma}\Gamma_t^{^3}/4
\right)\left(z+2\nu\Gamma_t^{^0}\right)^2/z^2,
\end{eqnarray*}
($k$) with constant magnetic field
\begin{eqnarray*}
d\left(z-2\nu\Gamma_s^{^0}\right)/d {l}&=&
-4g\nu\Gamma_t^{^3}\left(z-2\nu\Gamma_s^{^0}\right)^2/z^2,\\
d\left(z+2\nu\Gamma_t^{^0}\right)/d {l}&=&0,
\end{eqnarray*}
($l$) with magnetic impurities
\begin{eqnarray*}
d(z-2\nu\Gamma_s^{^0})/d
{l}&=&-2g\nu\left(2\Gamma_c^{^3}\frac{z}{z_c}
+\frac{z}{4}\right)\left(z-2\nu\Gamma_s^{^0}\right)^2/z^2. 
\end{eqnarray*}
Notice that in all the cases, when chirality is spoiled, $z_1$ is not
renormalized as already pointed out previously 
in Refs.\cite{Finkelstein,Castel84}.

\section{Conclusions}

In this work we have analyzed the role of disorder in $d$-wave 
superconductors, which have gapless Landau-Bogoliubov quasiparticle 
excitations.  We have considered several universality classes, including 
the chiral symmetry which occurs at half-filling for a two-sublattice 
model. 
In addition, we have studied in details the effects of the 
residual quasiparticle interaction. 
The main results of this work are summarized in the following. 

In the presence of non magnetic impurities the spin 
conductivity is suppressed by quantum interference corrections, in 
agreement with Ref. \cite{Fisher}. The density of states vanishes 
in the insulating regime.  

On the contrary, surprisingly, magnetic 
impurities gives a delocalization correction to the conductivity 
meanwhile enhancing the density of states \cite{FDC}.

If chiral symmetry is present, namely at half filling 
for a two-sublattice model, the spin stays delocalized in spite of disorder 
and the conductivity 
remains finite.  The DOS diverges in the absence of magnetic field and 
magnetic impurities \cite{FDC}.
 
The quasiparticle charge conductivity, namely the optical conductivity at small frequency, has in general the same behavior as 
the spin conductivity. 
However, when chiral symmetry holds and time reversal symmetry is broken,  
the dirty $d$-wave superconductor behaves like a spin metal but 
charge insulator (at finite small frequency, excluding the Drude peak of the superflow), manifesting a sort of spin-charge separation.

Charge fluctuations as well as fluctuations of the 
real part of the order parameter, assuming the average value to be real,  
are not diffusive in a 
superconductor. Therefore the residual quasiparticle interaction 
written in terms of the diffusive modes only contains the spin-triplet 
particle-hole channel and the Cooper channel representing 
$s$-wave fluctuations of the imaginary part of the order parameter. 

For repulsive interaction, particles and holes repel each other in the 
spin-triplet channel, hence opposing localization. In fact we find 
that a repulsive residual interaction      
gives a delocalizing correction to the conductivity and enhances the 
density of states. 
Namely the interaction can compete with quantum interferences
due to the disorder. 

According to the symmetry of the system, we have found that the interactions can be i) relevant, 
when both time reversal symmetry, $\hat{T}$, 
and spin rotational invariance, SU($2$), are preserved, ii) marginal, when 
$\hat{T}$ is broken and SU($2$) holds, iii) irrelevant, when both $\hat{T}$ 
and SU($2$) are broken.

We have also studied $(\pi,\pi)$ momentum transfer 
interactions, since they are coupled to diffusive staggered 
spin fluctuations at half-filling. We find that the corrections 
to the conductivity due to the interaction at $(\pi,\pi)$ have opposite 
sign than the corrections coming from the interaction at small 
momentum. Moreover, we notice that an analog of the Anderson theorem for
{\it s}-wave superconductors holds at half-filling for staggered density
fluctuations. Namely, since these modes are diffusive, the staggered
susceptibility remains log-divergent even in the presence of disorder. As a
result, for repulsive interaction, the Stoner instability towards a
spin-density wave is not destroyed by disorder.

In summary we have derived the complete sets of RG equations for many 
different symmetries at the lowest order in the small
parameter $g$, which does not depend on the disorder, and at all
orders in the interaction amplitudes, $\nu\Gamma_t$, $\nu\Gamma_c$, 
which are the real phenomenological 
parameters, related instead to the impurity concentration 
through the density of states
proportional to the scattering rate, 
$\nu\propto \Sigma=\frac{1}{2\tau}$. 
We have derived the RG equations 
even in the normal phase when nesting effects occur. In the latter
case we find that the compressibility is a renormalized quantity 
unlike the standard case \cite{Finkelstein,Castel84} 
when the chiral symmetry is not present.

\section*{Acknowledgments}
The author would like to thank Michele Fabrizio and Claudio Castellani 
for useful discussions.

\appendix
\section{Transverse modes}
In this Appendix we summarize the properties of the $W$ matrices for soft 
modes needed to evaluate the gaussian averages. S (A) and R (I) mean
symmetric (antisymmetric) in the replica space and real (imaginary) $s_i$,
$i=0,1,2,3$, components defined in Eqs.(\ref{Wspin}-\ref{W_a}). Notice that
the number of soft modes decreases by decreasing the symmetry since 
some of them become massive. 
\begin{enumerate}
\item {\bf Without sublattice symmetry}
\begin{enumerate}
\item {\bf Without superconducting order parameter}
\begin{enumerate}
\item {\bf With time reversal symmetry}:
  $\textrm{Sp}(4n)/\textrm{Sp}(2n)\times \textrm{Sp}(2n).$ 
\[
\begin{array}{||c||c|c||}\hline
W_0            &  s_1      &  s_2       \\ \hline\hline
W_{S0}        & {\rm A,R} & {\rm S,I}  \\ \hline
W_{S1}        & {\rm A,I} & {\rm S,R}  \\ \hline
W_{S2}        & {\rm A,I} & {\rm S,R}  \\ \hline
W_{S3}        & {\rm S,R} & {\rm A,I}  \\ \hline
\vec{W}_{T0}  & {\rm S,R} & {\rm A,I}  \\ \hline
\vec{W}_{T1}  & {\rm S,I} & {\rm A,R}  \\ \hline
\vec{W}_{T2}  & {\rm S,I} & {\rm A,R}  \\ \hline
\vec{W}_{T3}  & {\rm A,R} & {\rm S,I}  \\ \hline
\end{array}
\]
%
%
\item {\bf Without time reversal symmetry}: $\textrm{U}(4n)/
  \textrm{U}(2n)\times \textrm{U}(2n).$ 
\[
\begin{array}{||c||c|c||}\hline
W_0            &  s_1      &  s_2      \\ \hline\hline
W_{S0}        & {\rm A,R} & {\rm S,I} \\ \hline
W_{S3}        & {\rm S,R} & {\rm A,I} \\ \hline
\vec{W}_{T0}  & {\rm S,R} & {\rm A,I}   \\ \hline
\vec{W}_{T3}  & {\rm A,R} & {\rm S,I}   \\ \hline
\end{array}
\]
%
\item {\bf With magnetic field}: $\textrm{U}(2n)\times\textrm{U}(2n)/
  \textrm{U}(2n).$ 
\[
\begin{array}{||c||c|c||}\hline
W_0            &  s_1      &  s_2      \\ \hline\hline
W_{S0}        & {\rm A,R} & {\rm S,I} \\ \hline
W_{S3}        & {\rm S,R} & {\rm A,I} \\ \hline
{W}_{T_z0}  & {\rm S,R} & {\rm A,I}  \\ \hline
{W}_{T_z3}  & {\rm A,R} & {\rm S,I} \\ \hline
\end{array}
\]
%
\item {\bf With magnetic impurities}: $\textrm{U}(2n)/\textrm{U}(n)\times
  \textrm{U}(n).$ 
\[
\begin{array}{||c||c|c||}\hline
W_0            &  s_1      &  s_2       \\ \hline\hline
W_{S0}        & {\rm A,R} & {\rm S,I}  \\ \hline
W_{S3}        & {\rm S,R} & {\rm A,I}  \\ \hline
\end{array}
\]
\end{enumerate}
%
\item {\bf With superconducting order parameter}
\begin{enumerate}
\item {\bf With time reversal symmetry}: $\textrm{Sp}(2n)\times
  \textrm{Sp}(2n)/ \textrm{Sp}(2n).$ 
\[
\begin{array}{||c||c|c||}\hline
W_0            &  s_1      &  s_2       \\ \hline\hline
W_{S0}        & {\rm A,R} & {       }  \\ \hline
W_{S1}        & {       } & {\rm S,R}  \\ \hline
W_{S2}        & {\rm A,I} & {       }  \\ \hline
W_{S3}        & {       } & {\rm A,I}  \\ \hline
\vec{W}_{T0}  & {\rm S,R} & {       }  \\ \hline
\vec{W}_{T1}  & {       } & {\rm A,R}  \\ \hline
\vec{W}_{T2}  & {\rm S,I} & {       }  \\ \hline
\vec{W}_{T3}  & {       } & {\rm S,I}  \\ \hline
\end{array}
\]
%
\item {\bf Without time reversal symmetry}: $\textrm{Sp}(2n)/\textrm{U}(2n).$
\[
\begin{array}{||c||c|c||}\hline
W_0            &  s_1      &  s_2      \\ \hline\hline
W_{S0}        & {\rm A,R} & {       } \\ \hline
W_{S3}        & {       } & {\rm A,I} \\ \hline
\vec{W}_{T0}  & {\rm S,R} & {       } \\ \hline
\vec{W}_{T3}  & {       } & {\rm S,I} \\ \hline
\end{array}
\]
%
\item {\bf With magnetic field}:
  $\textrm{U}(2n)/\textrm{U}(n)\times\textrm{U}(n).$ 
\[
\begin{array}{||c||c|c||}\hline
W_0            &  s_1      &  s_2      \\ \hline\hline
W_{S0}        & {\rm A,R} & {       } \\ \hline
W_{S3}        & {       } & {\rm A,I} \\ \hline
{W}_{T_z0}  & {\rm S,R} & {       } \\ \hline
{W}_{T_z3}  & {       } & {\rm S,I} \\ \hline
\end{array}
\]
%
\item {\bf With magnetic impurities}: $\textrm{O}(2n)/ \textrm{U}(n).$
\[
\begin{array}{||c||c|c||}\hline
W_0           &  s_1      &  s_2      \\ \hline\hline
W_{S0}       & {\rm A,R} & {       } \\ \hline
W_{S3}       & {       } & {\rm A,I} \\ \hline
\end{array}
\]
%
\end{enumerate}
\end{enumerate}
\item{\bf With sublattice symmetry}
\begin{enumerate}
\item{\bf Without superconducting order parameter}
\begin{enumerate}
\item{\bf With time reversal symmetry}: $\textrm{U}(8n)/ \textrm{Sp}(4n).$
\begin{center}
\begin{minipage}{4cm}
\[
\begin{array}{||c||c|c||}\hline
W_0            &  s_1      &  s_2       \\ \hline\hline
W_{S0}        & {\rm A,R} & {\rm S,I}  \\ \hline
W_{S1}        & {\rm A,I} & {\rm S,R}  \\ \hline
W_{S2}        & {\rm A,I} & {\rm S,R}  \\ \hline
W_{S3}        & {\rm S,R} & {\rm A,I}  \\ \hline
\vec{W}_{T0}  & {\rm S,R} & {\rm A,I}  \\ \hline
\vec{W}_{T1}  & {\rm S,I} & {\rm A,R}  \\ \hline
\vec{W}_{T2}  & {\rm S,I} & {\rm A,R}  \\ \hline
\vec{W}_{T3}  & {\rm A,R} & {\rm S,I}  \\ \hline
\end{array}
\]
\end{minipage}
\hspace{1cm}
\begin{minipage}{4cm}
\[
\begin{array}{||c||c|c||}\hline
W_3 &             s_0      &  s_3      \\ \hline\hline
W_{S0}        & {\rm S,I} & {\rm S,I} \\ \hline
W_{S1}        & {\rm S,R} & {\rm S,R} \\ \hline
W_{S2}        & {\rm S,R} & {\rm S,R} \\ \hline
W_{S3}        & {\rm A,I} & {\rm A,I} \\ \hline
\vec{W}_{T0}  & {\rm A,I} & {\rm A,I} \\ \hline
\vec{W}_{T1}  & {\rm A,R} & {\rm A,R} \\ \hline
\vec{W}_{T2}  & {\rm A,R} & {\rm A,R} \\ \hline
\vec{W}_{T3}  & {\rm S,I} & {\rm S,I} \\ \hline
\end{array}
\]
\end{minipage}
\end{center}
\vspace{0.4cm}
%
%
\item{\bf Without time reversal symmetry}:
  $\textrm{U}(4n)\times\textrm{U}(4n)/ \textrm{U}(4n).$ 
\begin{center}
\begin{minipage}{4cm}
\[
\begin{array}{||c||c|c||}\hline
W_0            &  s_1      &  s_2      \\ \hline\hline
W_{S0}        & {\rm A,R} & {\rm S,I} \\ \hline
W_{S3}        & {\rm S,R} & {\rm A,I} \\ \hline
\vec{W}_{T0}  & {\rm S,R} & {\rm A,I}   \\ \hline
\vec{W}_{T3}  & {\rm A,R} & {\rm S,I}   \\ \hline
\end{array}
\]
\end{minipage}
\hspace{1cm}
\begin{minipage}{4cm}
\[
\begin{array}{||c||c|c||}\hline
W_3 &             s_0      &  s_3      \\ \hline\hline
W_{S0}        & {\rm S,I} & {\rm S,I} \\ \hline
W_{S3}        & {\rm A,I} & {\rm A,I} \\ \hline
\vec{W}_{T0}  & {\rm A,I} & {\rm A,I} \\ \hline
\vec{W}_{T3}  & {\rm S,I} & {\rm S,I} \\ \hline
\end{array}
\]
\end{minipage}
\end{center}
\vspace{0.4cm}
%
\item{\bf With magnetic field}: $\textrm{U}(4n)/
  \textrm{U}(2n)\times\textrm{U}(2n).$ 
\begin{center}
\begin{minipage}{4cm}
\[
\begin{array}{||c||c|c||}\hline
W_0            &  s_1      &  s_2      \\ \hline\hline
W_{S0}        & {\rm A,R} & {\rm S,I} \\ \hline
W_{S3}        & {\rm S,R} & {\rm A,I} \\ \hline
{W}_{T_z0}  & {\rm S,R} & {\rm A,I}  \\ \hline
{W}_{T_z3}  & {\rm A,R} & {\rm S,I} \\ \hline
\end{array}
\]
\end{minipage}
\hspace{1cm}
\begin{minipage}{4cm}
\[
\begin{array}{||c||c|c||}\hline
W_3 &             s_0    &  s_3      \\ \hline\hline
{W}_{T_x0}  & {\rm A,I} & {\rm A,I} \\ \hline
{W}_{T_x3}  & {\rm S,I} & {\rm S,I} \\ \hline
{W}_{T_y0}  & {\rm A,I} & {\rm A,I} \\ \hline
{W}_{T_y3}  & {\rm S,I} & {\rm S,I} \\ \hline
\end{array}
\]
\end{minipage}
\end{center}
\vspace{0.4cm}
%
\item{\bf With magnetic impurities}: $\textrm{Sp}(2n)/ \textrm{U}(2n).$
\begin{center}
\begin{minipage}{4cm}
\[
\begin{array}{||c||c|c||}\hline
W_0            &  s_1      &  s_2       \\ \hline\hline
W_{S0}        & {\rm A,R} & {\rm S,I}  \\ \hline
W_{S3}        & {\rm S,R} & {\rm A,I}  \\ \hline
\end{array}
\]
\end{minipage}
\hspace{1cm}
\begin{minipage}{4cm}
\[
\begin{array}{||c||c|c||}\hline
W_3 &             s_0      &  s_3      \\ \hline\hline
W_{S1}        & {\rm S,R} & {\rm S,R} \\ \hline
W_{S2}        & {\rm S,R} & {\rm S,R} \\ \hline
\end{array}
\]
\end{minipage}
\end{center}
%
\end{enumerate}
%
\vspace{0.4cm}
\item{\bf With superconducting order parameter}
\begin{enumerate}
\item{\bf With time reversal symmetry}: $\textrm{U}(4n)\times\textrm{U}(4n)/
  \textrm{U}(4n).$ 
\begin{center}
\begin{minipage}{4cm}
\[
\begin{array}{||c||c|c||}\hline
W_0            &  s_1      &  s_2       \\ \hline\hline
W_{S0}        & {\rm A,R} & {       }  \\ \hline
W_{S1}        & {       } & {\rm S,R}  \\ \hline
W_{S2}        & {\rm A,I} & {       }  \\ \hline
W_{S3}        & {       } & {\rm A,I}  \\ \hline
\vec{W}_{T0}  & {\rm S,R} & {       }  \\ \hline
\vec{W}_{T1}  & {       } & {\rm A,R}  \\ \hline
\vec{W}_{T2}  & {\rm S,I} & {       }  \\ \hline
\vec{W}_{T3}  & {       } & {\rm S,I}  \\ \hline
\end{array}
\]
\end{minipage}
\hspace{1cm}
\begin{minipage}{4cm}
\[
\begin{array}{||c||c|c|c|c||}\hline
W_3 &             s_0      &  s_3      \\ \hline\hline
W_{S0}        & {\rm S,I} & {       } \\ \hline
W_{S1}        & {       } & {\rm S,R} \\ \hline
W_{S2}        & {\rm S,R} & {       } \\ \hline
W_{S3}        & {       } & {\rm A,I} \\ \hline
\vec{W}_{T0}  & {\rm A,I} & {       } \\ \hline
\vec{W}_{T1}  & {       } & {\rm A,R} \\ \hline
\vec{W}_{T2}  & {\rm A,R} & {       } \\ \hline
\vec{W}_{T3}  & {       } & {\rm S,I} \\ \hline
\end{array}
\]
\end{minipage}
\end{center}
\vspace{0.4cm}
%
%
\item{\bf Without time reversal symmetry}: $\textrm{U}(4n)/ \textrm{O}(4n).$
\begin{center}
\begin{minipage}{4cm}
\[
\begin{array}{||c||c|c||}\hline
W_0            &  s_1      &  s_2      \\ \hline\hline
W_{S0}        & {\rm A,R} & {       } \\ \hline
W_{S3}        & {       } & {\rm A,I} \\ \hline
\vec{W}_{T0}  & {\rm S,R} & {       } \\ \hline
\vec{W}_{T3}  & {       } & {\rm S,I} \\ \hline
\end{array}
\]
\end{minipage}
\hspace{1cm}
\begin{minipage}{4cm}
\[
\begin{array}{||c||c|c||}\hline
W_3 &             s_0      &  s_3      \\ \hline\hline
W_{S0}        & {\rm S,I} & {       } \\ \hline
W_{S3}        & {       } & {\rm A,I} \\ \hline
\vec{W}_{T0}  & {\rm A,I} & {       } \\ \hline
\vec{W}_{T3}  & {       } & {\rm S,I} \\ \hline
\end{array}
\]
\end{minipage}
\end{center}
\vspace{0.4cm}
%
\item{\bf With magnetic field}: $\textrm{O}(4n) /
  \textrm{O}(2n)\times\textrm{O}(2n).$ 
\begin{center} 
\begin{minipage}{4cm}
\[
\begin{array}{||c||c|c||}\hline
W_0            &  s_1      &  s_2      \\ \hline\hline
W_{S0}        & {\rm A,R} & {       } \\ \hline
W_{S3}        & {       } & {\rm A,I} \\ \hline
{W}_{T_z0}  & {\rm S,R} & {       } \\ \hline
{W}_{T_z3}  & {       } & {\rm S,I} \\ \hline
\end{array}
\]
\end{minipage}
\hspace{1cm}
\begin{minipage}{4cm}
\[
\begin{array}{||c||c|c||}\hline
W_3 &             s_0      &  s_3      \\ \hline\hline
{W}_{T_x0}  & {\rm A,I} & {       } \\ \hline
{W}_{T_x3}  & {       } & {\rm S,I} \\ \hline
{W}_{T_y0}  & {\rm A,I} & {       } \\ \hline
{W}_{T_y3}  & {       } & {\rm S,I} \\ \hline
\end{array}
\]
\end{minipage}
\end{center}
\vspace{0.4cm}
%
\item{\bf With magnetic impurities}:
  $\textrm{U}(2n)/\textrm{U}(n)\times\textrm{U}(n).$ 
\begin{center}
\begin{minipage}{4cm}
\[
\begin{array}{||c||c|c||}\hline
W_0           &  s_1      &  s_2      \\ \hline\hline
W_{S0}       & {\rm A,R} & {       } \\ \hline
W_{S3}       & {       } & {\rm A,I} \\ \hline
\end{array}
\]
\end{minipage}
\hspace{1cm}
\begin{minipage}{4cm}
\[
\begin{array}{||c||c|c|c|c||}\hline
W_3 &             s_0      &  s_3      \\ \hline\hline
W_{S1}        & {       } & {\rm S,R} \\ \hline
W_{S2}        & {\rm S,R} & {       } \\ \hline
\end{array}
\]
\end{minipage}
\end{center}
\end{enumerate}
\end{enumerate}
\end{enumerate}

\section{Charge and spin Vector potentials}

Let us consider an operator diagonal in the Nambu space. Namely
\[
{\cal A}_{ij} =
\left(
\begin{array}{cc}
{\cal A}_{\uparrow,ij} & 0 \\
0 & {\cal A}_{\downarrow,ij} \\
\end{array}
\right),
\]
where the matrix elements are matrices in the retarded/advanced and replica
space.
If we take ${\cal A}_{ii}=0$, then such an operator corresponds to
\begin{eqnarray*}
\sum_{ij} c^\dagger_{i\uparrow}{\cal A}_{\uparrow,ij}
c^{\phantom{\dagger}}_{j\uparrow}
- c^\dagger_{i\downarrow}{\cal A}^t_{\downarrow,ji}
c^{\phantom{\dagger}}_{j\downarrow}
 = \sum_{ij} c^\dagger_i
\left[\frac{1}{2}\left( {\cal A}_{\uparrow,ij} - 
{\cal A}^t_{\downarrow,ji}\right)
+\frac{1}{2}\sigma_z \left(
{\cal A}_{\uparrow,ij} + {\cal A}^t_{\downarrow,ji}\right)
\right] c^{\phantom{\dagger}}_j.
\end{eqnarray*}
In the path integral formalism, a generic operator diagonal in the
Nambu space,
\[
\sum_{ij} \bar{\Psi}_i A_{ij} \Psi_j,
\]
with
\[
A_{ij} =
\left(
\begin{array}{cc}
A_{1,ij} & 0 \\
0 & A_{2,ij} \\
\end{array}
\right),
\]
corresponds instead to 
\[
\frac{1}{2}\sum_{ij} \bar{c}_i \left[ A_{1,ji}^t
+ \sigma_y A_{2,ij} \sigma_y \right] c_j.
\]
By comparison we have that 
\[
\left[ A_{1,ji}^t  
+ \sigma_y A_{2,ij} \sigma_y \right] =
\left[\left( {\cal A}_{\uparrow,ij} - 
{\cal A}^t_{\downarrow,ji}\right)
+\sigma_z \left(
{\cal A}_{\uparrow,ij} + {\cal A}^t_{\downarrow,ji}\right)
\right].
\]
Suppose that the operators in question are currents. Then 
${\cal A}_{ij} = - {\cal A}_{ji}$, and the above relation reads
\[
\left[ - A_{1,ij}^t
+ \sigma_y A_{2,ij} \sigma_y \right] =
\left[\left( {\cal A}_{\uparrow,ij} +
{\cal A}^t_{\downarrow,ij}\right)
+\sigma_z \left(
{\cal A}_{\uparrow,ij} - {\cal A}^t_{\downarrow,ij}\right)
\right].
\]
In general we can consider either a charge current, implying
$\cal{A}_\uparrow=\cal{A}_\downarrow=\cal{A}$, 
or a spin current, in which case
$\cal{A}_\uparrow=-\cal{A}_\downarrow=\cal{A}$.

In the former case
\[
\left[ - A_{1,ij}^t
+ \sigma_y A_{2,ij} \sigma_y \right] =
\left[\left( {\cal A}_{ij} +
{\cal A}^t_{ij}\right) 
+\sigma_z \left(
{\cal A}_{ij} - {\cal A}^t_{ij}\right)
\right],
\]
while in the latter
\[
\left[ - A_{1,ij}^t
+ \sigma_y A_{2,ij} \sigma_y \right] =
\left[\left( {\cal A}_{ij} -
{\cal A}^t_{ij}\right)
+\sigma_z \left(
{\cal A}_{ij} + {\cal A}^t_{ij}\right)
\right].
\]

We therefore see that, if $\cal{A}$ (we assume the same property holds
for $A$) is a symmetric matrix, the
charge current operator is proportional to the identity in spin space 
and
$
-A_{1,ij}+A_{2,ij}= 2{\cal A}_{ij},
$
namely
$
A_{2,ij}=-A_{1,ij}={\cal A}_{ij},
$
while the spin is proportional to $\sigma_z$ and
$
-A_{1,ij}+\sigma_y A_{2,ij}\sigma_y = 2\sigma_z {\cal A}_{ij},
$
implying
$
A_{2,ij}=A_{1,ij}=-\sigma_z {\cal A}_{ij}.
$

In the opposite
case of an antisymmetric $\cal{A}$, the charge current
multiplies $\sigma_z$ and
$
\left[ A_{1,ij}
+ \sigma_y A_{2,ij} \sigma_y \right] =
2\sigma_z {\cal A}_{ij},
$
leading to
$
A_{1,ij}=-A_{2,ij}=\sigma_z {\cal A}_{ij},
$
while the spin is proportional to the identity and
$
\left[ A_{1,ij}
+ \sigma_y A_{2,ij} \sigma_y \right] =
2 {\cal A}_{ij},
$
leading to
$
A_{1,ij}=A_{2,ij}={\cal A}_{ij}.
$

These relations imply that the charge current 
, if ${\cal A}$ is symmetric, is the identity in spin space,
otherwise is proportional to $\sigma_z$. For the spin current, the
opposite occurs. 
Let us now see which are the vector potentials for charge and spin.

\subsection{Current-current correlation function}
Let us suppose to calculate the current-current correlation function
$\langle J(R) J(R^{^{\prime}})\rangle$.
The current in Nambu spinor representation is
\be
\vec{J}(R)=-i\sum_{R_{1}}(\vec{R}-\vec{R}_{1})\bar{\Psi}_{R}{\mathcal{H}}%
_{RR_{1}}\Psi _{R_{1}}=\sum_{R_{1}R_{2}}\bar{\Psi}_{R_{1}}J_{R_{1}R_{2}}(R)%
\Psi _{R_{2}},
\ee
with
\be
J_{R_{1}R_{2}}(R)=-i(\vec{R}_{1}-\vec{R}_{2}){\mathcal{H}}_{R_{1}R_{2}}\delta
_{R_{1}R}.
\ee
In the case of spin current we have ${\mathcal{H}}=( t+i\Delta \tau
_{2}s_{1})\sigma _{z}$, while for the charge current ${\mathcal{H}}= t\tau
_{3}$. The correlation function becomes 
\be
\langle J(R)J(R^{^{\prime }})\rangle =\sum_{R_{1}R_{2}R_{3}R_{4}}\langle
\bar{\Psi}_{R_{1}}J_{R_{1}R_{2}}(R)\Psi _{R_{2}}\bar{\Psi}%
_{R_{3}}J_{R_{3}R_{4}}(R^{^{\prime }})\Psi _{R_{4}}\rangle.
\ee
Let us introduce for simplicity 
multilabels indicating 
replica, hole-particle, spin, energy and position indices. The
correlation function can be rewritten in this way 
\be
\langle \bar{\Psi}^{i}J_{ij}\Psi ^{j}\bar{\Psi}^{l}J_{lm}\Psi ^{m}\rangle
=-J_{ij}\langle \Psi ^{j}\bar{\Psi}^{l}\rangle J_{lm}\langle \Psi ^{m}\bar{%
\Psi}^{i}\rangle +J_{ij}\langle \Psi ^{j}\Psi ^{m}\rangle J_{lm}\langle \bar{%
\Psi}^{l}\bar{\Psi}^{i}\rangle,
\ee
or in matrix language in the following way
\be
\langle \bar{\Psi}J\Psi \bar{\Psi}J\Psi \rangle =-Tr\left( J\langle \Psi
\bar{\Psi}\rangle J\langle \Psi \bar{\Psi}\rangle \right) +Tr\left( J\langle
\Psi \Psi ^{t}\rangle J^{t}\langle \bar{\Psi}^{t}\bar{\Psi}\rangle \right).
\ee
Since
\be
\bar{\Psi}=(C\Psi )^{t},
\ee
with $C=i\tau _{1}\sigma _{y}$, the correlation function becomes
\be
-Tr\left( J\langle \Psi \bar{\Psi}\rangle J\langle \Psi \bar{\Psi}\rangle
+J\langle \Psi \bar{\Psi}\rangle \,CJ^{t}C^{t}\langle \Psi \bar{\Psi}\rangle
\right).
\ee
In terms of single particle Green's functions
\be
-Tr\left( JG(J+CJ^{t}C^{t})G\right).
\ee
Let us define the following quantities, 
for spin 
\be
\label{J^S}
J^S_{R_{1}R_{2}}=-i(R_{1}-R_{2})(t_{R_{1}R_{2}}+i\Delta _{R_{1}R_{2}}\tau
_{2}s_{1}),
\label{Js}
\ee
and for charge
\be
J^C_{R_{1}R_{2}}=-i(R_{1}-R_{2})t_{R_{1}R_{2}},
\label{Jc}
\ee
such that $J=J^S\sigma_z$ for spin and 
$J=J^C\tau_3$ for charge. In both the cases 
the following relation holds
\be
CJ^{t}C^{t}=J,
\ee
therefore the correlation function can be reduced to
\be
-2Tr\left( JGJG\right).
\ee
Introducing a vector potential which is coupled to the current vertex, 
we needs to evaluate 
\be
-Tr\left( J^{K} A G(J^{K} A+C(J^{K} A)^{t}C^{t})G\right),
\label{jagjag}
\ee
where $K=C$ for the charge current vertex and $K=S$ for the spin current 
vertex. 
In the case of charge ($J^K=J^C$), with $A=(A^0s_0 +A^1s_1)\tau_3$ and in the case of spin ($J^K=J^S$), with $A=(A^0s_0 +A^1s_1)\sigma_z$, the following
relation holds 
\be
C(J^{K}A)^{t}C^{t}=J^{K}A.
\ee
Applying to (\ref{jagjag}) the second derivative with respect to $A^0$ and 
to $A^1$, in the following combination
\be
\frac{\partial ^{2}}{\partial {A^{0}}^{2}}\Big|_{A=0}-\frac{\partial
    ^{2}}{\partial {A^{1}}^{2}}\Big|_{A=0}, 
\label{derivA0A1}
\ee 
we obtain in both the cases
\be
\label{proptosigma}
-\left( Tr(JG^{+}JG^{+})-Tr(JG^{-}JG^{+})\right) \propto \sigma,
\ee
which is the charge (if $J=J^C\tau_3$) or the spin 
(if $J=J^S\sigma_z$) conductivity.  
Instead of using (\ref{derivA0A1}), exploiting the $d$-wave symmetry, 
it is straightforward
to show that Eq. (\ref{proptosigma}) comes easily from Eq. 
(\ref{jagjag}) taking
simply $A=(s_0+i\,s_1)\tau_3$ for charge and $A=(s_0+i\,s_1)\sigma_z$ for spin.\\ 

Let us consider in what follows only the spin case, with $J^K=J^S$, 
Eq. (\ref{J^S}), since in the $d$-wave superconductors only spin conductivity is conserved.
If we now use the gauges $A=(A^{0} s_{0}+A^{1} s_{1})\tau_3$ and $A=A^{0} s_{3}\sigma_z+A^{1} s_{2}$ we obtain
\be
C(J_{RR^{^{\prime }}}A)^{t}C^{t}=-i(R-R^{^{\prime }})(t_{RR^{^{\prime
}}}-i\Delta _{RR^{^{\prime }}}\tau _{2}s_{1})A,
\ee
therefore the correlation function (\ref{jagjag}) becomes ($\delta =R-R^{^{\prime }}$)
\be
2Tr\left( \delta \,( t+i\Delta \tau _{2}s_{1}) A\,G\,\delta
\, t\,A\,G\right).
\label{strano}
\ee
Taking advantage of $d$-wave symmetry and using the relations
\begin{eqnarray}
\tau _{0}s_{2}G\tau _{0}s_{2} &=&\tau _{3}s_{1}G\tau _{3}s_{1}, \\
\tau _{0}s_{3}G\tau _{0}s_{3} &=&\tau _{3}s_{0}G\tau _{3}s_{0},
\end{eqnarray}
which come from the following structure of the Green's function at fixed
disorder 
\[
G=[(\tau _{0},\,\tau _{3})\otimes s_{0}+(\tau _{2},\,\tau
_{1})\otimes s_{1}]\otimes (\gamma _{1},\gamma _{2})+i\Sigma [
(\tau _{2},\,\tau
_{1})\otimes s_{2}\otimes \gamma _{3}+(\tau _{3},\,\tau _{0})\otimes s_{3}\otimes \gamma _{0})],
\]
(the terms $\tau _{3}s_{0}$, $\tau _{1}s_{1}$, $\tau _{1}s_{2}$, $\tau
_{3}s_{3}$ disappears if time reversal symmetry is preserved), 
we can find that Eq. (\ref{derivA0A1}) on Eq. (\ref{strano}) gives  
\be
\label{kubo_c}
Tr\left( \delta \, t\tau _{3}s_{0}G\,\delta \, t\tau _{3}s_{0}G\right)
-Tr\left( \delta \, t\tau _{3}s_{1}G\,\delta \, t\tau _{3}s_{1}G\right).
\ee
Reminding that $J^C=-i\delta t$ and $G^-=s_1G^+s_1$, we can recognize Eq. (\ref{kubo_c}) as the Kubo formula for charge conductivity.

We have shown that the gauges $A=\left(A^{0}s_{0}+A^{1}s_{1}\right)\tau_3$ 
and $A=\left(A^{0}s_{3}\sigma_z+A^{1}s_{2}\right)\tau_0$
are equivalent to generate charge conductivity. 
The same calculation can be done to prove the equivalence of 
$A=\left(A^{0}s_{0}+A^{1}s_{1}\right)\tau_0\,\sigma_z$ and 
$A=\left(A^{0}s_{3}+A^{1}s_{2}\sigma_z\right)\tau _{3}$
to produce spin conductivity. 
At the end, the possible choices for the vector potentials,  
taking into account also the replica space through 
a symmetric tensor, 
$\lambda_s
=\frac{1}{\sqrt{2}}(\delta_{a1}\delta_{b2}+\delta_{a2}\delta_{b1})$, 
or an antisymmetric one, $\lambda_a
=\frac{1}{\sqrt{2}}(\delta_{a1}\delta_{b2}-\delta_{a2}\delta_{b1})$, 
are collected in the following table.\\

\begin{tabular}{|c||c|c|c|c|c|c|c|c|}
\hline
&\multicolumn{4}{c}{Charge}\vline&\multicolumn{4}{c}{Spin}\vline\\
\hline
$A^0$ & $\lambda_s\,\tau_3\,s_0\,\sigma_0$ & $\lambda_a \tau_3\,s_0\,\sigma_z$ 
& $\lambda_s\,\tau_0\,s_3\sigma_z$ & $\lambda_a\,\tau_0\,s_3\,\sigma_0$
& $\lambda_s\,\tau_0\,s_0\,\sigma_z$ & $\lambda_a \tau_0\,s_0\,\sigma_0$ 
& $\lambda_s\,\tau_3\,s_3\sigma_0$ & $\lambda_a\,\tau_3\,s_3\,\sigma_z$\\
\hline
$A^1$ & $\lambda_s\,\tau_3\,s_1\,\sigma_0$ &$\lambda_a\,\tau_3\,s_1\,\sigma_z$ 
& $\lambda_s\,\tau_0\,s_2\,\sigma_0$ & $\lambda_a\,\tau_0\,s_2\,\sigma_z$
&$\lambda_s\,\tau_0\,s_1\,\sigma_z$ &$\lambda_a\,\tau_0\,s_1\,\sigma_0$ 
& $\lambda_s\,\tau_3\,s_2\,\sigma_z$ & $\lambda_a\,\tau_3\,s_2\,\sigma_0$\\
\hline
\end{tabular}

\section{The action with vector potentials}

\subsection{Charge vector potential}

Let us introduce a slow varying vector potential $\vec{A}\propto\tau
_{3}$, taking $A_0$ and $A_1$ from the table above. The fields
change in this way 
\bea
&&c_R\longrightarrow e^{i\frac{e}{c}\int_0^R
  \vec{A}_{R^{\prime}}\,d\vec{R^{\prime}}}\,c_R\simeq
e^{i\frac{e}{c}\vec{A}\vec{R}}\,c_R,\\ 
&&c^{\dagger}_R\longrightarrow c^{\dagger}_R e^{-i\frac{e}{c}\int_0^R
  \vec{A}_{R^{\prime}}\,d\vec{R^{\prime}}} \simeq
c^{\dagger}_R\,e^{-i\frac{e}{c}\vec{A}\vec{R}}, 
\eea
since $A$ is a slow varying function, nearly constant on the lattice length.
In the Hamiltonian this transformation is equivalent to changing the hopping
term in this way 
\be
t_{RR^{^{\prime }}}\longrightarrow t_{RR^{^{\prime }}}e^{-i\frac{e}{c}%
\vec{A}\cdot (\vec{R}^{^{\prime }}-\vec{R}%
)}\simeq t_{RR^{^{\prime }}}\left(1-i\frac{e}{c}\vec{\delta }\cdot
\vec{A}-\frac{e^{2}}{2c^{2}}(\vec{\delta }\cdot
\vec{A})^{2}\right),
\label{gaugehopping}
\ee
where  $\vec{\delta }=\vec{R}^{^{\prime }}-%
\vec{R}$\\
The interaction term, before the parametrization by $\Delta$, is unaffected by
this gauge transformation and so  
\be
\Delta_{RR^{^{\prime }}}\longrightarrow \Delta_{RR^{^{\prime }}}.
\label{gaugedelta}
\ee
The expressions in Eq. (\ref{gaugehopping}) and Eq. (\ref{gaugedelta}) 
represent the U$(1)$ symmetry breaking. Now, returning to 
Eq. (\ref{intermedio}), 
we should consider in $\widetilde{T}_{R}H^0_{RR^{^{\prime
    }}}T_{R^{^{\prime}}}$ the transformed hopping term 
\begin{eqnarray*}
\hspace{-0.8cm}&&\widetilde{T}_{R}t_{RR^{^{\prime
    }}}\left(1-i\frac{e}{c}\vec{\delta }%
\cdot \vec{A}-\frac{e^{2}}{2c^{2}}(\vec{\delta }\cdot
\vec{A})^{2}\right)T_{R^{^{\prime }}}^{\dagger }\simeq \\
\hspace{-0.8cm}&&t_{RR^{^{\prime }}}\left(1+T_{R}\vec{\delta }\cdot
\vec{\nabla }T_{R}^{\dagger }+\frac{1}{2}T_{R}(\vec{\delta }%
\cdot \vec{\nabla })^{2}T_{R}^{\dagger
}\right)-\widetilde{T}_{R}t_{RR^{^{\prime }}}\left(i\frac{e}{c}\vec{\delta
}\cdot \vec{A}+\frac{e^{2}}{2c^{2}}(\vec{\delta }\cdot
\vec{A})^{2}\right)T_{R^{^{\prime }}}^{\dagger }= \\ 
\hspace{-0.8cm}&&t_{RR^{^{\prime }}}+t_{RR^{^{\prime }}}\vec{\delta }\cdot
T_{R}%
\vec{\nabla }T_{R}^{\dagger }+\frac{1}{2}\sum_{ij}\delta
_{i}\delta _{j}T_{R}\partial _{ij}T_{R}^{\dagger }-i\frac{e}{c}\widetilde{T}%
_{R}t_{RR^{^{\prime }}}\vec{\delta }\cdot \vec{A}%
T_{R^{^{\prime }}}^{\dagger }
-\frac{e^{2}}{2c^{2}}\widetilde{T}_{R}t_{RR^{^{\prime }}}(%
\vec{\delta }\cdot \vec{A})^{2}T_{R^{^{\prime }}}^{\dagger }.
\end{eqnarray*}
Besides the standard terms 
(the second and the third above) which bring to the 
non-linear $\sigma$-model as seen before, defining $t=t_{1}\gamma
_{1}+t_{2}\gamma _{2}$, $\Delta =\Delta _{1}\gamma _{1}+\Delta 
_{2}\gamma _{2}$ and  
$G=g+i\frac{\Sigma}{E^{2}+\Sigma^{2}} s_{3}$ with $g(k)=-\frac{1}{%
E^{2}+\Sigma^{2}}[(t_{1}-i\Delta _{1}\tau _{2}s_{1})\gamma
_{1}+(t_{2}-i\Delta _{2}\tau _{2}s_{1})\gamma _{2}]$, 
we have the following additional terms

\begin{enumerate}
\item  $i\frac{e}{c}Tr\left(Gt\vec{\delta }\cdot
\vec{A}\right)$,

\item  $i\frac{e}{c}Tr\left(G\widetilde{T}t\vec{\delta }\cdot
\vec{A}\vec{\delta }\cdot \vec{\nabla }%
T^{\dagger }\right)$,

\item  $\frac{e^{2}}{2c^{2}}Tr\left(G\widetilde{T}t(\vec{\delta }\cdot
\vec{A})^{2}T^{\dagger }\right)$,

\item  $\frac{e^{2}}{2c^{2}}Tr\left(G\widetilde{T}t\vec{\delta }\cdot
\vec{A}T^{\dagger }G\widetilde{T}t\vec{\delta }\cdot
\vec{A}T^{\dagger }\right)$,

\item  $i\frac{e}{c}Tr\left(Gt\vec{\delta }\cdot T\vec{%
\nabla }T^{\dagger }G\widetilde{T}t\vec{\delta }\cdot 
\vec{A}T^{\dagger }\right)$.
\end{enumerate}
The first term is zero and using:
\[
\textrm{i)}\;Q=\widetilde{T}^{\dagger }\Sigma s_{3}T,\,\,\,\,\,\,\,\,
\textrm{ii)}\; g=\frac{1}{2}(G^{+}+G^{-}),\,\,\,\,\,\,\,\,
\textrm{iii)}\; \frac{\Sigma}{E^2+\Sigma^2} =\frac{1}{2i}(G^{+}-G^{-}),
\]
vi) Eq. (\ref{sigma}), v) the $d$-wave symmetry, implying that odd terms in  
$\Delta$ are zero under momentum integration and finally vi) the relation
$(t_1\Delta_1+\Delta_2 t_2)^2=(\Delta_1^2+\Delta_2^2)(t_1^2+t_2^2)$, 
due to $t_1\Delta_2=t_2\Delta_1$, 
we obtain, summing all terms and multiplying them for 
$-\frac{1}{2}$ (the coefficient in front of 
$Tr\ln(\varepsilon-H^0+iQ)$), 
the following additional contribution, depending on the vector potential, in the action 
\bea
\nonumber S(A)=\frac{\pi}{32\Sigma^{2}}\sigma_c Tr\left[ \left(
\nabla Q+i\frac{e}{c}[Q,A]\right) \left( \nabla Q^{\dagger }-i\frac{e}{c}[%
A,Q^{\dagger }]\right) -(\nabla Q\nabla Q^{\dagger })\right],
\label{SA}
\eea
with 
\begin{equation}
\label{sigmac}
\sigma_c=\sigma - \frac{\Sigma^2}{\pi V}\sum_{k}\left[ \frac{(\nabla
_{k}\Delta_k )^{2}}{(E^{2}+\Sigma ^{2})^{2}}\right], 
\end{equation}
where $\sigma$ is given by Eq. (\ref{sigma})

\subsubsection{Bare charge conductivity}

Let us suppose to have $A=A^0 s_0+A^1 s_1$, to recover the Kubo formula 
we have to evaluate
\be
\left( \frac{\partial ^{2}ln{\mathcal{Z}}}{{\partial A^{0}}^{2}}-\frac{%
\partial ^{2}ln{\mathcal{Z}}}{{\partial A^{1}}^{2}}\right) \Bigg|_{A=0},
\label{KuboZ}
\ee
with
\[
{\mathcal{Z}}(A)=\int DQe^{-S_{0}-S(A)}.
\]
Since $S(A=0)=0$, the generating function at zero vector potential is again
${\mathcal{Z}}(A=0)={\mathcal{Z}}_{0}=\int DQe^{-S_{0}}$ and, 
introducing the notation $\langle ...\rangle_0$ to denote the quantum average with weight $e^{-S_{0}}$, we have
\be
\frac{\partial ^{2}ln{\mathcal{Z}}}{{\partial A^{\alpha }}^{2}}\Bigg|_{A=0}%
=-\left\langle \frac{\partial S(A)}{\partial A^{\alpha }}\Bigg|%
_{A=0}\right\rangle _{0}^{2}-\left\langle \frac{\partial ^{2}S(A)}{{\partial
A^{\alpha }}^{2}}\Bigg|_{A=0}\right\rangle _{0}+\left\langle \left( \frac{%
\partial S(A)}{\partial A^{\alpha }}\Bigg|_{A=0}\right) ^{2}\right\rangle
_{0}.
\ee
The first term is zero, the second is the average of the operator
\be
\frac{\partial ^{2}S(A)}{{\partial A^{\alpha }}^{2}}\Bigg|_{A=0}= 
\frac{e^{2}\pi%
}{16c^{2}\Sigma ^{2}}\sigma_c Tr\left( [Q(R),\tau _{3}s_{\alpha
}][\tau _{3}s_{\alpha },Q(R)^{\dagger }]\right), 
\label{deriv2SA}
\ee
while the third is the average of the square value of the following operator
\be
\frac{\partial S(A)}{\partial A^{\alpha }}\Bigg|_{A=0}=i\frac{e}{c}%
\left( \frac{\sigma_c \pi}{8\Sigma ^{2}}Tr(\nabla Q(R)\,Q(R)^{\dagger
}\tau _{3}\,s_{\alpha })\right).
\label{derivSA}
\ee
At the saddle point, 
\[
Q(R)=Q_{sp}=\Sigma s_{3},
\]
Eq. (\ref{derivSA}) is zero and the action is simply
\[
S(A)=\frac{\pi e^{2}}{8c^{2}}\left( \sigma_c Tr({A^{1}}^{2})
\right).
\]
The bare conductivity is given by applying Eq. (\ref{KuboZ}) which yields
\be
-\frac{\partial ^{2}S(A)}{{\partial A^{0}}^{2}}\Bigg|_{A=0}+\frac{\partial
^{2}S(A)}{{\partial A^{1}}^{2}}\Bigg|_{A=0}=\frac{8\pi e^{2}}{c^{2}}\sigma_c,
\ee
where $\sigma_c$ is given by Eq. (\ref{sigmac}), more explicitly, 
using Eq. (\ref{sigma}), 
\be
\sigma_c =\frac{\Sigma^2}{\pi V}\sum_{k} \left[ \frac{(\nabla
_{k}\epsilon_k  )^{2}}{(E^{2}+\Sigma ^{2})^{2}}\right]\simeq \frac{1}{\pi^2}
\frac{v_1}{v_2}.
\ee
At the Born level the charge conductivity that we found is in perfect agreement with the diagrammatic approach \cite{Lee2}.

\subsection{Spin vector potential}

Let us now introduce a vector potential of this kind
$\vec{A}\propto\tau_{0}\sigma_z$. The fields change as follows 
\bea
&&c_{R\uparrow}\longrightarrow e^{i\frac{1}{2}\vec{A}\vec{R}}\,c_{R\uparrow},\\
&&c_{R\downarrow}\longrightarrow
e^{-i\frac{1}{2}\vec{A}\vec{R}}\,c_{R\downarrow}, 
\eea
The parameters of the Hamiltonian becomes 
\begin{eqnarray*}
&&t_{RR^{^{\prime }\longrightarrow }}t_{RR^{^{\prime }}}e^{-\frac{i}{2}%
\vec{A}\cdot (\vec{R}^{^{\prime }}-\vec{R}%
)}\simeq t_{RR^{^{\prime }}}\left(1-\frac{i}{2}\vec{\delta }%
\cdot \vec{A}-\frac{1}{4}(\vec{\delta }%
\cdot \vec{A})^{2}\right), \\
&&\Delta _{RR^{^{\prime }\longrightarrow }}\Delta _{RR^{^{\prime
    }}}\frac{1}{2}%
\Big(1+e^{-i\vec{A}\cdot (\vec{R}^{^{\prime }}-%
\vec{R})}\Big)\simeq \Delta _{RR^{^{\prime }}}\left(1-\frac{i}{2%
}\vec{\delta }\cdot \vec{A}-\frac{1}{4}(%
\vec{\delta }\cdot \vec{A})^{2}\right).
\end{eqnarray*}
Although the parameters change differently, the two expansions to second order in $A$ are equal and the additional term to the action due 
to the spin vector potential is 
\be
S(A)=\frac{\pi\sigma_s }{32\Sigma^{2}}Tr\left[ \left( \nabla Q+\frac{i}{2}[Q,A%
]\right) \left( \nabla Q^{\dagger }-\frac{i}{2}[A,Q^{\dagger }]\right)
-(\nabla Q\nabla Q^{\dagger })\right] ,
\ee
where now the bare spin conductivity is $\sigma_s=\sigma$, exactly 
the stiffness of spin fluctuations, given by Eq. (\ref{sigma}). 
\section{Useful formul\AE}

In the presence of chiral symmetry the gaussian propagator becomes
\bea
\nonumber&&\hspace{-0.5cm}\left\langle W^{ab}_{p\,{\cal
      S}\,i\,nm}(k)W^{cd}_{p\,{\cal
      S}\,i\,rq}(-k)\right\rangle=\frac{1}{2}\Big( 1-(-)^p\lambda_n\lambda_m
\Big)\\ 
\nonumber&&\hspace{-0.5cm}
\Big[
(-)^p(\pm)D^p_{nm}(k)\Big(\delta^{ac}_{nr}\delta^{bd}_{mq}(-)^p[\pm] 
\delta^{ad}_{nq}\delta^{bc}_{mr}(-)^i\delta^{ac}_{n-r}
\delta^{bd}_{m-q}(-)^p(-)^i[\pm]\delta^{ad}_{n-q}\delta^{bc}_{m-r}\Big)\\ 
&&\hspace{-0.5cm} +\Pi_{nr}(k)\delta_{p3}\delta_{{\cal
    S}S}\delta_{i,0}\delta^{ab}_{nm}\delta^{cd}_{rq}\Big], 
\eea
where $p=0,3$, depending on the $\gamma$-components, ${\cal S}=S,T$ for singlet or triplet component, $(\pm)$ are
related to real or imaginary matrix elements of 
$W_0$ (the $\gamma_0$-component), listed in Section \ref{sec:RGint_q=0}, 
$[\pm]$ for symmetric or antisymmetric matrix elements 
(always referring to $W_0$, the sign $(-)^p$ takes care of the sign differences, when they occur, between 
the two components in sublattice space, $W_0$ and $W_3$), 
$(-)^i$ the sign that $W$ acquires changing the signs of Matsubara frequencies
(this occurs only for modes proportional to $\tau_1$ and $\tau_3$), and
finally 
\bea
&&D^0_{nm}(k)=\frac{1}{4\pi\nu}\frac{1}{Dk^2+|\epsilon_n-\epsilon_m|},
\,\,\,\;\,\,\textrm{with}\,\,\,\lambda_n=-\lambda_m,\\ 
&&D^3_{nm}(k)=\frac{1}{4\pi\nu}\frac{1}{Dk^2+|\epsilon_n+\epsilon_m|},\,\,\,\;
\,\,\textrm{with}\,\,\,\lambda_n=\lambda_m,\\ 
&&\Pi_{n,r}(k)= D^3_{nn}(k) \frac{\Pi k^2}{2\nu(D k^2 +2|\epsilon_r|)},\,\,\,\,
\,\,\,\textrm{in the $0$ replica limit.} 
\eea

Using the charge conjugation of the unitary transformation $U$ 
\be
\tau_1 \sigma_y U^t \tau_1 \sigma_y=\widetilde{U}^{\dagger},\\
\ee
we can derive the following relation
\be
Tr \left(U_{m_{1}m_{2}}^{ab}\tau_l \sigma \widetilde{U}_{m_{3}m_{4}}^{\dagger
    cd} \tau_i \sigma_j\right)=  
\eta \,[\pm]\,Tr
\left(U_{m_{4}m_{3}}^{dc}\tau_l \sigma \widetilde{U}_{m_{2}m_{1}}^{\dagger ba}
  \tau_i \sigma_j\right) ,
\ee
with
\begin{displaymath}
\eta=\left\{ \ba{rl}
-(-)^l& \textrm{in p-h singlet channel, }\,(l=0,3,\, \sigma=\sigma_0),\\
(-)^l& \textrm{in p-h triplet channel, }\,(l=0,3,\, \sigma=\vec{\sigma}),\\
-\phantom{)^l}& \textrm{in p-p Cooper channel, }\,(l=1,2, \, \sigma=\sigma_0),
\ea \right.
\end{displaymath}
where $[\pm]$ refers to the symmetric or antisymmetric $\tau_i$-$\sigma_j$-component of $W$, with $i=0,1,2,3$, $j=0,x,y,z$ (the same sign, $[\pm]$, 
which appear in the corresponding gaussian propagator).

For the following derivative operator 
$A=\nabla \widetilde{U}\widetilde{U}^{\dagger}$, 
whose charge conjugation condition is
\be
\tau_1 \sigma_y A^t \tau_1 \sigma_y= -\gamma_1 A \gamma_1,
\ee
where $\gamma_1$ is the first Pauli matrix on sublattice space, we can derive 
the following property, under the trace, 
\be
Tr\left(\tau_i\sigma_j\gamma_q\tau_{i^{\prime}}\sigma_{j^{\prime}}\gamma_p
  A_{nm}^{ab}\right) 
=-(-)^p (-)^q [\pm]_{ij}[\pm]_{i^{\prime}j^{\prime}}
Tr\left(A_{mn}^{ba}\tau_{i^{\prime}}\sigma_{j^{\prime}}\gamma_p
  \tau_i\sigma_j\gamma_q\right), 
\ee
useful to evaluate $\langle S_2 S^{1}_{int} \rangle$ and
$\frac{1}{2}\langle S_2 S_2 S^{1}_{int} \rangle$, 
where $\gamma_q$ and $\gamma_p$ can be $\gamma_0$ and $\gamma_3$, matrices in sublattice
space, while $\sigma_j$ and $\sigma_{j^{\prime}}$ are identities or Pauli
matrices in spin space.

Defining the quaternions $\bar{\tau}_i=\tau_0, i\tau_1, i\tau_2, i\tau_3$ and
$\bar{\sigma}_j=\sigma_0, i\sigma_x, i\sigma_y, i\sigma_z$, we have also this
sum rule 
\be
\sum_{i,j}(\pm)_{ij}[\pm]_{ij}Tr\left(M
  \bar{\tau}_i\bar{\sigma}_j\right)Tr\left(N
  \bar{\tau}_i\bar{\sigma}_j\right)=-4Tr(MN). 
\ee
where $M$ and $N$ are generic $4\times 4$ matrices. 

Applying the relations 
seen above, we can obtain, for instance, the generic expression 
for the mean value
of $S_{int}^{1}$ in the presence of chiral symmetry
\bea
&&\nonumber\left\langle S_{int}^{1}\right\rangle=-{\frac{\pi^2\nu}{8}}\sum
\nu\Gamma^{q} D_{m_{1}m_{2}}^{k_{1}}
\Big(1-(-)^{k_{1}}\lambda 
_{m_{1}}\lambda _{m_{2}}\Big)\,\delta ({n_{1}}\mp {n_{2}}\pm
{n_{3}}-{n_{4}})\\ 
\nonumber&&\phantom{\left\langle S_{int}^{1}\right\rangle} \Big\{  Tr
\Big(\lambda _{m_{2}} tr (U_{m_{2}n_{2}}^{gd}\tau _{l}\sigma 
\widetilde{U}_{n_{1}m_{1}}^{\dagger de}\gamma _{k_{1}}\gamma _{q})\lambda
_{m_{1}} tr (U_{m_{1}n_{4}}^{ed}\tau _l\sigma %
\widetilde{U}_{n_{3}m_{2}}^{\dagger dg}\gamma _{k_{1}}\gamma
_{q})\Big)\\
\nonumber&&\phantom{\left\langle S_{int}^{1}\right\rangle} -(-)^l  Tr \Big(\lambda _{m_{2}} tr (U_{m_{2}n_{2}}^{gd}\tau _{l}\sigma 
\widetilde{U}_{n_{1}m_{1}}^{\dagger de}\gamma _{k_{1}}\gamma _{q})\lambda
_{m_{1}} tr (U_{m_{1}-n_{4}}^{ed}\tau _{l}\sigma %
\widetilde{U}_{-n_{3}m_{2}}^{\dagger dg}\gamma _{k_{1}}\gamma _{p})\Big)\Big\},
\eea
where we have dropped, for simplicity, the momentum dependences and 
where $ Tr $ means trace 
over all degrees of freedom except in sublattice space over which we shall
trace by $ tr $.

\end{document}